\documentclass[reqno, a4paper]{amsart}
\usepackage{amsmath}
\usepackage{amssymb}
\usepackage{amsthm}

\usepackage[scale=0.9]{geometry}

\usepackage{natbib}
\usepackage{bibentry} 

\usepackage[english]{babel}
\usepackage[utf8]{inputenc}

\usepackage{subfig}
\usepackage{graphicx}
\usepackage{color}

\usepackage[unicode]{hyperref}
\usepackage[active]{srcltx} 

\usepackage{mathbbol}
\usepackage{bm} 
\usepackage{MnSymbol} 
\usepackage{gensymb}
\usepackage{eurosym}

\usepackage{units}
\usepackage{tensor}
\usepackage{accents}

\usepackage{enumitem}

\usepackage{lineno}

\addtocontents{toc}{\protect\begin{multicols}{2}} 
\usepackage{multicol}

\usepackage{placeins}

\usepackage{todonotes}

\usepackage{xcolor}
\usepackage{mdframed}
\usepackage{newfloat}

\usepackage{booktabs}

\DeclareMathOperator{\divergence}{div}

\DeclareMathOperator{\Tr}{Tr}

\newcommand{\reference}{\mathrm{ref}}

\newcommand{\bydefinition}{\mathrm{def}}
\newcommand{\traceless}[1]{{#1}_{\delta}}

\newcommand{\diff}{\mathrm{d}}

\renewcommand{\vec}[1]{\ensuremath{\mathbf{#1}}}
\newcommand{\greekvec}[1]{\ensuremath{\boldsymbol{#1}}}
\makeatletter
\@ifpackageloaded{bm}%
{\renewcommand{\vec}[1]{\ensuremath{\bm{#1}}}%
\renewcommand{\greekvec}[1]{\ensuremath{\bm{#1}}}%
}{%
\relax
}
\makeatother
\newcommand{\tensorq}[1]{\ensuremath{\mathbb{#1}}}      

\newcommand{\transpose}[1]{#1^\top}
\newcommand{\transposei}[1]{#1^{-\top}}
\newcommand{\inverse}[1]{#1^{-1}}

\newcommand{\identity}{\ensuremath{\tensorq{I}}}

\newcommand{\cstress}{\tensorq{T}}

\newcommand{\fpstress}{\tensorq{T}_{\mathrm{R}}}

\newcommand{\deformation}{\greekvec{\chi}}

\newcommand{\fgrad}{\tensorq{F}}

\newcommand{\detfgrad}{J}

\newcommand{\rcg}{\tensorq{C}}

\newcommand{\lcg}{\tensorq{B}}

\newcommand{\lcgb}{\overline{\lcg}} 

\makeatletter
\@ifpackageloaded{bm}%
{%
}{%

}

\@ifpackageloaded{bm}%
{%
 
}{%

}

\@ifpackageloaded{bm}%
{%
}{%

}
\makeatother

\newcommand{\generictensor}{{\tensorq{A}}}

\newcommand{\vecv}{\ensuremath{\vec{v}}}

\newcommand{\gradsym}{\ensuremath{\tensorq{D}}}
\newcommand{\dgradsymsymb}{\ensuremath{\gradsym_{\delta}}}
\newcommand{\gradvl}{\ensuremath{\tensorq{L}}}


\newcommand{\Young}{\mathrm{E}}
\newcommand{\Poisson}{\mathrm{\nu}}

\newcommand{\ienergy}{\ensuremath{e}} 
\newcommand{\fenergy}{\ensuremath{\psi}} 
\newcommand{\entropy}{\ensuremath{\eta}} 

\newcommand{\temp}{\ensuremath{\theta}} 
\newcommand{\Temp}{\ensuremath{\Theta}} 

\newcommand{\mns}{\ensuremath{m}} 

\newcommand{\cheatvolref}{\ensuremath{c_{\mathrm{V}, \reference}}} 

\newcommand{\hfluxc}{\vec{j}_{q}}     
\newcommand{\hflux}{\vec{J}_{q}}     

\newcommand{\entfluxc}{\vec{j}_{\entropy}} 

\newcommand{\entprodc}{\xi} 
\newcommand{\entprodctemp}{\zeta} 

\newcommand{\pd}[2]{\ensuremath{\frac{\partial {#1}}{\partial {#2}}}}
\newcommand{\ppd}[2]{\ensuremath{\frac{\partial^2 {#1}}{\partial {#2^2}}}}
\newcommand{\dd}[2]{\ensuremath{\frac{\diff {#1}}{\diff {#2}}}}

\newcommand{\ddd}[2]{\ensuremath{\frac{\diff^2 {#1}}{\diff {#2}^2}}}

\makeatletter
\@ifpackageloaded{tensor}
{

}{%

}
\makeatother

\makeatletter
\@ifpackageloaded{tensor}
{

}{%

}
\makeatother

\newcommand{\absnorm}[1]{\ensuremath{\left|#1\right|}}

\makeatletter
\@ifundefined{volume}{%
}%
{%
}
\makeatother

\makeatletter
\@ifpackageloaded{MnSymbol} 
{
\newcommand{\tensordot}[2]{\ensuremath{#1 \vdotdot #2}} 
}{%
\newcommand{\tensordot}[2]{\ensuremath{#1 : #2}} 
}
\makeatother

\newcommand{\vectordot}[2]{\ensuremath{#1 \bullet #2}}

\DeclareFloatingEnvironment[fileext=sum,placement={!ht},name=Summary]{summaryflt}

\newenvironment{summary}[1][]
    {
    \begin{summaryflt}[tb]
        \begin{summarycont}[#1] 
    }
    {
        \end{summarycont}
        \end{summaryflt}
    }

\mdfdefinestyle{summarystyle}{%
linecolor=black,linewidth=3pt,%
frametitlerule=true,%
frametitlebackgroundcolor=gray!20,
innertopmargin=\topskip,
}

\mdtheorem[style=summarystyle]{summarycont}{Summary}

\newmdenv[style=scholionstyle]{scholion}
\mdfdefinestyle{scholionstyle}{
  skipabove=1em,
  skipbelow=1em,
  backgroundcolor=gray!10,
  roundcorner=10pt
}

\numberwithin{equation}{section}

\title[Heat generation at the tip of a cutout]{Temperature field and heat generation at the tip of a cutout in a viscoelastic solid body undergoing loading}

\author{V\'{\i}t Pr\r{u}\v{s}a}
\date{\today}
\address{
Faculty of Mathematics and Physics\\
Charles University\\
Sokolovsk\'a 83\\
Praha 8 -- Karl\'{\i}n\\
CZ 186\;75\\
Czech Republic
}
\email[Corresponding author]{prusv@karlin.mff.cuni.cz}

\author{Karel T\r{u}ma}
\address{
Faculty of Mathematics and Physics\\
Charles University\\
Sokolovsk\'a 83\\
Praha 8 -- Karl\'{\i}n\\
CZ 186\;75\\
Czech Republic
}
\email{ktuma@karlin.mff.cuni.cz}

\thanks{V\'{\i}t Pr\r{u}\v{s}a and Karel T\r{u}ma thank the Czech Science Foundation, grant number 18-12719S, for its support.}

\begin{document}

\begin{abstract}
Using the finite element method we quantitatively analyse temperature field evolution in a viscoelastic solid undergoing a loading--unloading process. In particular we investigate the temperature field inside a Kelvin--Voigt type viscoelastic body with a thin cutout. We find that the viscosity significantly contributes to the temperature field changes, and that the temperature field changes initiated by the loading--unloading process are strongly concentrated at the tip of the thin cutout. The predicted temperature field qualitatively corresponds to the temperature field observed in experiments focused on simultaneous heat and strain measurements at the crack tip inside materials such as the filled rubber.


\end{abstract}

\maketitle

\tableofcontents



\section{Introduction}
\label{sec:introduction}
Rubber is perceived to be an epitome of an elastic material, hence most mathematical models for its behaviour are developed in such a way that they predict zero entropy production due to mechanical processes, see for example a recent overview by~\cite{destrade.m.saccomandi.g.ea:methodical} and also a non-traditional approach recently discussed by~\cite{muliana.a.rajagopal.kr.ea:determining}. It however turns out that inelastic (entropy producing) processes are of importance as well especially for particle-reinforced rubbers, and various models for such an inelastic behaviour have been proposed, see for example~\cite{ogden.rw.roxburgh.dg:pseudo-elastic}, \cite{dorfmann.a.ogden.rw:constitutive}, \cite{wineman.a:nonlinear} or \cite{rajagopal.kr.srinivasa.ar:implicit*1}. Although some of these models have reached high degree of complexity, see for example~\cite{devendiran.vk.mohankumar.kv.ea:thermodynamically,devendiran.vk.mohankumar.kv.ea:validation}, and they describe very well the mechanical inelastic response, the available models largely do not deal with the associated thermal effects. Furthermore, even if the thermal effects are taken into account the corresponding models are rarely used especially in the context numerical simulations of \emph{non-homogeneous finite deformations} and \emph{spatially non-uniform temperature distributions}.

Since the inelastic response is by definition associated with the entropy production due to mechanical processes (dissipation), such a response is likely to leave a thermal signature, that eventually accompanies or even superimposes the well-known Gough--Joule effect in purely elastic materials, see for example~\cite{treloar.lrg:physics}. In other words, the inelastic deformation of material should lead to the heat generation, and hence in general to a spatially non-uniform temperature field in the material. The corresponding thermal signature might be small, but if it can be measured, it could be used for validation of the proposed model for the inelastic response; the mathematical model for the inelastic response must be able to correctly predict both the mechanical and thermal response.

Interestingly, recently developed experimental techniques allow one to simultaneously measure---with a sufficient spatio-temporal resolution and accuracy---both strain and temperature fields in a material, see for example~\cite{toussaint.e.balandraud.x.ea:combining}, \cite{martinez.jrs.cam.jbl.ea:filler}, \cite{martinez.jrs.toussaint.e.ea:heat}, \cite{wang.xg.liu.ch.ea:simultaneous}, \cite{di-cesare.n.corvec.g.ea:tearing} and~\cite{charles.s.le-cam.j:inverse}. (For the early developments see also~\cite{chrysochoos.a.louche.h:infrared}, \cite{boulanger.t.chrysochoos.a.ea:calorimetric} and \cite{chrysochoos.a:infrared}.) In particular, the experimental data reported by~\cite{martinez.jrs.toussaint.e.ea:heat} clearly quantify the heat generated at the crack tip of filled rubber under cyclic loadings. The question is whether the available models for inelastic response have the capability to explain, at least qualitatively, such combined thermo-mechanical experimental data.

\begin{figure}[h]
  \centering
  \subfloat[Map of temperature changes with respect to a background temperature, maximum global stretch ratio.\label{fig:martinez-a}]{\includegraphics[width=0.3\textwidth]{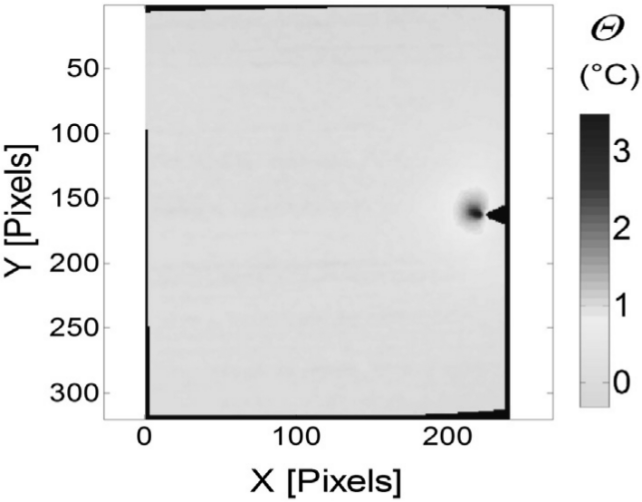}}
  \qquad
  \subfloat[Sample geometry, front and side view.\label{fig:martinez-c}]{\includegraphics[width=0.5\textwidth]{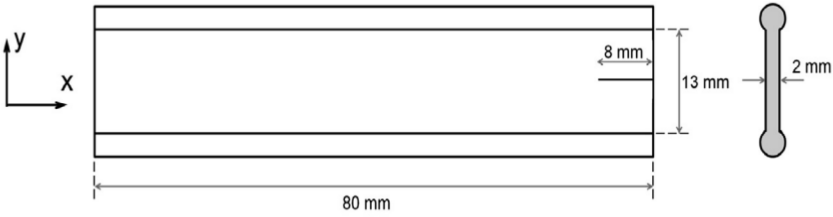}}
  \caption{Measured temperature field and sample geometry; black-and-white conversions of Figure~1 and Figure~7 from~\cite{martinez.jrs.toussaint.e.ea:heat}. (Reprinted from~\cite{martinez.jrs.toussaint.e.ea:heat} with permission from Elsevier.)}
  \label{fig:martinez}
\end{figure}

We investigate this question for a \emph{viscoelastic rate-type model}. Namely we consider a variant of the classical Kelvin--Voigt model, see for example~\cite{wineman.as.rajagopal.kr:mechanical}, that is formulated for finite deformations, and that allows one to work with temperature dependent material coefficients. We show that the model is consistent with the first and second law of thermodynamics, and then we perform numerical simulations of the material response in a setting that resembles the experimental setting used by~\cite{martinez.jrs.toussaint.e.ea:heat}. In this experiment, the authors have measured the temperature and deformation fields in a cracked specimen, see Figure~\ref{fig:martinez-c}, made of carbon black filled styrene butadiene rubber, while the specimen has been subject to cyclic loading. The objective is to check whether the model can predict the increase of the temperature in the vicinity of the crack tip (sharp thin cutout), see Figure~\ref{fig:martinez-a}. 

The paper is organised as follows. In Section~\ref{sec:model} we derive the generalised Kelvin--Voigt type model with temperature dependent material parameters, and then in Section~\ref{sec:governing-equations} we formulate the corresponding governing equations. 
Next we proceed with numerical simulations regarding the cracked specimen problem. In Section~\ref{sec:problem-sescription} we describe the problem geometry and parameter values for the numerical simulations reported in Section~\ref{sec:numerical-simulation}.

\section{Model}
\label{sec:model}
Regarding the mathematical model for a rubber-like material, we opt for a generalisation of the classical Kelvin--Voigt model for one-dimensional response of viscoelastic solids. This model is based on a spring--dashpot analogue, see for example~\cite{wineman.as.rajagopal.kr:mechanical}, wherein a spring and a dashpot are connected in parallel. We generalise this model to the setting of finite deformations, and we consider a variant with temperature dependent material parameters. (For various other approaches to the generalisation of the Kelvin--Voigt model see for example~\cite{rajagopal.kr:note} or~\cite{bulcek.m.malek.j.ea:on*3}, \cite{bulcek.m.kaplicky.p.ea:on} and~\cite{sengul.y:viscoelasticity}, \cite{erbay.ha.sengul.y:thermodynamically} and references therein. In these works the interested reader also finds references to available mathematical results regarding the viscoelastic solids. In this respect the reader is also referred to~\cite{kruzk.m.roubcek.t:mathematical}.) The derivation of the model is done in the Eulerian description, and at the end we convert the corresponding system of governing equations to the Lagrangian description, which is in the current case more convenient for a computational treatment.

The derivation of the model in principle follows the approach developed by~\cite{rajagopal.kr.srinivasa.ar:thermodynamic} in the context of viscoelastic fluids, see also~\cite{malek.j.rajagopal.kr.ea:on}, \cite{hron.j.milos.v.ea:on}, \cite{malek.j.rajagopal.kr.ea:derivation}, \cite{malek.j.prusa.v:derivation} or~\cite{rehor.m.gansen.a.ea:comparison}. This approach is based on the assumption that \emph{each material is characterised by its energy storage ability and entropy production ability}.

\subsection{Energy storage and entropy production mechanisms}
\label{sec:energy-stor-entr}
The energy storage ability of the material is characterised by the choice of the Helmholtz free energy function. (Other thermodynamic potentials can be however used as well, see for example~\cite{rajagopal.kr.srinivasa.ar:gibbs-potential-based}, \cite{narayan.spa.little.dn.ea:nonlinear}, \cite{gokulnath.c.saravanan.u.ea:representations} or \cite{prusa.v.rajagopal.kr.ea:gibbs}.) We assume that the specific Helmholtz free energy $\fenergy$ is given as
\begin{subequations}
  \label{eq:fenergy-ansatz}
  \begin{equation}
  \label{eq:fenergy-ansatz-mechanical}
  \fenergy(\temp, \detfgrad, \lcgb) =_{\bydefinition}
  \tilde{\fenergy}(\temp)
  +
  \frac{K(\temp)}{\rho_R\beta^2}\left(\beta \ln{\detfgrad}+ \detfgrad^{-\beta}-1\right)
  +
  \frac{\mu_1(\temp)}{2\rho_R}\left(\Tr\lcgb-3\right).
\end{equation}
(Note that we use the \emph{specific} Helmholtz free energy that is the Helmholtz free energy per unit mass, $[\fenergy] = \unitfrac{J}{kg}$.) The formula for the specific Helmholtz free energy coincides---in the isothermal case---with a frequently used model for slightly compressible rubber-like materials introduced by~\cite{ogden.rw:large}. If needed other popular models for slightly compressible solids might be easily used here as well. A list of models for slightly compressible solids is given for example in~\cite{horgan.co.saccomandi.g:constitutive}, and regarding various choices for the Helmholtz free energy for elastic solids the reader is also referred to~\cite{destrade.m.saccomandi.g.ea:methodical}.

The notation used in~\eqref{eq:fenergy-ansatz-mechanical} is the standard one. The symbol $\temp$ in~\eqref{eq:fenergy-ansatz-mechanical} denotes the thermodynamic temperature, $\rho_R$ denotes the density in the reference configuration, and
\begin{equation}
  \label{eq:45}
\detfgrad =_{\bydefinition} \det \fgrad,\quad \lcgb=_{\bydefinition} \frac{\lcg}{ \detfgrad^{\frac{2}{3}}},
\end{equation}
denote the determinant of the deformation gradient $\fgrad$ and the rescaled left Cauchy--Green tensor respectively, $\lcg =_{\bydefinition} \fgrad \transpose{\fgrad}$. See for example~\cite{horgan.co.saccomandi.g:constitutive} for the rationale behind the use of the rescaled left Cauchy--Green tensor. (We note that using the exponent $\frac{2}{3}$ we implicitly assume that we are working in a three-dimensional setting.)

Unlike in the theory of hyperelastic solids (Green elastic solids) we also explicitly include the purely thermal part of the Helmholtz free energy~$\tilde{\fenergy}(\temp)$, which takes the standard form
\begin{equation}
  \label{eq:fenergy-thermal}
  \tilde{\fenergy}(\temp)
  =_{\bydefinition}
  -\cheatvolref \temp \left[ \ln \left( \frac{\temp}{\temp_{\reference}} \right) - 1 \right].
\end{equation}
\end{subequations}
This choice would, in the absence of other terms in~\eqref{eq:fenergy-ansatz-mechanical}, lead to a material with a constant specific heat at constant volume~$\cheatvolref$. (Symbol $\temp_{\reference}$ denotes a reference temperature that has no effect on the dynamics.) Functions $K(\temp)$ and $\mu_1(\temp)$ in~\eqref{eq:fenergy-ansatz-mechanical} represent the temperature dependent bulk and shear moduli respectively, while $\beta$ is a (dimensionless) exponent. Particular formulae for functions~$K(\temp)$ and $\mu_1(\temp)$ are given later is Section~\ref{sec:problem-sescription}. Once the Helmholtz free energy is specified, the ``elastic'' properties of the given visco-\emph{elastic} material are fixed.


Regarding the entropy production $\entprodc$, which is the other fundamental characterisation of the given material, we assume that
\begin{subequations}
  \label{eq:entprod-ansatz}
  \begin{align}
    \label{eq:3}
    \entprodc
    &=_{\bydefinition}
      \frac{1}{\temp}
      \entprodctemp_{\mathrm{mech}}
      +
      \kappa (\temp)\frac{\vectordot{\nabla \temp}{\nabla \temp}}{\temp^2}
      ,
    \\
    \label{eq:5}
    \entprodctemp_{\mathrm{mech}}
    &=_{\bydefinition}
      2\nu(\temp)\absnorm{\dgradsymsymb}^2
      +
      \frac{2\nu(\temp)+3\lambda(\temp)}{3}(\divergence\vecv)^2,
  \end{align}
\end{subequations}
where $\nu(\temp)$ and $\lambda(\temp)$ are temperature dependent viscosities, $\kappa(\temp)$ is the temperature dependent thermal conductivity, $\gradsym =_{\bydefinition} \frac{1}{2} \left( \nabla \vec{v} + \transpose{\left(\nabla \vec{v}\right)} \right)$ denotes the symmetric part of the velocity gradient, $\traceless{\generictensor} =_{\bydefinition} \generictensor - \frac{1}{3} \left(\Tr \generictensor \right) \identity$ denotes the traceless part (deviatoric part) of the corresponding tensor, and $\absnorm{\generictensor}$ denotes the standard Frobenius matrix norm. (We note that using the factor $\frac{1}{3}$ in the formula for the deviatoric part we implicitly assume that we are working in a three-dimensional setting.) Particular formulae for the functions~$\nu(\temp)$, $\lambda(\temp)$ and $\kappa(\temp)$ are again given later is Section~\ref{sec:problem-sescription}.

Clearly, if the functions $\nu(\temp)$, $\lambda(\temp)$ and $\kappa(\temp)$ are non-negative, then the entropy production is non-negative and the second law of thermodynamics is automatically satisfied. (We recall that the entropy production $\entprodc$ is the source term in the evolution equation for the entropy, that is $\rho \dd{\entropy}{t} + \divergence \entfluxc = \entprodc$, where $\entropy$ denotes the specific entropy, $\entfluxc$ denotes the entropy flux and $\dd{}{t}$ denotes the material time derivative.) The choice of the entropy production in the form~\eqref{eq:entprod-ansatz} means that the material dissipates the mechanical energy in the same manner as a compressible viscous fluid, and that the heat conduction contributes to the entropy production as in a material that obeys the Fourier law of heat conduction with a temperature dependent heat conductivity, see for example~\cite{malek.j.prusa.v:derivation} for details. The choice of the entropy production specifies the ``viscous'' properties of the given \emph{visco}-elastic material.

Following~\cite{rajagopal.kr.srinivasa.ar:on*7} we briefly show that the specification of the two \emph{scalar functions}~\eqref{eq:fenergy-ansatz} and~\eqref{eq:entprod-ansatz} is indeed sufficient to find constitutive relations for the Cauchy stress \emph{tensor} and the heat flux \emph{vector}. The general evolution equation for the specific internal energy $\ienergy$ of a continuous medium in the Eulerian description, see for example~\cite{truesdell.c.noll.w:non-linear*1}, reads
\begin{equation}
  \label{eq:1}
  \rho \dd{\ienergy}{t} = \tensordot{\cstress}{\gradsym} - \divergence \hfluxc,
\end{equation}
where $\rho$ denotes the density in the current configuration, $\dd{}{t}=_{\bydefinition} \pd{}{t} + \vectordot{\vecv}{\nabla}$ denotes the material time derivative, $\cstress$ denotes the Cauchy stress tensor, $\hfluxc$ denotes the heat flux, and the symbol $\tensordot{\generictensor_1}{\generictensor_2}=_{\bydefinition} \Tr \left(\generictensor_1 \transpose{ \generictensor_2}\right)$ denotes the standard Frobenius inner product on the space of matrices. Taking into account the standard relation between the specific Helmholtz free energy $\fenergy$, the temperature $\temp$, the specific entropy $\entropy$ and the specific internal energy $\ienergy$, 
\begin{equation}
  \label{eq:2}
  \fenergy(\temp, \detfgrad, \lcgb) = \left. \left[ \ienergy(\entropy, \detfgrad, \lcgb) - \temp \entropy \right] \right|_{\entropy = \entropy( \temp, \detfgrad, \lcgb)},
\end{equation}
and the standard definition of the thermodynamic temperature $\temp =_{\bydefinition} \pd{\ienergy}{\entropy}(\entropy, \detfgrad, \lcgb)$, we see that the application of the chain rule in~\eqref{eq:1} in fact gives us an evolution equation for the specific entropy
\begin{equation}
  \label{eq:4}
  \rho \temp \dd{\entropy}{t}
  +
  \divergence\hfluxc
  =
  \tensordot{
    \cstress
  }
  {
    \gradsym
  }
  -
  \rho
  \pd{\fenergy}{\detfgrad}\dd{\detfgrad}{t}
  -
  \rho
  \tensordot{
    \pd{\fenergy}{\lcgb}
  }
  {
    \dd{\lcgb}{t}
  }
  .
\end{equation}
(We have also used the formula $\entropy = - \pd{\fenergy}{\temp}(\temp, \detfgrad, \lcgb)$ linking the specific entropy and the specific Helmholtz free energy.) Now we use kinematic identities $\dd{\detfgrad}{t} = \detfgrad \divergence \vecv$ and $\dd{\fgrad}{t} = \gradvl \fgrad$, where $\gradvl=_{\bydefinition} \nabla \vecv$ denotes the gradient of the Eulerian velocity field $\vecv$, and we compute the material time derivative of $\lcgb$, 
\begin{equation}
  \label{eq:6}
  \dd{\lcgb}{t}
  =
  \dd{}{t}
  \left(
    \frac{\lcg}{\detfgrad^{\frac{2}{3}}}
  \right)
  =
  -\frac{2}{3}
  \lcgb
  \divergence \vec{v}
  +
  \gradvl
  \lcgb
  +
  \lcgb
  \transpose{\gradvl}
  .
\end{equation}
Using~\eqref{eq:6} in~\eqref{eq:4} yields---after some algebraic manipulation---the equation
\begin{equation}
  \label{eq:7}
  \rho \temp \dd{\entropy}{t}
  +
  \divergence\hfluxc
  =
  \tensordot{
    \cstress
  }
  {
    \gradsym
  }
  -
  \rho
  \pd{\fenergy}{\detfgrad} \detfgrad \divergence{\vec{v}}
  +
  \frac{2}{3}
  \rho
  \tensordot{
    \pd{\fenergy}{\lcgb}
  }
  {
    \lcgb
  }
  \divergence \vec{v}
  -
  2
  \rho
  \tensordot{
    \pd{\fenergy}{\lcgb}
    \lcgb
  }
  {
    \gradsym
  }
  ,
\end{equation}
where we have exploited the cyclic property of the trace and the fact that $\pd{\fenergy}{\lcgb}$ commutes with $\lcgb$. Now we split the Cauchy stress tensor to its deviatoric and spherical part $\cstress =_{\bydefinition} \mns \identity + \traceless{\cstress}$, where $\mns =_{\bydefinition} \frac{1}{3} \left(\Tr \cstress\right) \identity$, and we further manipulate~\eqref{eq:7} into the form
\begin{equation}
  \label{eq:8}
  \rho \temp \dd{\entropy}{t}
  +
  \divergence\hfluxc
  =
  \tensordot{
    \left[
      \traceless{\cstress}
      -
      2
      \rho
      \traceless{
        \left(
          \pd{\fenergy}{\lcgb}
          \lcgb
        \right)
      }
    \right]
  }
  {
    \traceless{\gradsym}
  }
  +
  \left[
    \mns
    -
    \rho_R
    \pd{\fenergy}{\detfgrad}
  \right]
  \divergence \vec{v}
  ,
\end{equation}
where we have also used the balance of mass $\rho_R = \rho \detfgrad.$ Finally, we manipulate the flux term, and we get the sought evolution equation for the entropy,
\begin{equation}
  \label{eq:10}
  \rho\dd{\entropy}{t}
  +
  \divergence \left(\frac{\hfluxc}{\temp}\right)
  =
  \frac{1}{\temp}
  \left\{
    \tensordot{
      \left[
        \traceless{\cstress}
        -
        2
        \rho
        \traceless{
          \left(
            \pd{\fenergy}{\lcgb}
            \lcgb
          \right)
        }
      \right]
    }
    {
      \traceless{\gradsym}
    }
    +
    \left[
      \mns
      -
      \rho_R
      \pd{\fenergy}{\detfgrad}
    \right]
    \divergence \vec{v}
  \right\}
  -
  \frac{\vectordot{\hfluxc}{\nabla \temp}}{\temp^2}
  .
\end{equation}
This equation has the desired flux--production structure
\begin{equation}
  \label{eq:9}
  \rho \dd{\entropy}{t} + \divergence \entfluxc = \entprodc,
\end{equation}
where $\entfluxc$ denotes the entropy flux and $\entprodc$ denotes the entropy production. Equation~\eqref{eq:10} reveals the structure of the entropy production $\entprodc$ that is implied by the chosen formula for the Helmholtz free energy.

\subsection{Constitutive relations for the heat flux and the Cauchy stress tensor}
\label{sec:const-relat-heat}
The constitutive relations for the heat flux $\hfluxc$ and the Cauchy stress tensor $\cstress$ are then---in principle---determined by comparison of the entropy production implied by~\eqref{eq:10} and the desired entropy production~\eqref{eq:entprod-ansatz}. (The actual procedure might be more involved, see~\cite{rajagopal.kr.srinivasa.ar:on*7}, but we, for the sake of clarity of the presentation, opt for a simplified argument.) The comparison of the right-hand side of~\eqref{eq:10} with the desired entropy production~\eqref{eq:entprod-ansatz} yields
\begin{subequations}
  \label{eq:constitutive-relations}
  \begin{align}
    \label{eq:12}
    \hfluxc
    &=
      - \kappa(\temp) \nabla \temp, \\
    \label{eq:13}
    \mns
    &=
      \rho_R\pd{\fenergy}{\detfgrad}
      +
      \frac{2\nu(\temp)+3\lambda(\temp)}{3}\divergence\vecv,
    \\
    \label{eq:14}
    \traceless{\cstress}
    &=
      2
      \rho
      \traceless{
      \left(
      \pd{\fenergy}{\lcgb}
      \lcgb
      \right)
      }
      +
      2\nu(\temp) \dgradsymsymb
      ,
  \end{align}
\end{subequations}
which for the given Helmholtz free energy~\eqref{eq:fenergy-ansatz} yields
\begin{subequations}
  \label{eq:constitutive-relations-specific}
  \begin{align}
    \label{eq:17}
    \hfluxc
    &=
      - \kappa(\temp) \nabla \temp, \\
    \label{eq:18}
    \mns
    &=
      \frac{K(\temp)}{\beta \detfgrad} \left(1 - \detfgrad^{-\beta}\right)
      +
      \frac{2\nu(\temp)+3\lambda(\temp)}{3}\divergence\vecv,
    \\
    \label{eq:19}
    \traceless{\cstress}
    &=
      \frac{\mu_1(\temp)}{\detfgrad}
      \traceless{
      \lcgb
      }
      +
      2\nu(\temp) \dgradsymsymb
      .
  \end{align}
\end{subequations}

Concerning the procedure outlined above, few comments are at hand. \emph{If we assume that all entropy production mechanisms except the heat conduction are inactive}, that is if $\nu$ and $\lambda$ and equal to zero, which implies that $\entprodctemp_{\mathrm{mech}} = 0$, then we recover the constitutive relations for the standard hyperelastic solid (Green elastic solid). Furthermore, we note that if $K$ and $\mu_1$ are linear functions of the temperature, and if we use~\eqref{eq:2}, then we find that the internal energy is a function of the temperature only. In other words, in this case we will be dealing the so-called \emph{entropic elasticity}, see~\cite{ericksen.jl:introduction}. We note that such a theory is sufficient for a qualitative explanation of the Gough--Joule effect, see~\cite{gough.j:description} and~\cite{joule.jp:on*2}. (The reader interested in the discussion of the classical Gough--Joule effect is, for example, referred to~\cite{anand.l:constitutive}.)  

\section{Governing equations}
\label{sec:governing-equations}
The last step in the derivation of the complete system of governing equations for the coupled thermal and mechanical processes is the derivation of the evolution equation for the temperature. So far we have derived only the evolution equation for the entropy, which is an inconvenient quantity to work with. However, once we have an evolution equation for the entropy and a formula for the Helmholtz free energy, it is straightforward to derive an evolution equation for the temperature.

We exploit the fact that the entropy is given as the derivative of the Helmholtz free energy with respect to the temperature,
\begin{equation}
  \label{eq:11}
  \entropy = - \pd{\fenergy}{\temp}(\temp, \detfgrad, \lcgb),
\end{equation}
and we use~\eqref{eq:11} in~\eqref{eq:10}. This manipulation yields
\begin{equation}
  \label{eq:15}
  - \rho \temp \dd{}{t} \left( \pd{\fenergy}{\temp} \right)
  =
  \divergence
  \left(
    \kappa(\temp)
    \nabla \temp
  \right)
  +
  2\nu(\temp)\absnorm{\dgradsymsymb}^2
  +
  \frac{2\nu(\temp)+3\lambda(\temp)}{3}(\divergence\vecv)^2,
\end{equation}
where we have used the already known constitutive relation for the heat flux~\eqref{eq:12} and the formula for the entropy production~\eqref{eq:entprod-ansatz}. Having a formula for the Helmholtz free energy~\eqref{eq:fenergy-ansatz} we see that
\begin{equation}
  \label{eq:16}
  \pd{\fenergy}{\temp}
  =
  \dd{\tilde{\fenergy}}{\temp}
  +
  \frac{1}{\rho_R\beta^2} \dd{K}{\temp} \left(\beta \ln{\detfgrad}+ \detfgrad^{-\beta}-1\right)
  +
  \frac{1}{2\rho_R}\dd{\mu_1}{\temp}\left(\Tr\lcgb-3\right)
  ,
\end{equation}
and consequently
\begin{multline}
  \label{eq:20}
  \dd{}{t}
  \left(
    \pd{\fenergy}{\temp}
  \right)
  =
  \left(
    \ddd{\tilde{\fenergy}}{\temp}
    +
    \frac{1}{\rho_R\beta^2} \ddd{K}{\temp} \left(\beta \ln{\detfgrad}+ \detfgrad^{-\beta}-1\right)
    +
    \frac{1}{2\rho_R}\ddd{\mu_1}{\temp}\left(\Tr\lcgb-3\right)
  \right)
  \dd{\temp}{t}
  +
  \frac{1}{\rho_R\detfgrad\beta} \dd{K}{\temp} \left(1- \detfgrad^{-\beta}\right)
  \dd{\detfgrad}{t}
  +
  \frac{1}{2\rho_R}\dd{\mu_1}{\temp}\Tr\left(\dd{\lcgb}{t}\right)
  .
\end{multline}
Next we make use of~\eqref{eq:20} in~\eqref{eq:15}, and we get
\begin{multline}
  \label{eq:21}
  \rho
  \left(
    \cheatvolref
    -
    \frac{\temp}{\rho_R\beta^2} \ddd{K}{\temp} \left(\beta \ln{\detfgrad}+ \detfgrad^{-\beta}-1\right)
    -
    \frac{\temp}{2\rho_R}\ddd{\mu_1}{\temp}\left(\Tr\lcgb-3\right)
  \right)
  \dd{\temp}{t}
  \\
  =
  \divergence
  \left(
    \kappa(\temp)
    \nabla \temp
  \right)
  +
  2\nu(\temp)\absnorm{\dgradsymsymb}^2
  +
  \frac{2\nu(\temp)+3\lambda(\temp)}{3}(\divergence\vecv)^2
  +
  \frac{\rho \temp}{\rho_R\beta} \dd{K}{\temp} \left(1- \detfgrad^{-\beta}\right)
  \divergence \vecv
  +
  \frac{\rho \temp}{\rho_R}\dd{\mu_1}{\temp} \tensordot{\traceless{\lcgb}}{\dgradsymsymb}
  ,
\end{multline}
where we have used explicit formulae for the material time derivative of $\detfgrad$ and $\lcgb$, see~\eqref{eq:6}. Now we are ready to write down the evolution equations for the unknown Eulerian fields $\rho$, $\vec{v}$, $\lcgb$ and $\temp$.

\subsection{Eulerian description}
\label{sec:eulerian-description}
The system of evolution equations in the Eulerian description in the absence of body forces reads
\begin{subequations}
  \label{eq:evolution-equations-eulerian}
  \begin{align}
    \label{eq:23}
    \dd{\rho}{t} + \rho \divergence \vec{v}
    &= 0,
    \\
    \label{eq:24}
    \rho \dd{\vec{v}}{t}
    &=
      \divergence \cstress,
    \\
    \label{eq:27}
    \cstress &= \mns \identity + \traceless{\cstress},
    \\
    \label{eq:28}
    \mns
    &=
      \frac{K(\temp)}{\beta \detfgrad} \left(1 - \detfgrad^{-\beta}\right)
      +
      \frac{2\nu(\temp)+3\lambda(\temp)}{3}\divergence\vecv
    \\
    \label{eq:29}
    \traceless{\cstress}
    &=
      \frac{\mu_1(\temp)}{\detfgrad}
      \traceless{
      \lcgb
      }
      +
      2\nu(\temp) \dgradsymsymb
    \\
    \label{eq:25}
    \dd{\lcgb}{t}
    &=
    -\frac{2}{3}
    \lcgb
    \divergence \vec{v}
    +
    \gradvl
    \lcgb
    +
    \lcgb
    \transpose{\gradvl}
    ,
  \end{align}
  and
  \begin{multline}
    \label{eq:30}
    \frac{1}{\detfgrad}
    \left(
      \rho_R
      \cheatvolref
      -
      \frac{\temp}{\beta^2} \ddd{K}{\temp} \left(\beta \ln{\detfgrad}+ \detfgrad^{-\beta}-1\right)
      -
      \frac{\temp}{2}\ddd{\mu_1}{\temp}\left(\Tr\lcgb-3\right)
    \right)
    \dd{\temp}{t}
    \\
  =
  \divergence
  \left(
    \kappa(\temp)
    \nabla \temp
  \right)
  +
  2\nu(\temp)\absnorm{\dgradsymsymb}^2
  +
  \frac{2\nu(\temp)+3\lambda(\temp)}{3}(\divergence\vecv)^2
  +
  \frac{\temp}{\detfgrad \beta} \dd{K}{\temp} \left(1- \detfgrad^{-\beta}\right)
  \divergence \vecv
  +
  \frac{\temp}{\detfgrad}\dd{\mu_1}{\temp} \tensordot{\traceless{\lcgb}}{\dgradsymsymb},
  \end{multline}
\end{subequations}
where $\detfgrad = \frac{\rho_R}{\rho}$.

Although the Eulerian description has been convenient for the derivation of the governing equations, it is inconvenient from the perspective of numerical simulations since the governing equations~\eqref{eq:evolution-equations-eulerian} must be solved in a deforming domain (current configuration). This drawback is however easy to mitigate via the transformation of the governing equations to the Lagrangian description. If we do so, we will be working in the reference configuration, and the computational domain will remain fixed.

\begin{summary}[Governing equations -- Eulerian description]
  \label{summary:governing-equations-eulerian}
  Specific Helmholtz free energy $\fenergy$ and entropy production $\entprodc$:
  \begin{align*}
    \fenergy
    &=_{\bydefinition}
    -
    \cheatvolref \temp \left[ \ln \left( \frac{\temp}{\temp_{\reference}} \right) - 1 \right]
    +
    \frac{K(\temp)}{\rho_R\beta^2}\left(\beta \ln{\detfgrad}+ \detfgrad^{-\beta}-1\right)
    +
    \frac{\mu_1(\temp)}{2\rho_R}\left(\Tr\lcgb-3\right)
    \\
    \entprodc
    &=_{\bydefinition}
      \frac{1}{\temp}
      \left(
      2\nu(\temp)\absnorm{\dgradsymsymb}^2
      +
      \frac{2\nu(\temp)+3\lambda(\temp)}{3}(\divergence\vecv)^2
      \right)
      +
      \kappa (\temp)\frac{\vectordot{\nabla \temp}{\nabla \temp}}{\temp^2}
  \end{align*}
  Constitutive relations for the Cauchy stress tensor $\cstress$ and the heat flux $\hfluxc$: 
  \begin{align*}
    \cstress &= \mns \identity + \traceless{\cstress}
               \\
    \mns
    &=
      \frac{K(\temp)}{\beta \detfgrad} \left(1 - \detfgrad^{-\beta}\right)
      +
      \frac{2\nu(\temp)+3\lambda(\temp)}{3}\divergence\vecv
    \\
    \traceless{\cstress}
    &=
      \frac{\mu_1(\temp)}{\detfgrad}
      \traceless{
      \lcgb
      }
      +
      2\nu(\temp) \dgradsymsymb
    \\
    \hfluxc
    &=
      - \kappa(\temp) \nabla \temp
  \end{align*}

  For unknown fields~$\rho(\vec{x},t)$, $\vecv(\vec{x},t)$, $\lcgb(\vec{x},t)$ and $\temp(\vec{x},t)$ solve:
  \begin{align*}
    \dd{\rho}{t} + \rho \divergence \vec{v}
    &= 0
    \\
    \rho \dd{\vec{v}}{t}
    &=
      \divergence \cstress
    \\
    \dd{\lcgb}{t}
    &=
    -\frac{2}{3}
    \lcgb
    \divergence \vec{v}
    +
    \gradvl
    \lcgb
    +
    \lcgb
      \transpose{\gradvl}
  \end{align*}
  \begin{multline*}
        \frac{1}{\detfgrad}
    \left(
      \rho_R
      \cheatvolref
      -
      \frac{\temp}{\beta^2} \ddd{K}{\temp} \left(\beta \ln{\detfgrad}+ \detfgrad^{-\beta}-1\right)
      -
      \frac{\temp}{2}\ddd{\mu_1}{\temp}\left(\Tr\lcgb-3\right)
    \right)
    \dd{\temp}{t}
    \\
  =
  \divergence
  \left(
    \kappa(\temp)
    \nabla \temp
  \right)
  +
  2\nu(\temp)\absnorm{\dgradsymsymb}^2
  +
  \frac{2\nu(\temp)+3\lambda(\temp)}{3}(\divergence\vecv)^2
  +
  \frac{\temp}{\detfgrad \beta} \dd{K}{\temp} \left(1- \detfgrad^{-\beta}\right)
  \divergence \vecv
  +
  \frac{\temp}{\detfgrad}\dd{\mu_1}{\temp} \tensordot{\traceless{\lcgb}}{\dgradsymsymb}
  \end{multline*}
\end{summary}

\subsection{Lagrangian description}
\label{sec:lagr-descr}
The transformation to the Lagrangian description is straightforward, see the standard reference books such as~\cite{ciarlet.pg:mathematical*2} or \cite{truesdell.c.noll.w:non-linear*1}, and we comment on it for the sake of completeness of our presentation. First, the Lagrangian velocity field $\vec{V}(\vec{X},t)$ is related to the Eulerian velocity field $\vec{v}(\vec{x},t)$ via the equality
\begin{equation}
  \label{eq:31}
  \vec{v}(\deformation(\vec{X},t), t) = \vec{V}(\vec{X},t),
\end{equation}
where $\deformation$ denotes the deformation function, and $\vec{x}$ and $\vec{X}$ denote the position of the given material point in the current and reference configuration respectively, $\vec{x} = \deformation(\vec{X},t)$. Consequently, we get
\begin{equation}
  \label{eq:32}
  \left. \left[ \nabla \vec{v} (\vec{x}, t) \right] \right|_{\vec{x} = \deformation(\vec{X},t)} = \left[ \nabla \vec{V} (\vec{X},t) \right] \inverse{\fgrad}(\vec{X},t),
\end{equation}
where the gradient on the left-hand-side is taken with respect to $\vec{x}$, while the gradient on the right-hand-side is taken with respect to $\vec{X}$. Equality~\eqref{eq:32} implies that
\begin{equation}
  \label{eq:33}
  \left. \left[ \gradsym (\vec{x}, t) \right] \right|_{\vec{x} = \deformation(\vec{X},t)} = \frac{1}{2} \left( \left[ \nabla \vec{V} \right] (\vec{X},t) \inverse{\fgrad}(\vec{X},t) +  \transposei{\fgrad}(\vec{X},t) \transpose{\left[\nabla \vec{V}\right]} (\vec{X},t) \right),
\end{equation}
which motivates us to introduce the notation $ \gradsym_{\vec{X}} (\vec{X},t) =_{\bydefinition} \left. \left[ \gradsym (\vec{x}, t) \right] \right|_{\vec{x} = \deformation(\vec{X},t)}$ that is
\begin{equation}
  \label{eq:37}
  \gradsym_{\vec{X}}
  =_{\bydefinition}
  \frac{1}{2}
  \left(
    \left[\nabla \vec{V}\right]  \inverse{\fgrad}
    +
    \transposei{\fgrad} \transpose{\left[\nabla \vec{V}\right]}
  \right),
\end{equation}
where the gradients on the right-hand-side are taken with respect to $\vec{X}$.

Using the transformation rules~\eqref{eq:31} and~\eqref{eq:37}, we can proceed with the transformation of the governing equations~\eqref{eq:evolution-equations-eulerian}. First, the balance of mass~\eqref{eq:23} takes the form $\rho \det \fgrad = \rho_R$. Second, the balance of linear momentum~\eqref{eq:24} is transformed as follows
\begin{equation}
  \label{eq:34}
  \rho_R \pd{\vec{V}}{t} (\vec{X}, t)
  =
  \divergence \fpstress (\vec{X}, t),
\end{equation}
where the divergence is taken with respect to the $\vec{X}$ variable and
\begin{equation}
  \label{eq:35}
  \fpstress (\vec{X},t)
  =
  \left[
    \det \fgrad (\vec{X},t)
  \right]
  \left.
    \cstress(\vec{x},t)
  \right|_{\vec{x} = \deformation (\vec{X}, t)}
  \transposei{\fgrad} (\vec{X},t),
\end{equation}
denotes the first Piola--Kirchhoff stress tensor. Relation~\eqref{eq:35} is a consequence of the Piola transformation, and if no confusion can arise, we simply write it as $\fpstress = \left(\det \fgrad\right) \cstress \transposei{\fgrad}$. The particular expression for the Cauchy stress tensor $\cstress$ is in our case~\eqref{eq:27}, hence we get
\begin{equation}
  \label{eq:36}
  \left.
    \cstress(\vec{x},t)
  \right|_{\vec{x} = \deformation (\vec{X}, t)}
  =
  \frac{K(\Temp)}{\beta \detfgrad} \left(1 - \detfgrad^{-\beta}\right)
  +
  \frac{2\nu(\Temp)+3\lambda(\Temp)}{3}
  \Tr
  \gradsym_{\vec{X}}
  +
  2 \nu(\Temp)   \gradsym_{\vec{X}}
  +
  \frac{\mu_1(\Temp)}{\detfgrad}
  \traceless{
    \lcgb
  }
  ,
\end{equation}
where the right-hand-side is interpreted as a function of $\vec{X}$ and $t$, and all the spatial derivatives are taken with respect to~$\vec{X}$. (In particular the symbol $\Temp$ denotes the Lagrangian temperature field $\Temp(\vec{X}, t) = \temp(\deformation(\vec{X}, t), t)$.) Third, unlike in the Eulerian description the field $\lcgb$ is not in the Lagrangian description interpreted as an independent variable with its own rate-type evolution equation, but is a known function of the deformation since $\lcg =_{\bydefinition} \fgrad \transpose{\fgrad}$. Fourth, we need to transform the evolution equation for the temperature. This transformation is a little bit more elaborate, we need to transform the heat flux vector $\hfluxc$ to the referential heat flux vector $\hflux$ using the transformation
\begin{equation}
  \label{eq:22}
  \hflux(\vec{X},t) = \left[\det \fgrad(\vec{X},t) \right] \inverse{\fgrad}(\vec{X},t) \left. \hfluxc(\vec{x},t) \right|_{\vec{x} = \deformation (\vec{X}, t)}.
\end{equation}
Next we use the Fourier law $\hfluxc = - \kappa(\temp) \nabla \temp$, which is naturally set in the Eulerian description, and we apply the chain rule in order to get the spatial derivatives with respect to~$\vec{X}$, which yields
\begin{equation}
  \label{eq:41}
  \left. \hfluxc(\vec{x}, t) \right|_{\vec{x} = \deformation (\vec{X}, t)} = - \kappa(\Temp(\vec{X},t)) \transposei{\fgrad}(\vec{X},t) \nabla \Temp(\vec{X},t). 
\end{equation}
(In~\eqref{eq:41} the gradient of $\Temp(\vec{X},t)$ is again the gradient with respect to~$\vec{X}$.) Consequently, for the referential heat flux we get
\begin{equation}
  \label{eq:26}
  \hflux(\vec{X},t)
  =
  -
  \kappa(\Temp(\vec{X},t))
  \left[\det \fgrad(\vec{X},t)  \right] \inverse{\rcg}(\vec{X},t) \nabla \Temp(\vec{X},t),
\end{equation}
where $\rcg =_{\bydefinition} \transpose{\fgrad} \fgrad$ denotes the right Cauchy--Green tensor, and we can conclude that the divergence term in~\eqref{eq:30} transforms as
\begin{equation}
  \label{eq:38}
  \left[\det \fgrad (\vec{X},t)\right]
  \left[ \divergence \left( \kappa(\temp) \nabla \temp (\vec{x},t) \right)  \right]_{\vec{x} = \deformation (\vec{X}, t)}
  =
  \divergence \left( \kappa(\Temp(\vec{X},t)) \left[\det \fgrad (\vec{X},t)\right] \inverse{\rcg}(\vec{X},t) \nabla \Temp(\vec{X},t) \right),
\end{equation}
where the derivatives on the left-hand-side are the derivatives with respect to $\vec{x}$, while the derivatives on the right-hand-side are the derivatives with respect to $\vec{X}$. If no confusion can arise, equation \eqref{eq:38} is simply written as $\left(\det \fgrad \right) \divergence \left( \kappa(\temp) \nabla \temp \right) =  \divergence \left( \kappa(\Temp) \left(\det \fgrad\right) \inverse{\rcg} \nabla \Temp \right)$. The remaining volumetric terms in~\eqref{eq:30} are straightforward to transform, and the Lagrangian version of~\eqref{eq:30} reads 
\begin{multline}
  \label{eq:39}
      \left(
      \rho_R
      \cheatvolref
      -
      \frac{\Temp}{\beta^2} \ddd{K}{\Temp} \left(\beta \ln{\detfgrad}+ \detfgrad^{-\beta}-1\right)
      -
      \frac{\Temp}{2}\ddd{\mu_1}{\Temp}\left(\Tr\lcgb-3\right)
    \right)
    \pd{\Temp}{t}
  =
  \divergence
  \left(
    \kappa(\Temp) \detfgrad \inverse{\rcg} \nabla \Temp
  \right)
  \\
  +
  2 \nu(\Temp)\detfgrad \absnorm{\traceless{\left(\gradsym_{\vec{X}}\right)}}^2
  +
  \frac{2\nu(\Temp)+3\lambda(\Temp)}{3}\detfgrad(\Tr \gradsym_{\vec{X}})^2
  +
  \frac{\Temp}{\beta} \dd{K}{\Temp} \left(1- \detfgrad^{-\beta}\right)
  \Tr \gradsym_{\vec{X}}
  +
  \Temp\dd{\mu_1}{\Temp} \tensordot{\traceless{\lcgb}}{\traceless{\left(\gradsym_{\vec{X}}\right)}}
  .
\end{multline}

Consequently, if we want to work in Lagrangian description, then the evolution equations for the unknown deformation~$\deformation(\vec{X},t)$ field and the unknown temperature field $\Temp(\vec{X},t)$ read
\begin{subequations}
  \label{eq:evolution-equations-lagrangian}
  \begin{equation}
    \label{eq:42}
    \rho_R \ppd{\deformation}{t}
    =
    \divergence
    \left(
      \left[
        \frac{K(\Temp)}{\beta} \left(1 - \detfgrad^{-\beta}\right)
        +
        \frac{2\nu(\Temp)+3\lambda(\Temp)}{3}
        \detfgrad
        \Tr
        \gradsym_{\vec{X}}
        +
        2 \nu(\Temp) \detfgrad   \gradsym_{\vec{X}}
        +
        \mu_1(\Temp)
        \traceless{
          \lcgb
        }
      \right]
      \transposei{\fgrad}
    \right)
  \end{equation}
  and
  \begin{multline}
    \label{eq:40}
          \left(
      \rho_R
      \cheatvolref
      -
      \frac{\Temp}{\beta^2} \ddd{K}{\Temp} \left(\beta \ln{\detfgrad}+ \detfgrad^{-\beta}-1\right)
      -
      \frac{\Temp}{2}\ddd{\mu_1}{\Temp}\left(\Tr\lcgb-3\right)
    \right)
    \pd{\Temp}{t}
  =
  \divergence
  \left(
    \kappa(\Temp) \detfgrad \inverse{\rcg} \nabla \Temp
  \right)
  \\
  +
  2 \nu(\Temp)\detfgrad \absnorm{\traceless{\left(\gradsym_{\vec{X}}\right)}}^2
  +
  \frac{2\nu(\Temp)+3\lambda(\Temp)}{3}\detfgrad(\Tr \gradsym_{\vec{X}})^2
  +
  \frac{\Temp}{\beta} \dd{K}{\Temp} \left(1- \detfgrad^{-\beta}\right)
  \Tr \gradsym_{\vec{X}}
  +
  \Temp\dd{\mu_1}{\Temp} \tensordot{\traceless{\lcgb}}{\traceless{\left(\gradsym_{\vec{X}}\right)}}
  .
  \end{multline}
\end{subequations}
The boundary and initial conditions are specified as normal in the solid mechanics, the specific boundary and initial conditions for the problem we are going to solve are given in Section~\ref{sec:problem-sescription}.

\begin{summary}[Governing equations -- Lagrangian description]
  \label{summary:governing-equations-lagrangian}
  Specific Helmholtz free energy $\fenergy$ and entropy production $\entprodc$:
  \begin{align*}
    \fenergy
    &=_{\bydefinition}
    -
    \cheatvolref \temp \left[ \ln \left( \frac{\temp}{\temp_{\reference}} \right) - 1 \right]
    +
    \frac{K(\temp)}{\rho_R\beta^2}\left(\beta \ln{\detfgrad}+ \detfgrad^{-\beta}-1\right)
    +
    \frac{\mu_1(\temp)}{2\rho_R}\left(\Tr\lcgb-3\right)
    \\
    \entprodc
    &=_{\bydefinition}
      \frac{1}{\temp}
      \left(
      2\nu(\temp)\absnorm{\dgradsymsymb}^2
      +
      \frac{2\nu(\temp)+3\lambda(\temp)}{3}(\divergence\vecv)^2
      \right)
      +
      \kappa (\temp)\frac{\vectordot{\nabla \temp}{\nabla \temp}}{\temp^2}
  \end{align*}
  Constitutive relations for the Cauchy stress tensor $\cstress$ and the heat flux $\hfluxc$: 
  \begin{align*}
    \cstress
    &=
    \frac{K(\Temp)}{\beta \detfgrad} \left(1 - \detfgrad^{-\beta}\right) \identity
    +
    \frac{2\nu(\Temp)+3\lambda(\Temp)}{3}
    \Tr
    \gradsym_{\vec{X}}
    +
    2 \nu(\Temp) \traceless{\left( \gradsym_{\vec{X}} \right)}
    +
    \frac{\mu_1(\Temp)}{\detfgrad}
    \traceless{
      \lcgb
      }
    \\
    \hfluxc
    &=
      - \kappa(\temp) \nabla \temp
  \end{align*}
Auxiliary definitions/notation:
\begin{equation*}
  \fgrad =_{\bydefinition} \nabla \deformation
  \quad
  \detfgrad =_{\bydefinition} \det \fgrad
  \quad
  \lcg =_{\bydefinition} \fgrad \transpose{\fgrad}
  \quad
  \lcgb =_{\bydefinition} \frac{\lcg}{\detfgrad^{\frac{2}{3}}}
  \quad 
  \vec{V}=_{\bydefinition} \pd{\deformation}{t}
  \quad
  \gradsym_{\vec{X}}=_{\bydefinition}
  \frac{1}{2}
  \left(
    \left[\nabla \vec{V}\right]  \inverse{\fgrad}
    +
    \transposei{\fgrad} \transpose{\left[\nabla \vec{V}\right]}
  \right)
\end{equation*}

  For unknown fields~$\deformation(\vec{X},t)$ and $\Temp(\vec{X},t)$ solve:
  \begin{equation*}
    \rho_R \ppd{\deformation}{t}
    =
    \divergence
    \left(
      \detfgrad
      \cstress
      \transposei{\fgrad}
    \right)
  \end{equation*}
  \begin{multline*}
              \left(
      \rho_R
      \cheatvolref
      -
      \frac{\Temp}{\beta^2} \ddd{K}{\Temp} \left(\beta \ln{\detfgrad}+ \detfgrad^{-\beta}-1\right)
      -
      \frac{\Temp}{2}\ddd{\mu_1}{\Temp}\left(\Tr\lcgb-3\right)
    \right)
    \pd{\Temp}{t}
  =
  \divergence
  \left(
    \kappa(\Temp) \detfgrad \inverse{\rcg} \nabla \Temp
  \right)
  \\
  +
  2 \nu(\Temp)\detfgrad \absnorm{\traceless{\left(\gradsym_{\vec{X}}\right)}}^2
  +
  \frac{2\nu(\Temp)+3\lambda(\Temp)}{3}\detfgrad(\Tr \gradsym_{\vec{X}})^2
  +
  \frac{\Temp}{\beta} \dd{K}{\Temp} \left(1- \detfgrad^{-\beta}\right)
  \Tr \gradsym_{\vec{X}}
  +
  \Temp\dd{\mu_1}{\Temp} \tensordot{\traceless{\lcgb}}{\traceless{\left(\gradsym_{\vec{X}}\right)}}
\end{multline*}
(All spatial derivatives are taken with respect to $\vec{X}$.)
\end{summary}

\section{Problem description}
\label{sec:problem-sescription}
We now proceed with numerical solution of the corresponding governing equations in a specific setting. We consider a \emph{two-dimensional problem} that resembles the experimental setting used in~\cite{martinez.jrs.toussaint.e.ea:heat}, see Figure~\ref{fig:martinez}.

\subsection{Geometry}
\label{sec:geometry}
\begin{figure}[h]
  \centering
  \subfloat[\label{fig:geometry-a}Specimen geometry and location of virtual temperature probes. (Not on scale.)]{\includegraphics[width=0.45\textwidth]{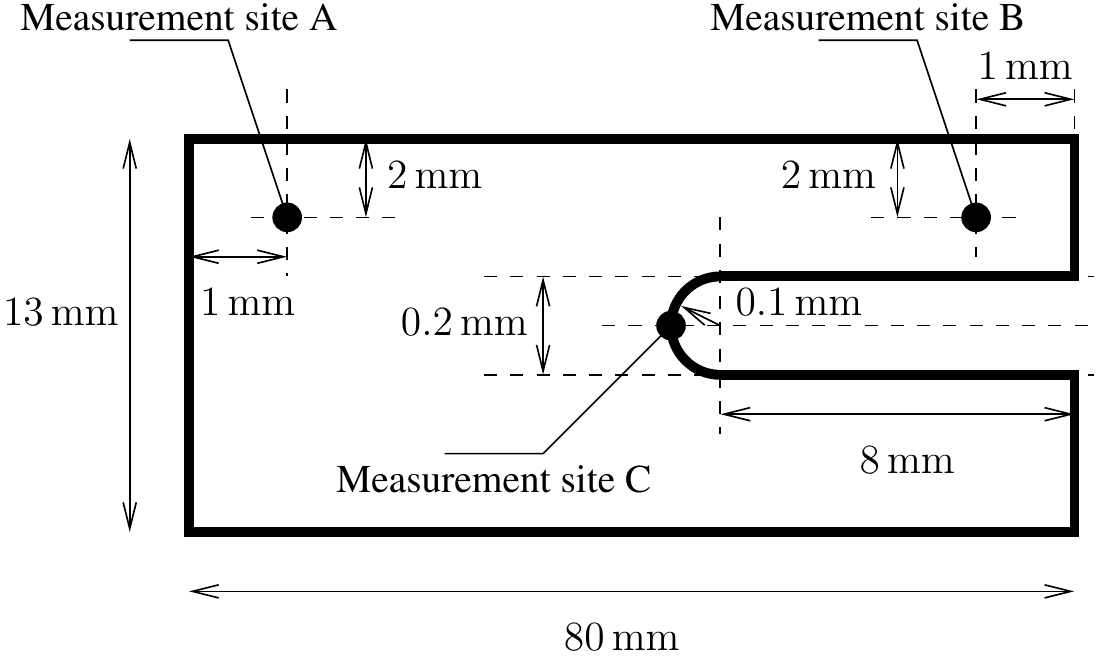}}
  \qquad
  \subfloat[\label{fig:geometry-b}Boundary conditions. (Not on scale.)]{\includegraphics[width=0.45\textwidth]{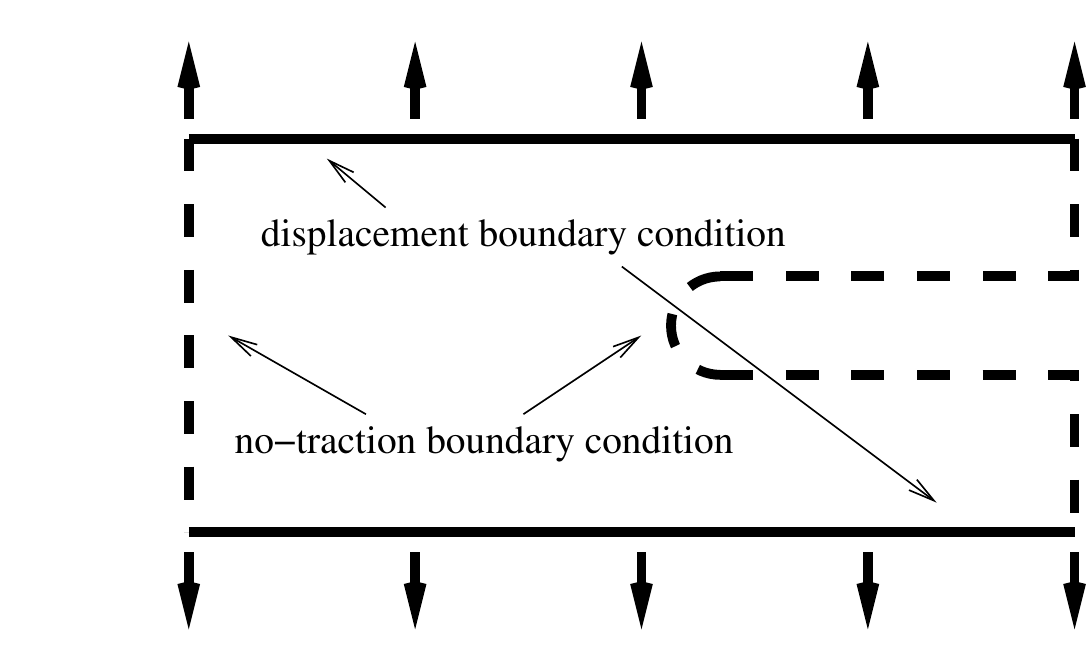}}
  \\
  \subfloat[\label{fig:geometry-c}Finite element mesh in the vicinity of the cutout tip.]{\includegraphics[width=0.4\textwidth]{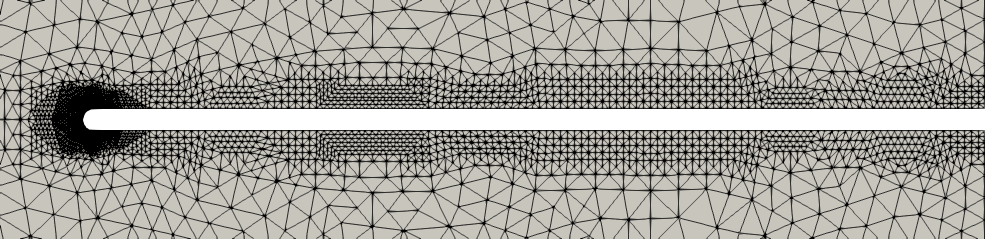}}
  \caption{Problem geometry.}
  \label{fig:geometry}
\end{figure}

The initial shape of the sample is shown in~Figure~\ref{fig:geometry}. The material sample has a rectangular shape $\unit[80]{mm} \times \unit[13]{mm}$ with a thin cutout in the middle of its right face. The cutout is $\unit[8]{mm}$ long and $\unit[0.2]{mm}$ wide, and the cutout ends with a semicircle of radius $\unit[0.1]{mm}$.

Later on we report temperature values measured at three different points in the sample. These points are denoted $A$, $B$ and $C$, while the point $C$ is located at the tip of the cutout. The remaining points are located close to the top left and top right corner of the sample, see Figure~\ref{fig:geometry} for the exact location of the measurement sites. This initial shape of the sample coincides with the computational domain $\Omega$ used for the Lagrangian description, see Section~\ref{sec:lagr-descr}, hence the measurement points are tracked during the evolution. (We measure temperature at the same material point, but this material point occupies at different times different positions in space.)   

\subsection{Material parameters}
\label{sec:parameter-values}
Material parameter values that are \emph{fixed in all numerical simulations} reported below are shown in Table~\ref{tab:parameter-values}.
(Density $\rho_R$, specific heat capacity at constant volume $\cheatvolref$ and thermal conductivity $\kappa$ correspond to the values used in~\cite{martinez.jrs.toussaint.e.ea:heat}.)
\begin{table}[h]
  \begin{tabular}[h]{lll}
    \toprule
    Parameter & Unit & Value \\
    \midrule
    $\rho_R$ & $\unitfrac{kg}{m^3}$ & 1101 \\
    $\beta$ & $-$ & 2 \\
    $\cheatvolref$ & $\unitfrac{J}{kg \cdot K}$ & 1591\\
    $\kappa$ & $\unitfrac{W}{m \cdot K}$ & 0.317  \\
    \bottomrule
  \end{tabular}
  \medskip
  \caption{Material parameters.}
  \label{tab:parameter-values}
\end{table}
The remaining material parameters in the numerical simulations are varied. In particular we systematically vary the material parameters that characterise the energy storage mechanisms in the material (elastic moduli) and the entropy entropy production mechanisms (viscosities) in the material. Regarding the elastic moduli, we consider two settings, namely the setting wherein the elastic moduli are \emph{constant}, that is
\begin{subequations}
  \label{eq:43}
  \begin{align}
    \label{eq:46}
    \mu_1 &= \mu_{1, \reference}, \\
    \label{eq:47}
    K &= K_{\reference}, 
  \end{align}
\end{subequations}
and the setting wherein the elastic moduli are \emph{linear functions of temperature}, that is
\begin{subequations}
  \label{eq:48}
  \begin{align}
    \label{eq:49}
    \mu_1 &= \mu_{1, \reference} \frac{\temp}{\temp_{\reference}}, \\
    \label{eq:50}
    K &= K_{\reference} \frac{\temp}{\temp_{\reference}}.
  \end{align}
\end{subequations}
(This temperature dependence corresponds to the classical entropic elasticity, see for example~\cite{ericksen.jl:introduction}.) The values of constants $\mu_{1, \reference}$, $K_{\reference}$ and $\temp_{\reference}$ are given in Table~\ref{tab:parameter-values-elastic}. The values used for the shear and bulk modulus, that is for $\mu_{1, \reference}$ and $K_{\reference}$, correspond---to the order of magnitude---to values for a generic rubber-like substance. (The Young modulus $\Young$ and Poisson $\Poisson$ ratio are, $\Young = \frac{9 K_{\reference} \mu_{1, \reference}}{3K_{\reference} + \mu_{1, \reference}} \approx \unit[3 \times 10^6]{Pa}$, $\Poisson = \frac{3 K_{\reference}  - 2\mu_{1, \reference}}{2 \left(3K_{\reference} + \mu_{1, \reference}\right)}=0.4995$.)

\begin{table}[h]
  \begin{tabular}[h]{lll}
    \toprule
    Parameter & Unit & Value \\
    \midrule
    $\mu_{1, \reference}$ & $\unit{Pa}$ & $1 \times 10^6$ \\
    $K_{\reference}$ & $\unit{Pa}$ & $1 \times 10^9$\\
    $\temp_{\reference}$ & $\unit{K}$ & 300\\
    \bottomrule
  \end{tabular}
  \medskip
  \caption{Parameter values in formulae~\eqref{eq:43} and~\eqref{eq:48} for the elastic moduli.}
  \label{tab:parameter-values-elastic}
\end{table}

Finally, we consider several values of the viscosities $\lambda$ and $\nu$, see Table~\ref{tab:parameter-values-viscosity}. The viscosities used in the numerical simulations deliberately differ by several orders of magnitude, and we choose such a vast range of viscosities in order to demonstrate the impact of the choice of viscosity on the overall quantitative behaviour of the material. (Some of these viscosity values are definitely too large from the perspective of real materials. However, using such large viscosities allows us to easily document the trends implied by the decrease/increase of viscosity.) Regarding the viscosity values, we also use the verbal description outlined in Table~\ref{tab:parameter-values-viscosity}.

\begin{table}[h]
  \begin{tabular}[h]{lllllllll}
    \toprule
    Parameter & Unit  & \multicolumn{6}{c}{Value} \\
    \multicolumn{2}{c}{} & tiny & small & medium & large & s-large & x-large \\
    \midrule
    $\lambda$ & $\unit{Pa \cdot s}$ & $1 \times 10^2$ & $1 \times 10^3$ & $1 \times 10^4$ & $1 \times 10^5$ & $5 \times 10^5$ & $1 \times 10^6$  \\
    $\nu$ & $\unit{Pa \cdot s}$ & $1 \times 10^2$ & $1 \times 10^3$ & $1 \times 10^4$ & $1 \times 10^5$ & $5 \times 10^5$ & $1 \times 10^6$  \\
    \bottomrule
  \end{tabular}
  \medskip
  \caption{Material parameters -- viscosity.}
  \label{tab:parameter-values-viscosity}
\end{table}

\subsection{Initial and boundary conditions}
\label{sec:boundary-conditions}
Regarding the boundary conditions for the temperature field, we assume that the whole sample is thermally isolated, hence we prescribe the no-flux boundary condition $\left. \vectordot{\hfluxc}{\vec{n}} \right|_{\deformation(\partial \Omega)} = 0$, where $\vec{n}$ denotes the unit outward normal to the boundary (in the current configuration). This means that the boundary condition for the temperature field $\Temp$ in the computational domain $\Omega$ reads
\begin{equation}
  \label{eq:44}
  \left.
    \vectordot{\nabla \Temp}{\vec{N}}
  \right|_{\partial \Omega}
  =
  0
  ,
\end{equation}
where $\vec{N}$ denotes the unit outward normal to the computational domain $\Omega$. The initial temperature field is homogeneous in space and the initial temperature is fixed as $\Temp = \temp_{\reference}$.

Regarding the deformation $\deformation$ we prescribe the displacement on the top and bottom boundary of the sample, see Figure~\ref{fig:geometry-b}, while on the remaining parts of the boundary we prescribe the no-traction boundary condition, $\left. \fpstress \vec{N} \right|_{\partial \Omega} = \vec{0}$. The displacement is prescribed in such a way that we can investigate two loading/unloading protocols---\emph{single loading-unloading} protocol and~\emph{oscillatory loading-unloading} protocol. The oscillatory loading protocol, see below, resembles the experimental setting used in~\cite{martinez.jrs.toussaint.e.ea:heat}.

\begin{figure}[h]
  \centering
  \subfloat[\label{fig:loading-protocol-a}Single loading-unloading.]{\includegraphics[width=0.4\textwidth]{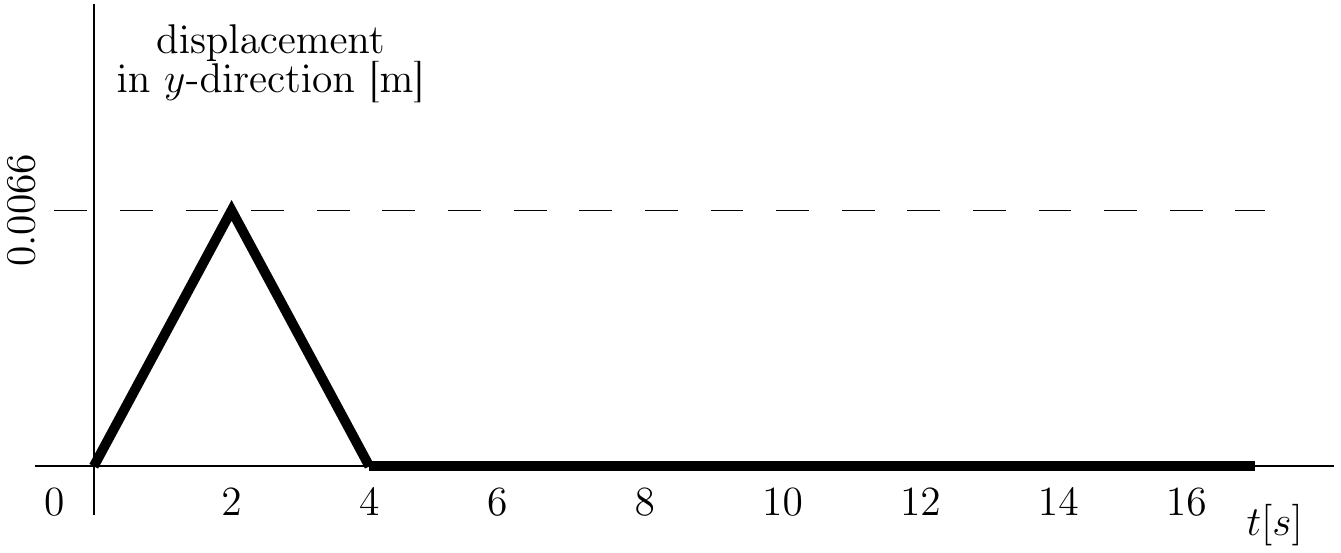}}
  \qquad
  \subfloat[\label{fig:loading-protocol-b}Oscillatory loading-unloading.]{\includegraphics[width=0.4\textwidth]{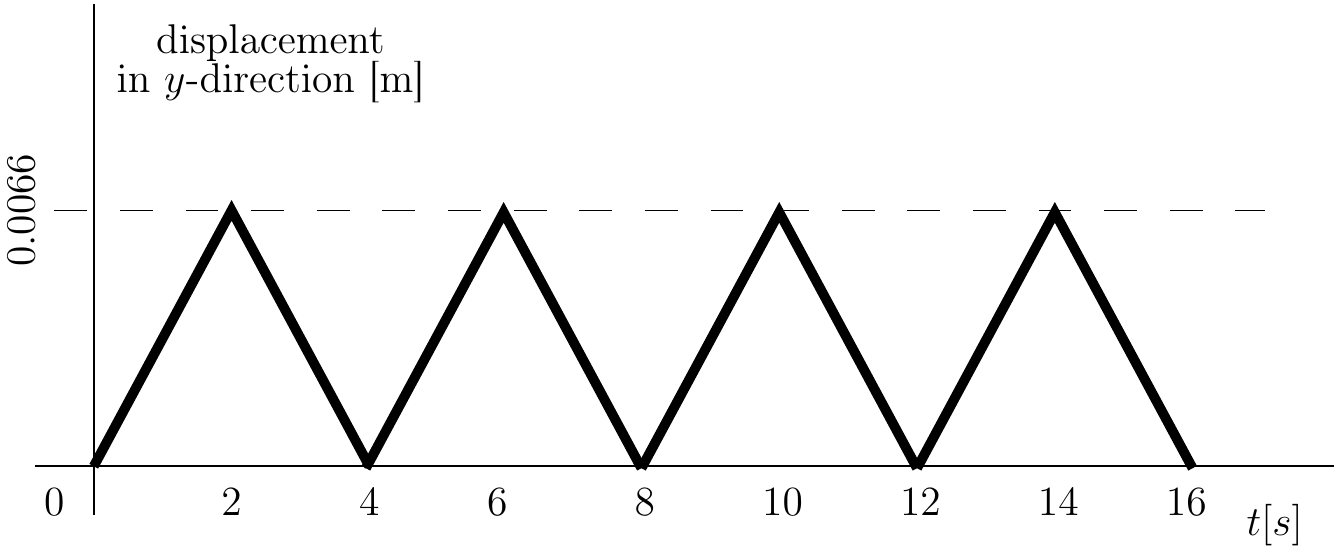}}
  \caption{Loading protocols.}
  \label{fig:loading-protocol}
\end{figure}

In the case of \emph{single loading-unloading} protocol we prescribe the zero displacement in the $x$-direction, and the displacement in the $y$-direction is given by the function depicted in Figure~\ref{fig:loading-protocol-a}. The top and bottom boundaries are from time $t= \unit[0]{s}$ to time $t= \unit[2]{s}$ moving with the constant velocity $\pm \unitfrac[0.0033]{m}{s}$ in the $y$-direction, and from time $t= \unit[2]{s}$ to time $t= \unit[4]{s}$ the top and bottom boundaries are moving backwards with the same (magnitude of) velocity. After $t=\unit[4]{s}$ the top and bottom boundaries no longer move.

Next we consider \emph{oscillatory loading-unloading} protocol. In this case we prescribe the zero displacement in the $x$-direction, and the displacement in the $y$-direction is given by the function depicted in Figure~\ref{fig:loading-protocol-a}. The top and bottom boundaries are from time $t= \unit[0]{s}$ to time $t= \unit[2]{s}$ moving with the constant velocity $\pm \unitfrac[0.0033]{m}{s}$ in the $y$-direction, and from time $t= \unit[2]{s}$ to time $t= \unit[4]{s}$ the top and bottom boundaries are moving backwards with the same (magnitude of) velocity. This loading/unloading cycle is then periodically repeated.

\section{Results and discussion}
\label{sec:numerical-simulation}
The governing equations~\eqref{eq:evolution-equations-lagrangian} subject to boundary conditions discussed in Section~\ref{sec:boundary-conditions} have been solved using the finite element method. For the displacement and the temperature fields we have used Lagrange elements of degree one, the time stepping has been done using an implicit Euler scheme, the arising nonlinear variational problem has been solved using the Newton method with automatic differentiation, and, finally, the arising systems of linear equations have been solved with the direct solver MUMPS, \cite{amestoy.pr.duff.is.ea:fully}. The numerical algorithm has been implemented in FEniCS, see~\cite{aln-s.m.blechta.j.ea:fenics}.

The finite element mesh has been manually refined near the cutout tip, see Figure~\ref{fig:geometry-c}, and the time step has been fixed to $\diff t = \unit[0.002]{s}$ (oscillatory loading-unloading) . The chosen finite element mesh lead to a problem with approximately $75\, 000$ degrees of freedom in each time step.  Regarding the single loading--unloading cycle the same time step $\diff t = \unit[0.002]{s}$  has been used for the first $\unit[4.4]{s}$, and then it has been manually increased in order to handle \unit[20\,000]{s} long time interval.

\subsection{Single loading--unloading cycle}
\label{sec:single-load-unlo}

\subsubsection{Models with temperature dependent elastic moduli versus models with constant elastic moduli}
\label{sec:comp-temp-valu}

In the first set of numerical simulations we compare temperature fields predicted by the models with \emph{constant elastic moduli} and the models with \emph{temperature dependent elastic moduli} (linear dependence on temperature), see Section~\ref{sec:parameter-values} for the full specification of the models.

\begin{figure}[h]
  \subfloat[Measurement site A.]{\includegraphics[width=0.32\textwidth]{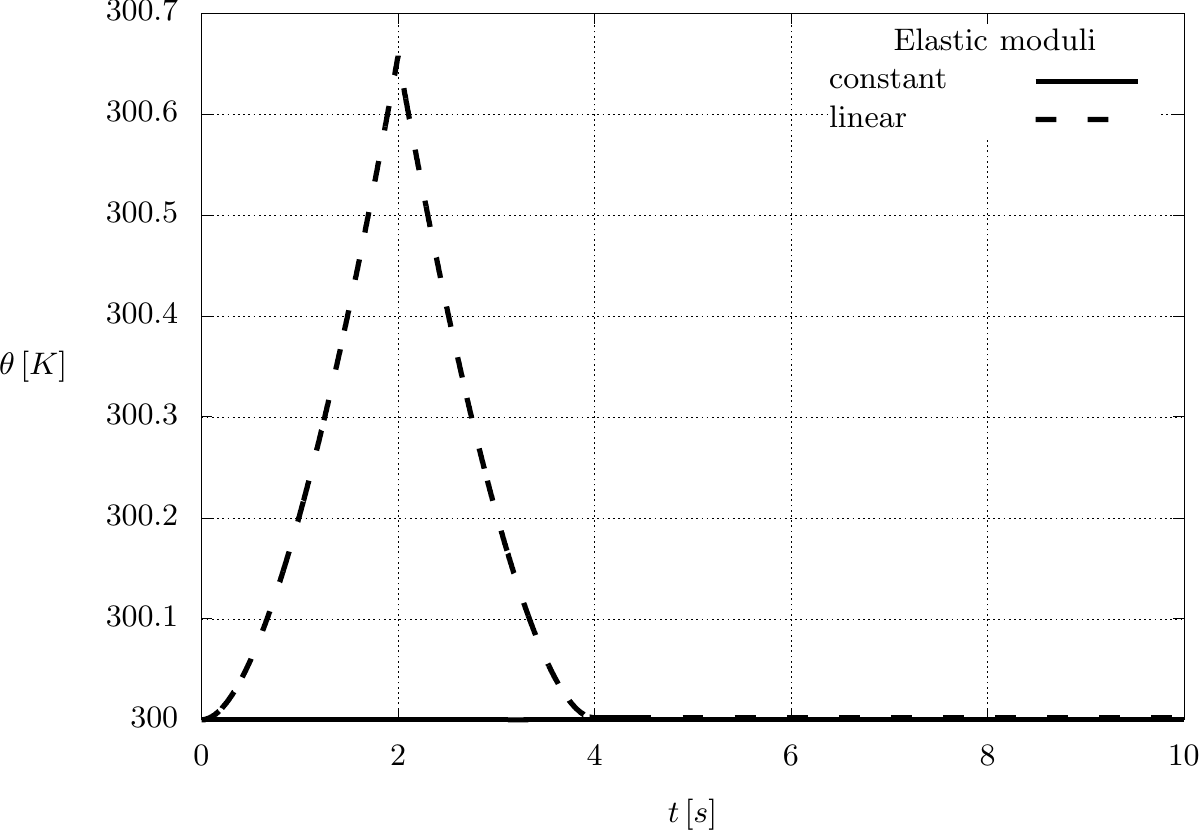}}
  \quad
  \subfloat[Measurement site B.]{\includegraphics[width=0.32\textwidth]{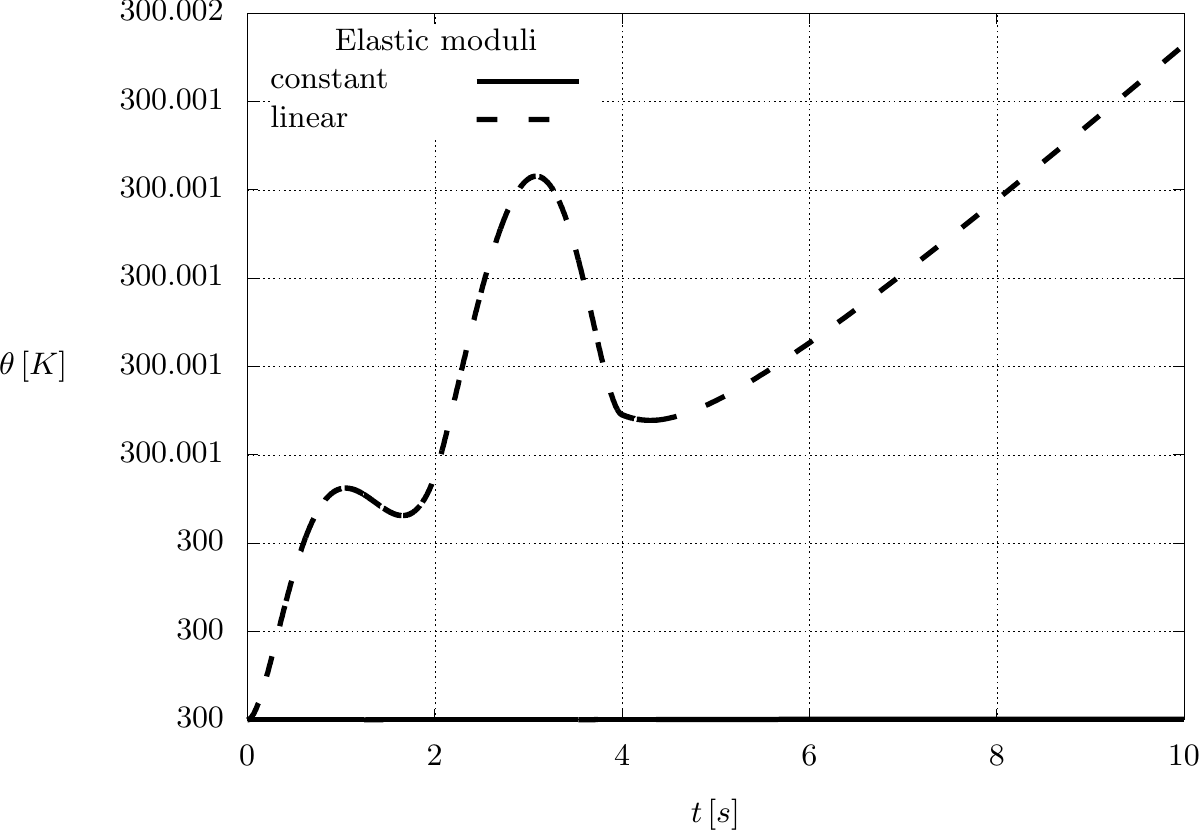}}
  \quad
  \subfloat[Measurement site C.]{\includegraphics[width=0.32\textwidth]{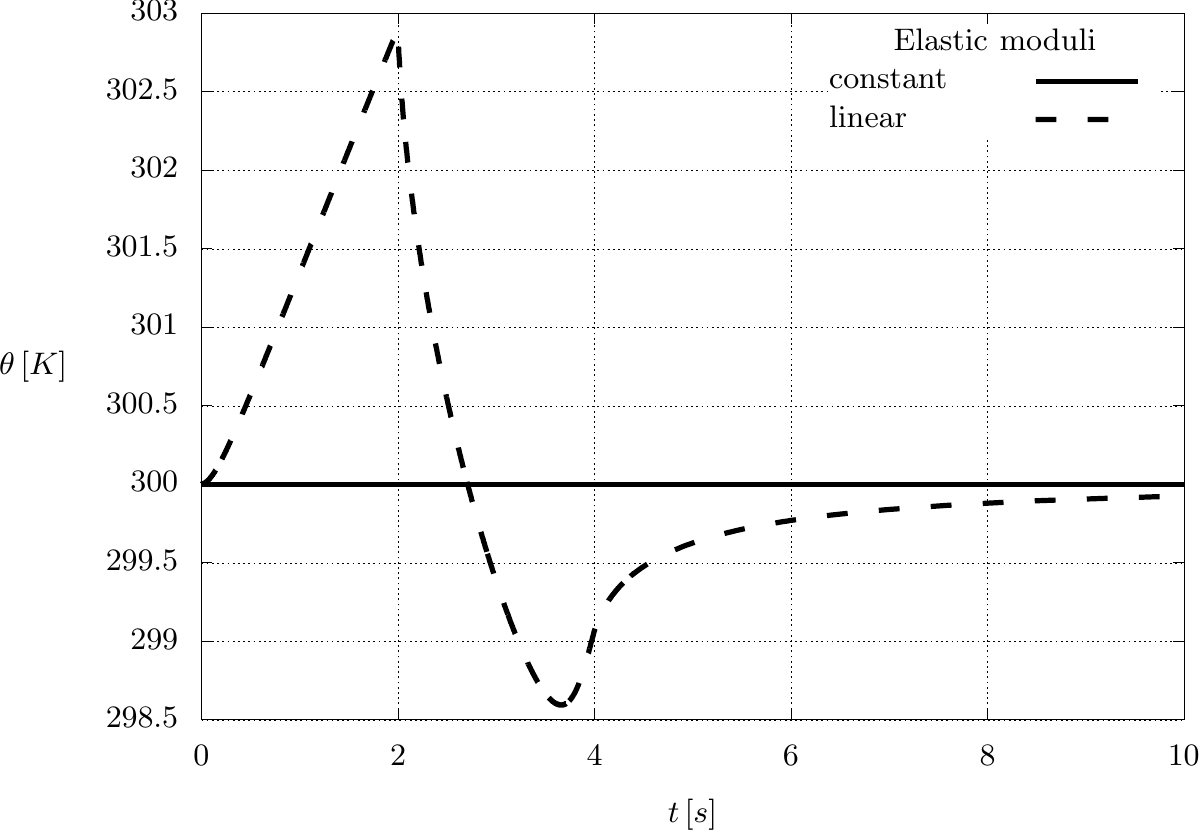}}
  \caption{Temperature values at given measurement sites. Comparison of models with \emph{constant} elastic moduli and \emph{temperature dependent} elastic moduli; \emph{tiny viscosity}. Nomenclature for viscosity values is described in Table~\ref{tab:parameter-values-viscosity}, remaining material parameters are given in Table~\ref{tab:parameter-values} and Table~\ref{tab:parameter-values-elastic}.}
  \label{fig:temperature-sites-linear}
\end{figure}

\begin{figure}[h]
  \subfloat[Measurement site A.]{\includegraphics[width=0.32\textwidth]{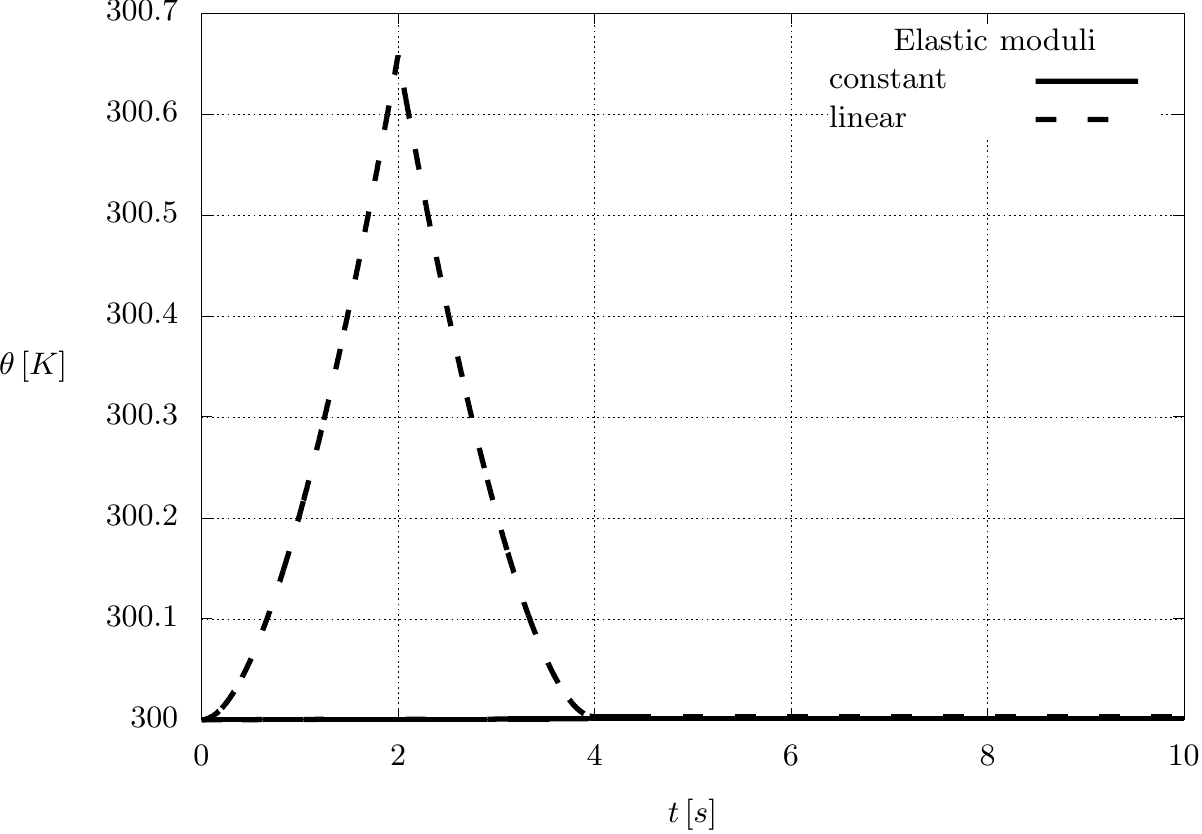}}
  \quad
  \subfloat[Measurement site B.]{\includegraphics[width=0.32\textwidth]{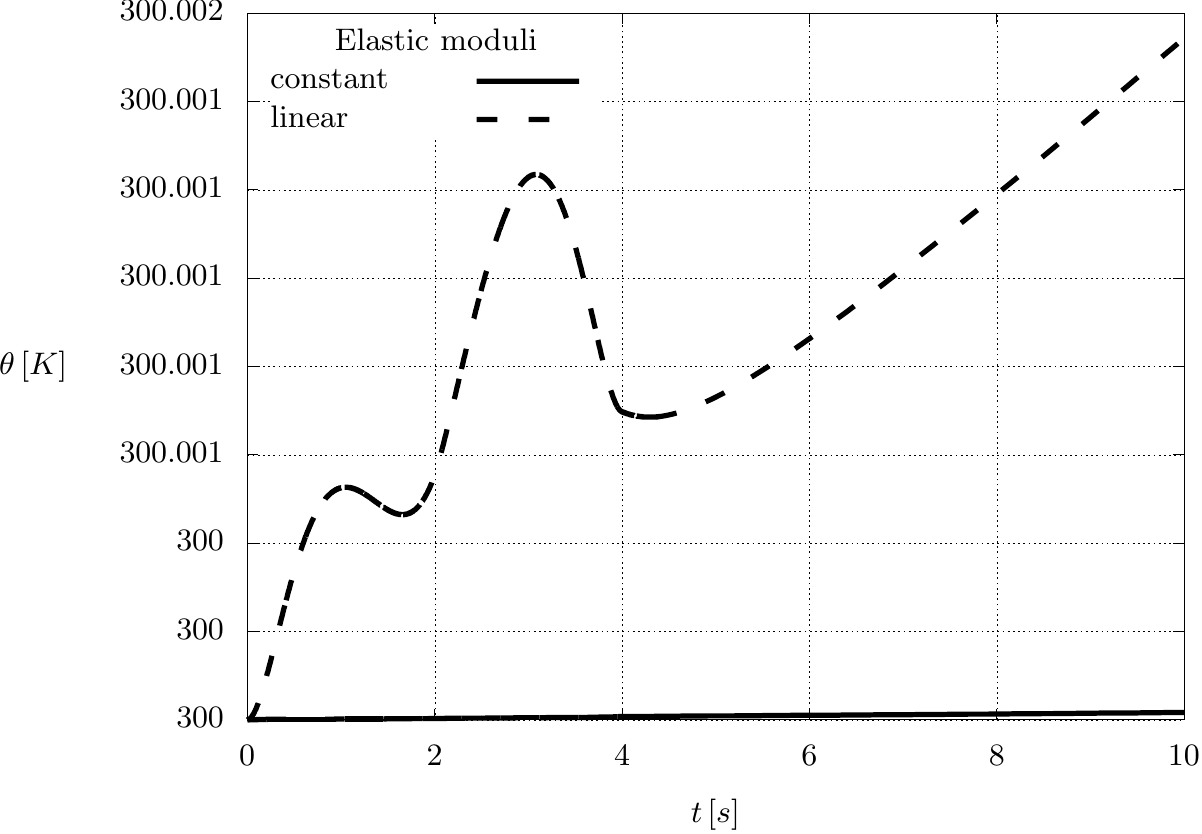}}
  \quad
  \subfloat[Measurement site C.]{\includegraphics[width=0.32\textwidth]{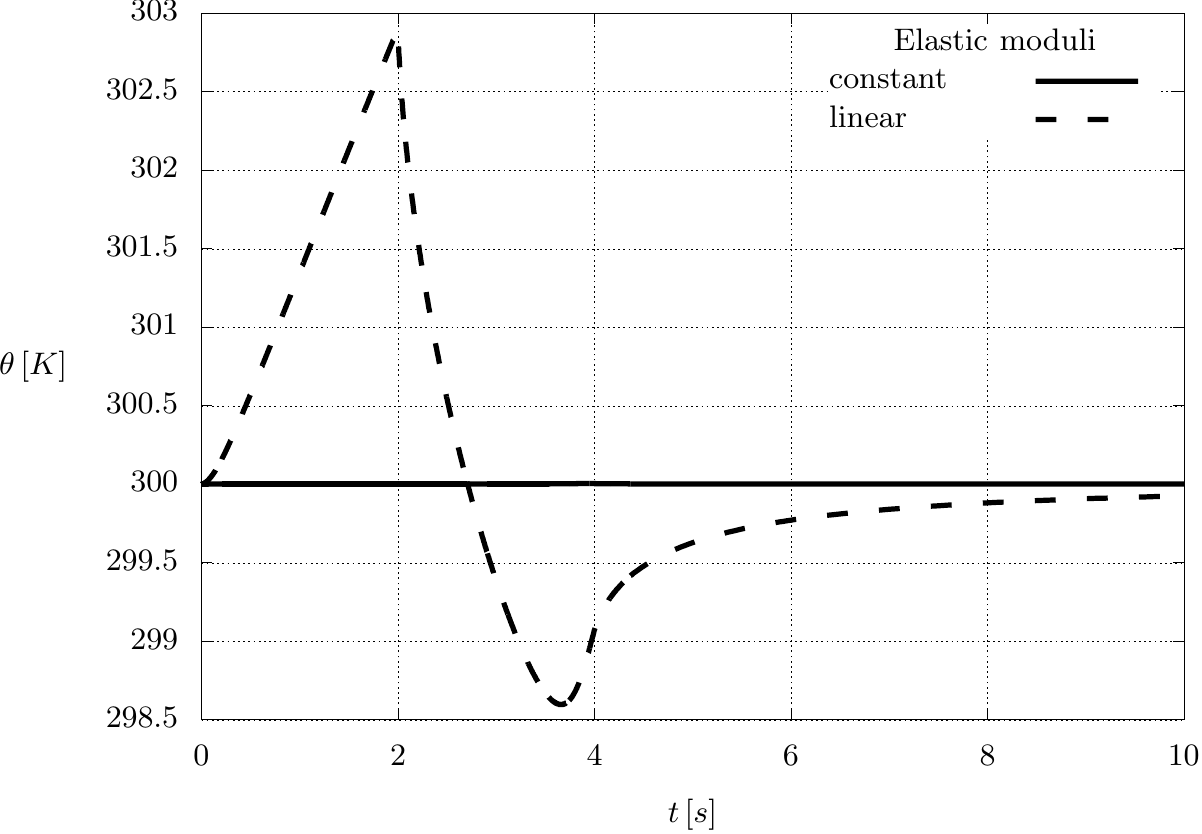}}
  \caption{Temperature values at given measurement sites. Comparison of models with \emph{constant} elastic moduli and \emph{temperature dependent} elastic moduli; \emph{small viscosity}. Nomenclature for viscosity values is described in Table~\ref{tab:parameter-values-viscosity}, remaining material parameters are given in Table~\ref{tab:parameter-values} and Table~\ref{tab:parameter-values-elastic}.}
  \label{fig:temperature-sites-linear-small}
\end{figure}

\begin{figure}[h]
  \subfloat[Measurement site A.]{\includegraphics[width=0.32\textwidth]{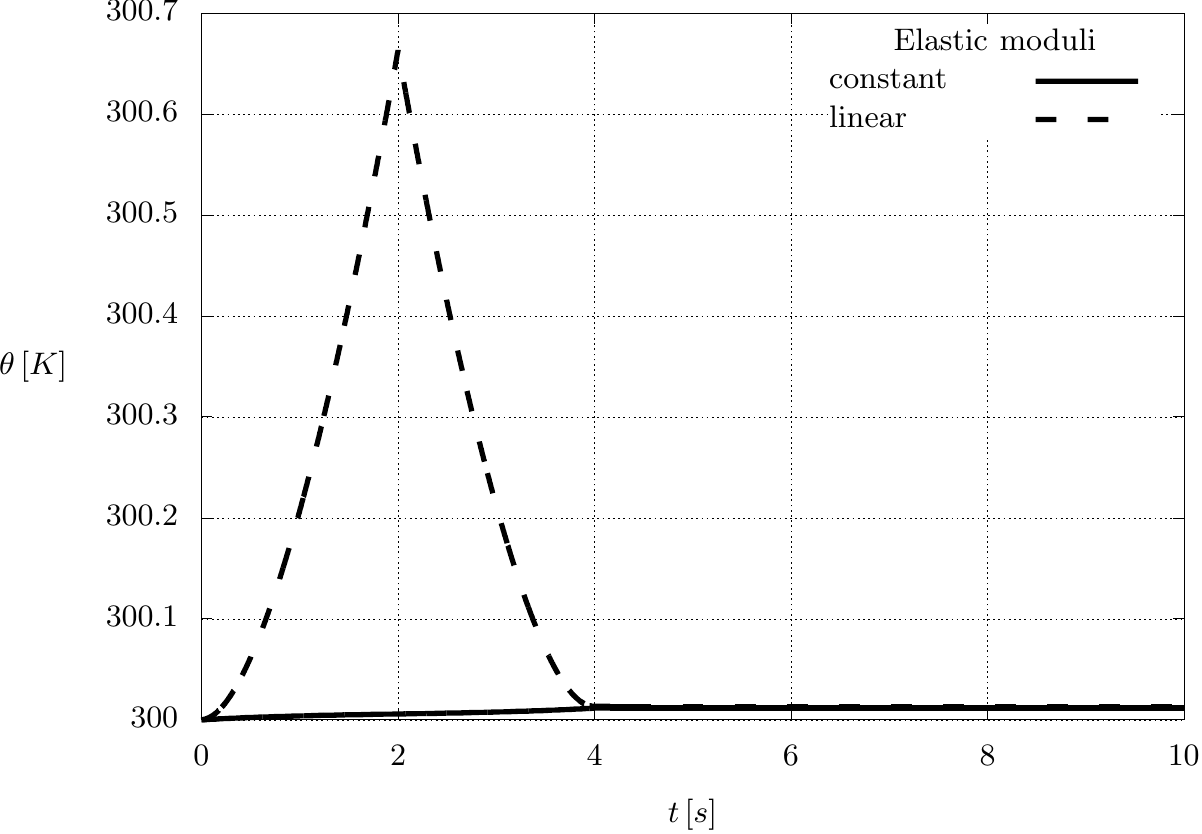}}
  \quad
  \subfloat[Measurement site B.]{\includegraphics[width=0.32\textwidth]{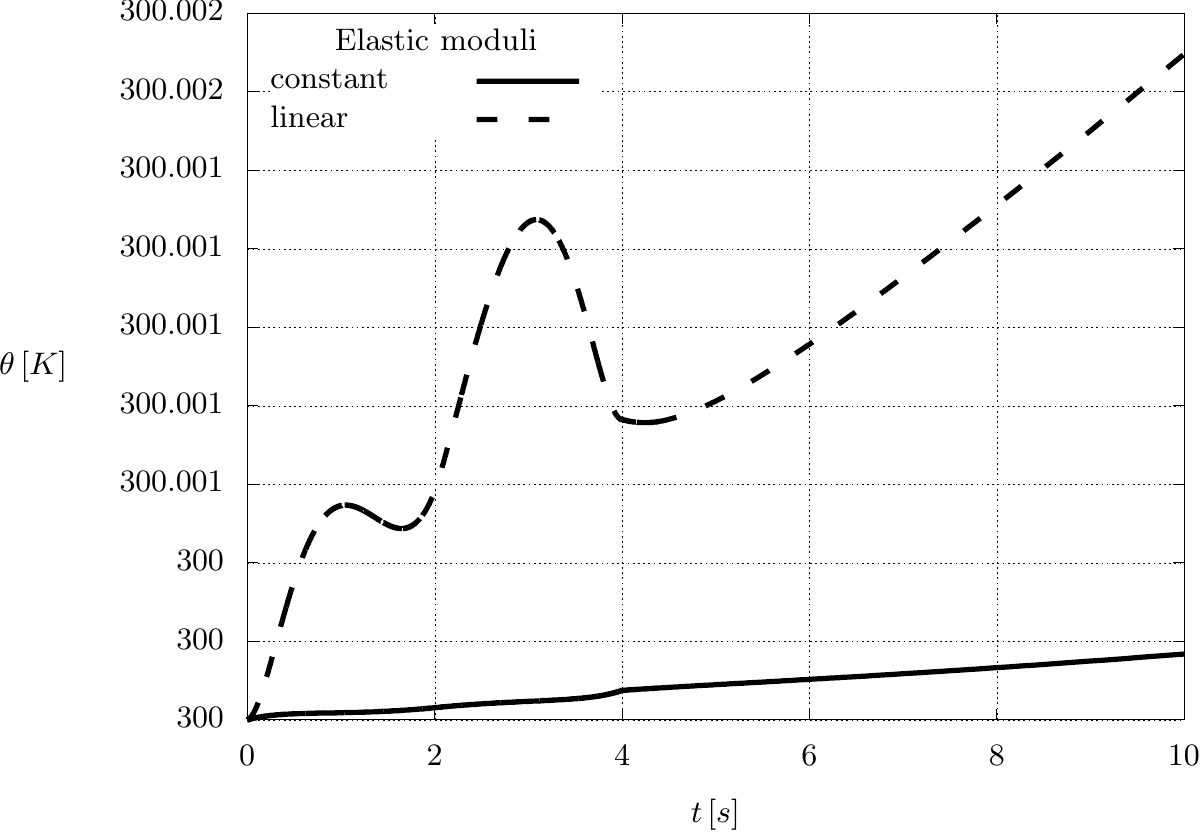}}
  \quad
  \subfloat[Measurement site C.]{\includegraphics[width=0.32\textwidth]{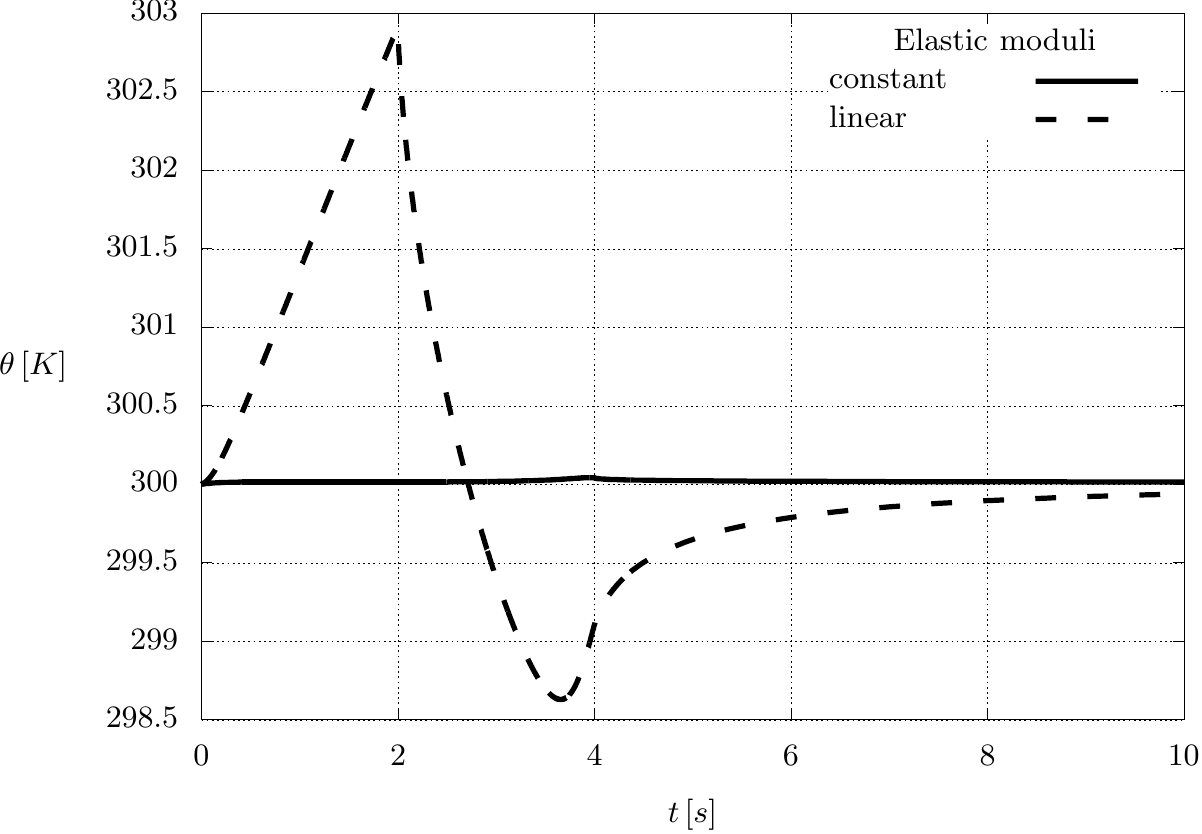}}
  \caption{Temperature values at given measurement sites. Comparison of models with \emph{constant} elastic moduli and \emph{temperature dependent} elastic moduli; \emph{medium viscosity}. Nomenclature for viscosity values is described in Table~\ref{tab:parameter-values-viscosity}, remaining material parameters are given in Table~\ref{tab:parameter-values} and Table~\ref{tab:parameter-values-elastic}.}
  \label{fig:temperature-sites-linear-medium}
\end{figure}

\begin{figure}[h]
  \subfloat[Measurement site A.]{\includegraphics[width=0.32\textwidth]{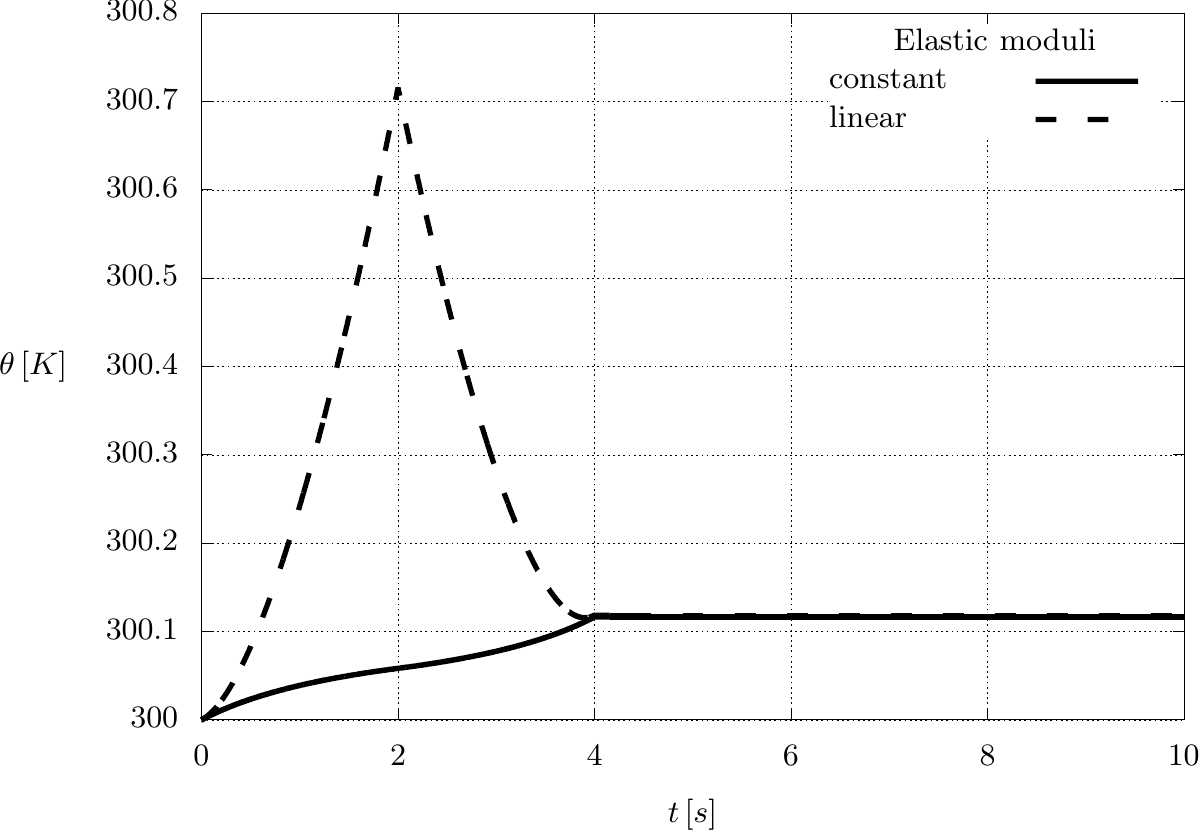}}
  \quad
  \subfloat[Measurement site B.]{\includegraphics[width=0.32\textwidth]{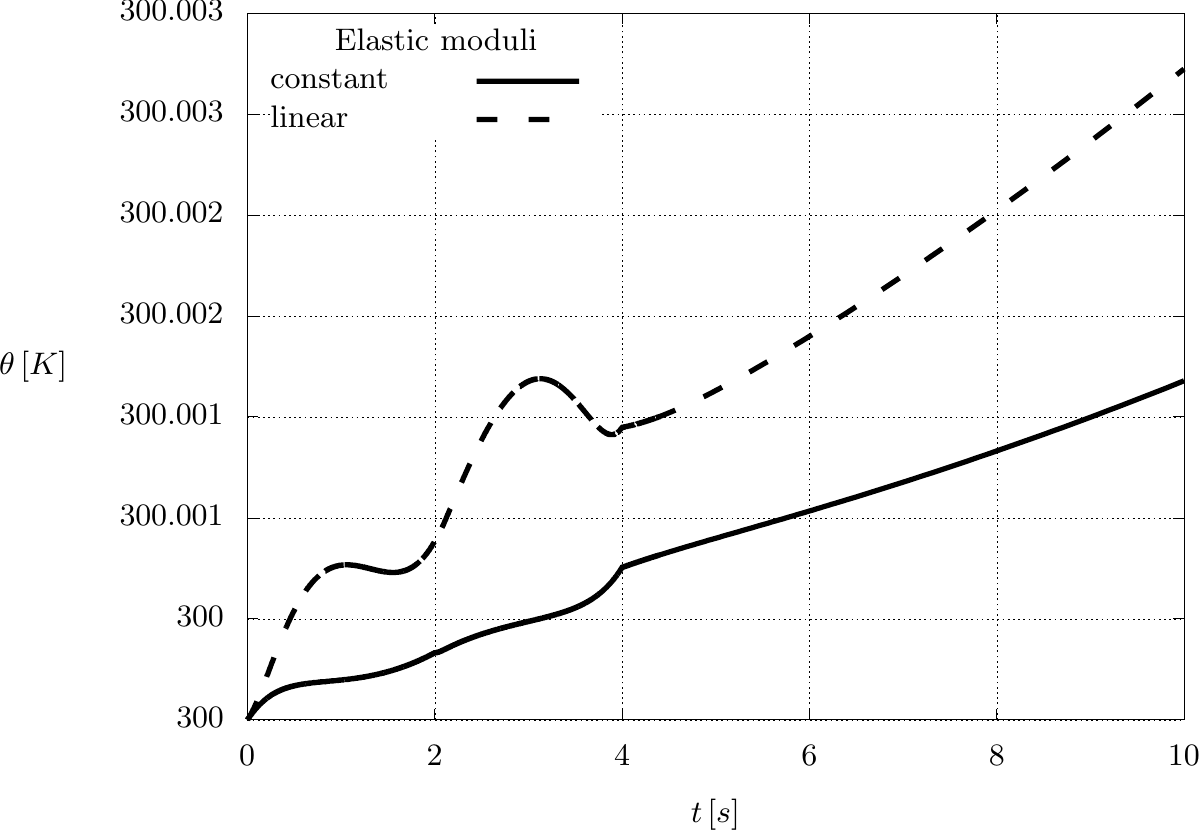}}
  \quad
  \subfloat[Measurement site C.]{\includegraphics[width=0.32\textwidth]{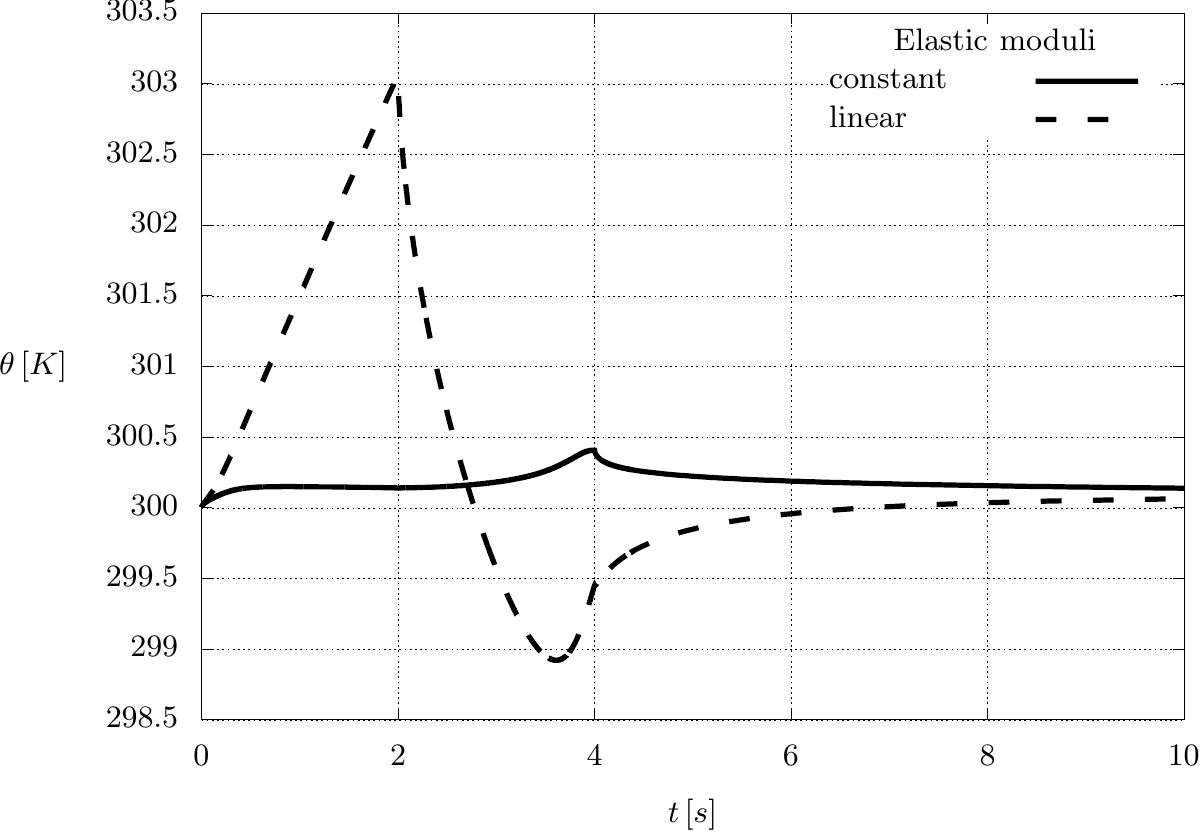}}
  \caption{Temperature values at given measurement sites. Comparison of models with \emph{constant} elastic moduli and \emph{temperature dependent} elastic moduli; \emph{large viscosity}. Nomenclature for viscosity values is described in Table~\ref{tab:parameter-values-viscosity}, remaining material parameters are given in Table~\ref{tab:parameter-values} and Table~\ref{tab:parameter-values-elastic}.}
  \label{fig:temperature-sites-linear-large}
\end{figure}

\begin{figure}[h]
  \subfloat[Measurement site A.]{\includegraphics[width=0.32\textwidth]{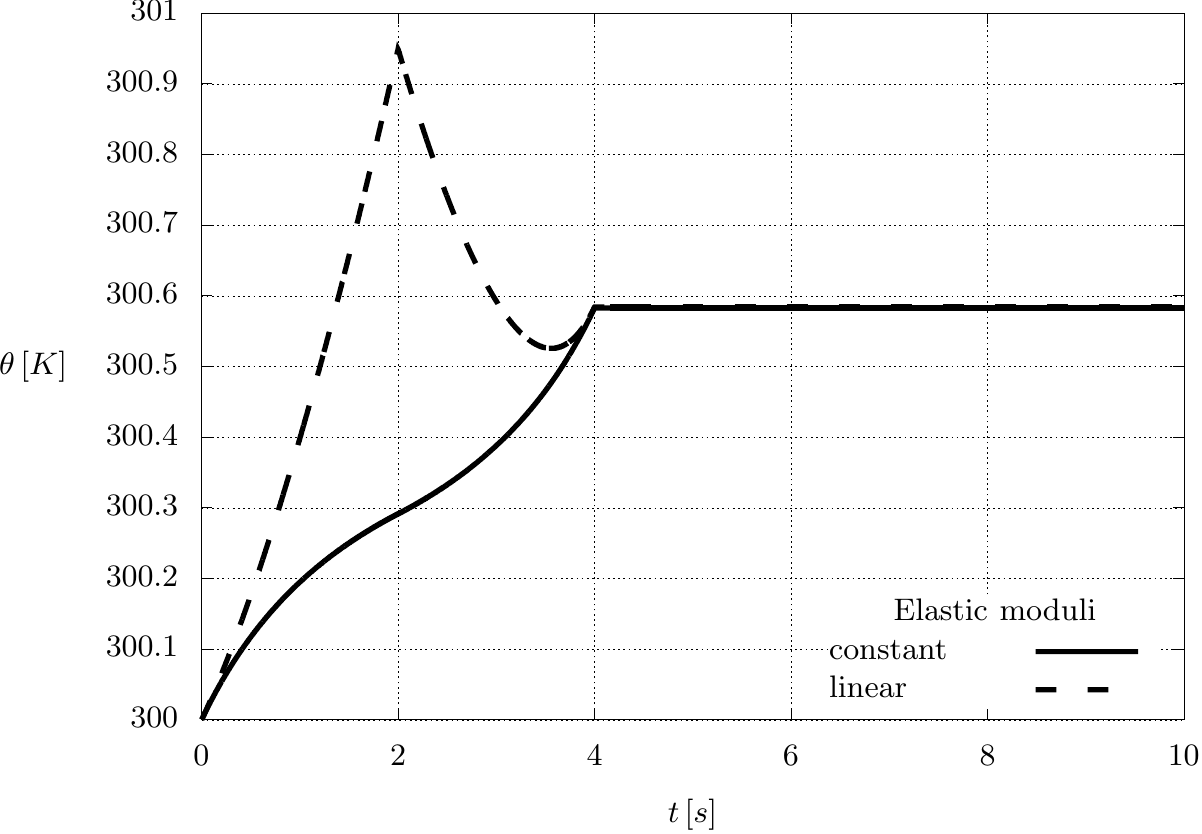}}
  \quad
  \subfloat[Measurement site B.]{\includegraphics[width=0.32\textwidth]{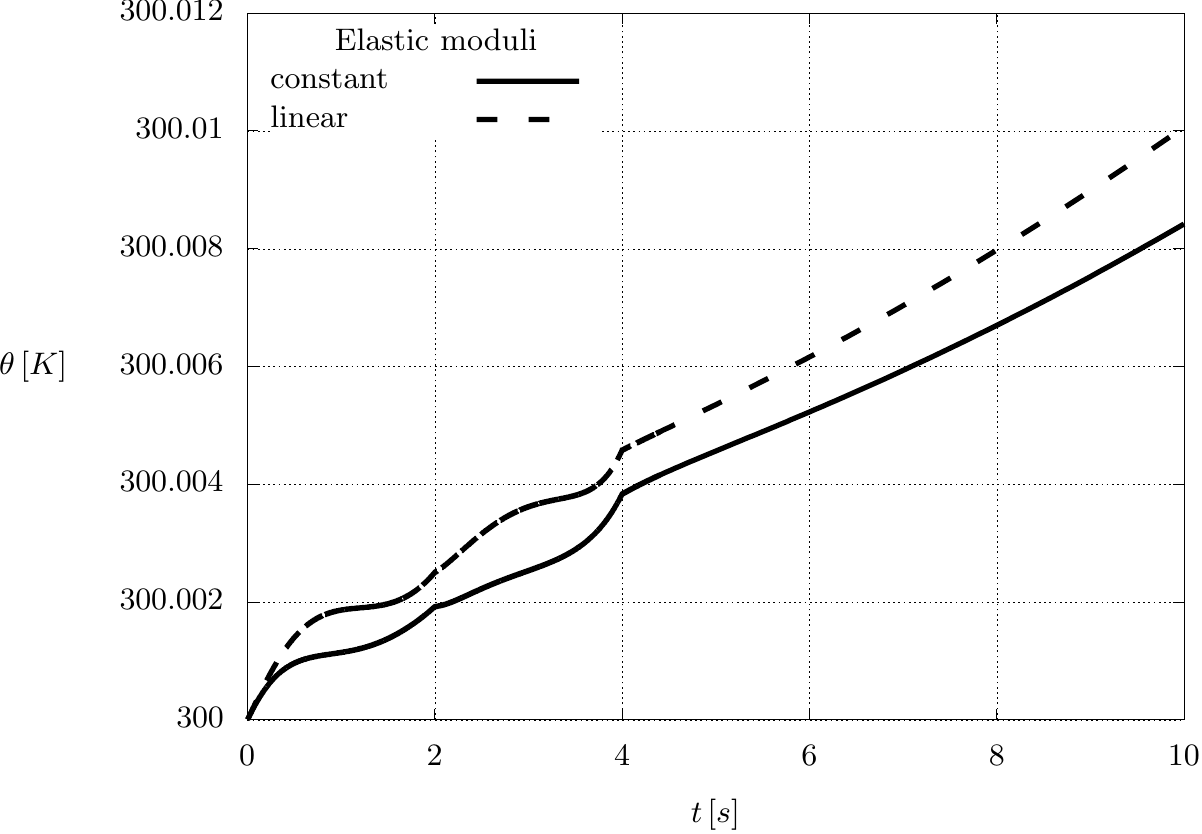}}
  \quad
  \subfloat[Measurement site C.]{\includegraphics[width=0.32\textwidth]{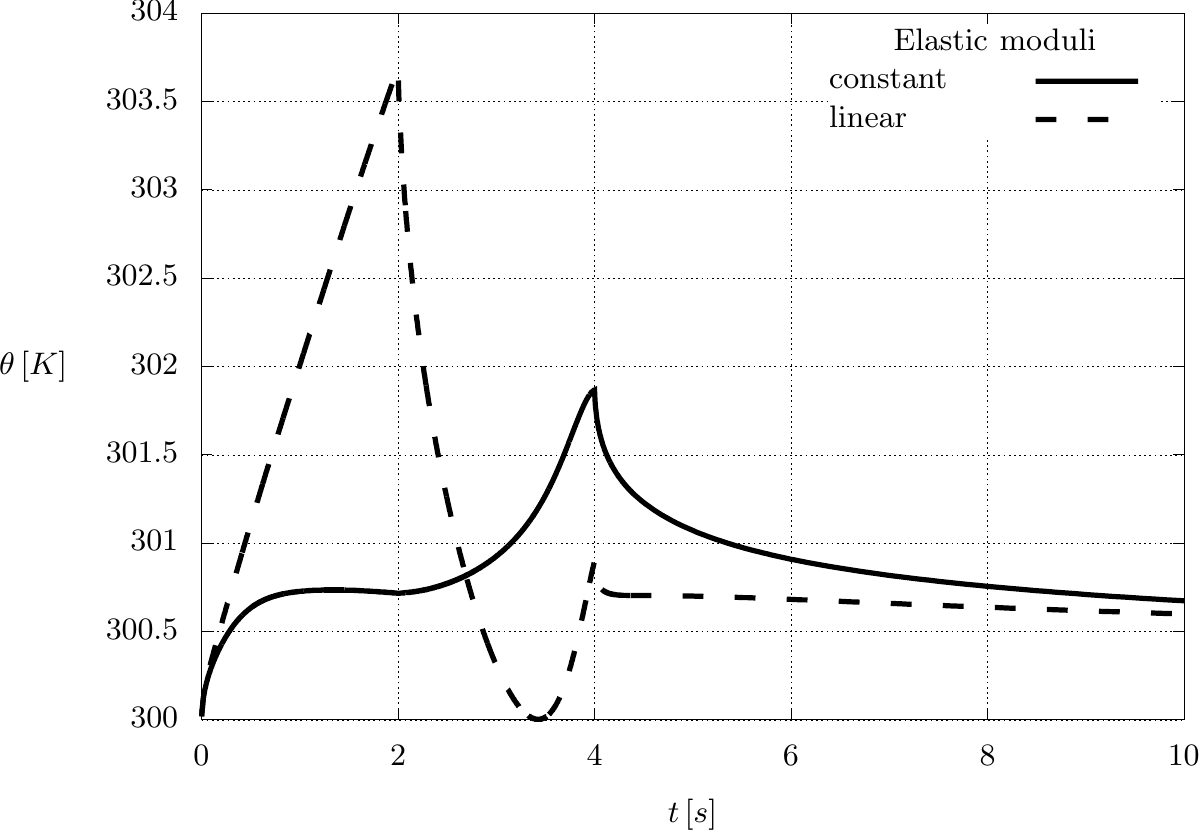}}
  \caption{Temperature values at given measurement sites. Comparison of models with \emph{constant} elastic moduli and \emph{temperature dependent} elastic moduli; \emph{s-large viscosity}. Nomenclature for viscosity values is described in Table~\ref{tab:parameter-values-viscosity}, remaining material parameters are given in Table~\ref{tab:parameter-values} and Table~\ref{tab:parameter-values-elastic}.}
  \label{fig:temperature-sites-linear-s-large}
\end{figure}

\begin{figure}[h]
  \subfloat[Measurement site A.]{\includegraphics[width=0.32\textwidth]{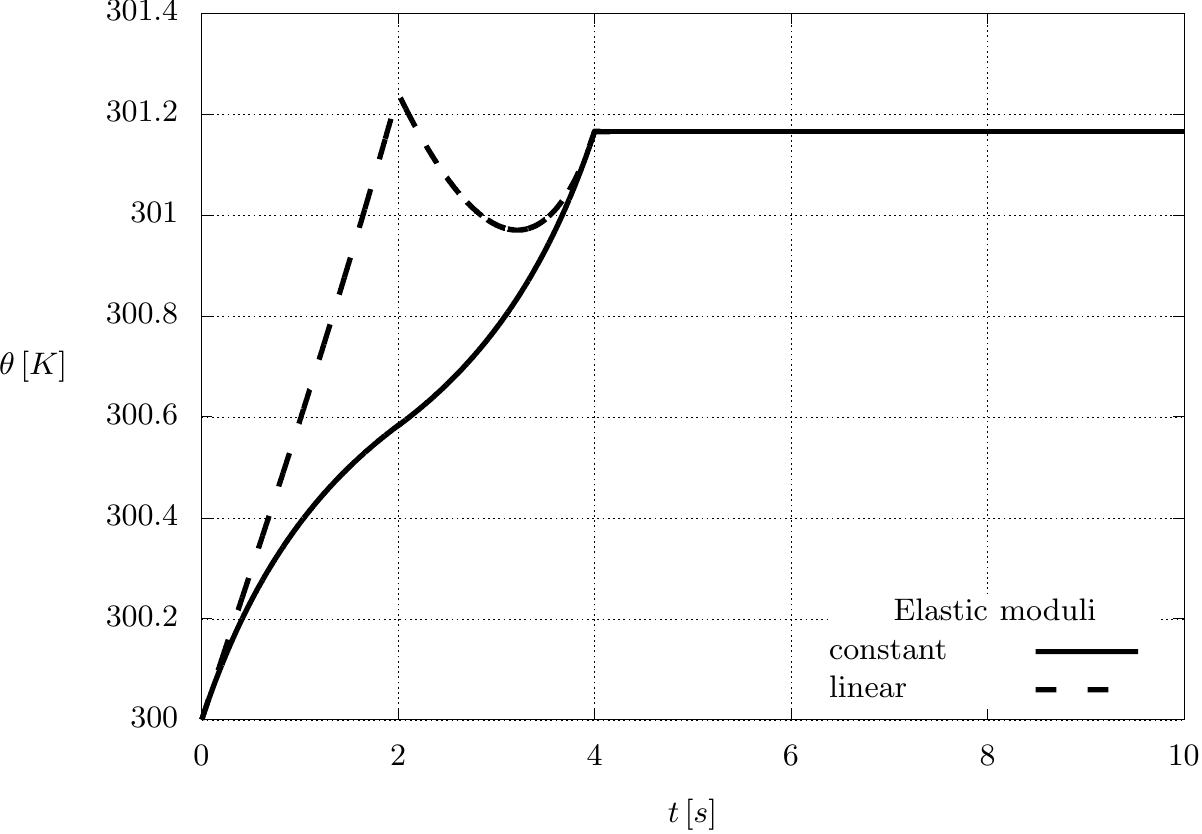}}
  \quad
  \subfloat[Measurement site B.]{\includegraphics[width=0.32\textwidth]{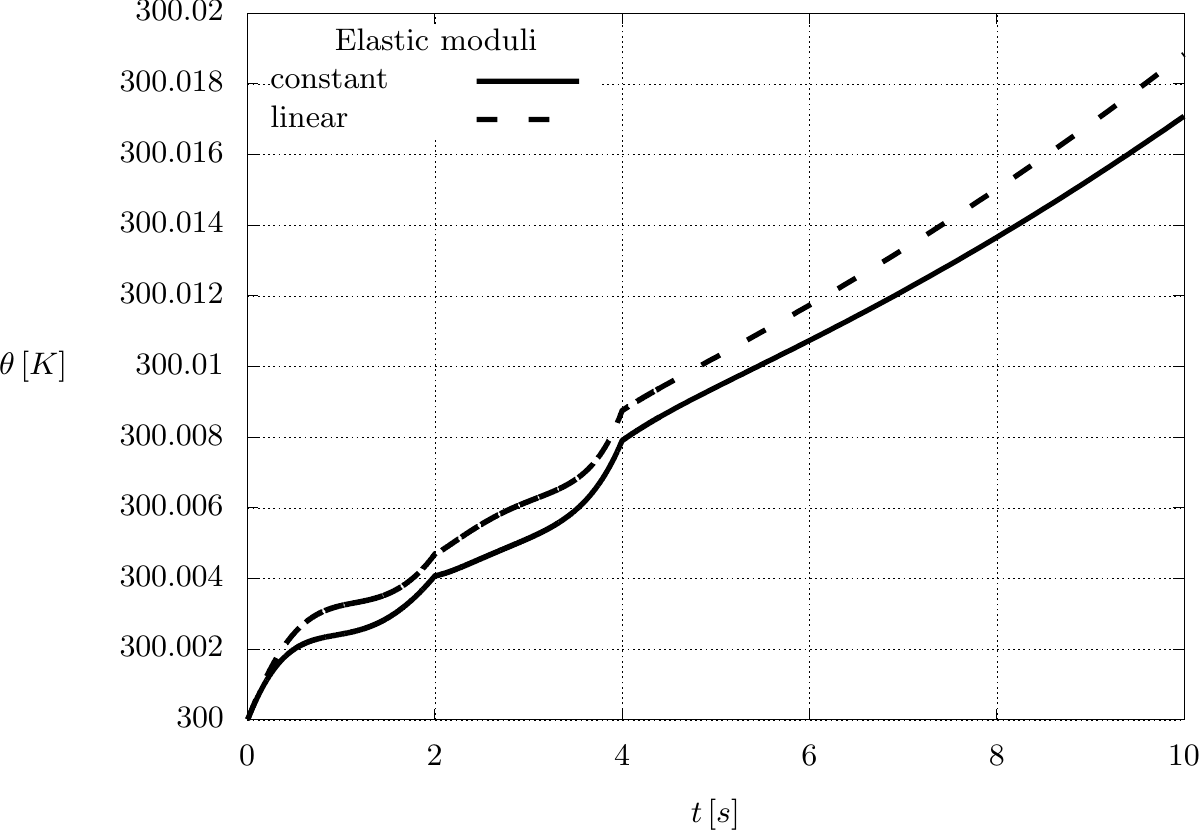}}
  \quad
  \subfloat[Measurement site C.]{\includegraphics[width=0.32\textwidth]{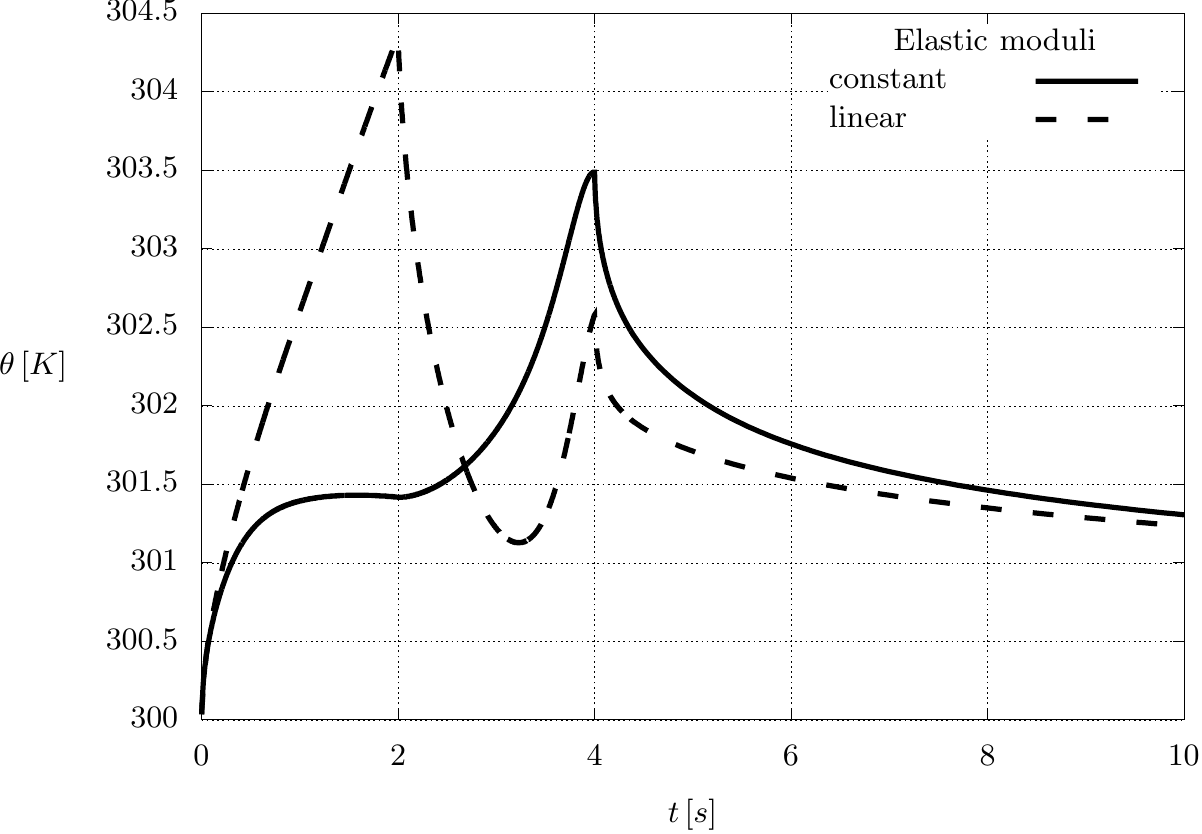}}
  \caption{Temperature values at given measurement sites. Comparison of models with \emph{constant} elastic moduli and \emph{temperature dependent} elastic moduli; \emph{x-large viscosity}. Nomenclature for viscosity values is described in Table~\ref{tab:parameter-values-viscosity}, remaining material parameters are given in Table~\ref{tab:parameter-values} and Table~\ref{tab:parameter-values-elastic}.}
  \label{fig:temperature-sites-linear-x-large}
\end{figure}

Figure~\ref{fig:temperature-sites-linear} shows the temperature values at measurement sites A, B and C for ``tiny'' viscosity value. The difference between the temperature values predicted by the model with \emph{constant elastic moduli} and the model with \emph{temperature dependent elastic moduli} (linear dependence on temperature) is for the given parameter values substantial. The same finding holds also for higher viscosity values, see Figure~\ref{fig:temperature-sites-linear}--Figure~\ref{fig:temperature-sites-linear-x-large}. In particular, if the viscosity is small, and if one considers models with \emph{constant elastic moduli} the temperature changes are in fact negligible and perhaps experimentally undetectable. Yet we can consistently compute and detect such small temperature variations in our numerical experiments.

On the other hand the temperature changes predicted by the models with~\emph{temperature dependent elastic moduli} are quite strong, and these temperature changes occur even for small viscosity values. The numerical experiment therefore shows that the temperature field is for these viscosities predominantly influenced by the non-dissipative mechanisms. (This means that the last two source terms in the temperature evolution equation~\eqref{eq:40}, which could change the temperature field even in a purely elastic material, are dominated by the remaining source terms on the right-hand side of~\eqref{eq:40}.) The dissipative heating, that is the terms in~\eqref{eq:40} that depend on the symmetric part of the velocity gradient $\gradsym_{\vec{X}}$, starts to influence the temperature evolution only for higher viscosities, see for example Figure~\ref{fig:temperature-sites-linear-s-large} and Figure~\ref{fig:temperature-sites-linear-x-large}. (We recall that the \emph{x-large} viscosity is rather extreme.)

\FloatBarrier

\begin{figure}[h]
  \subfloat[Measurement site A.]{\includegraphics[width=0.32\textwidth]{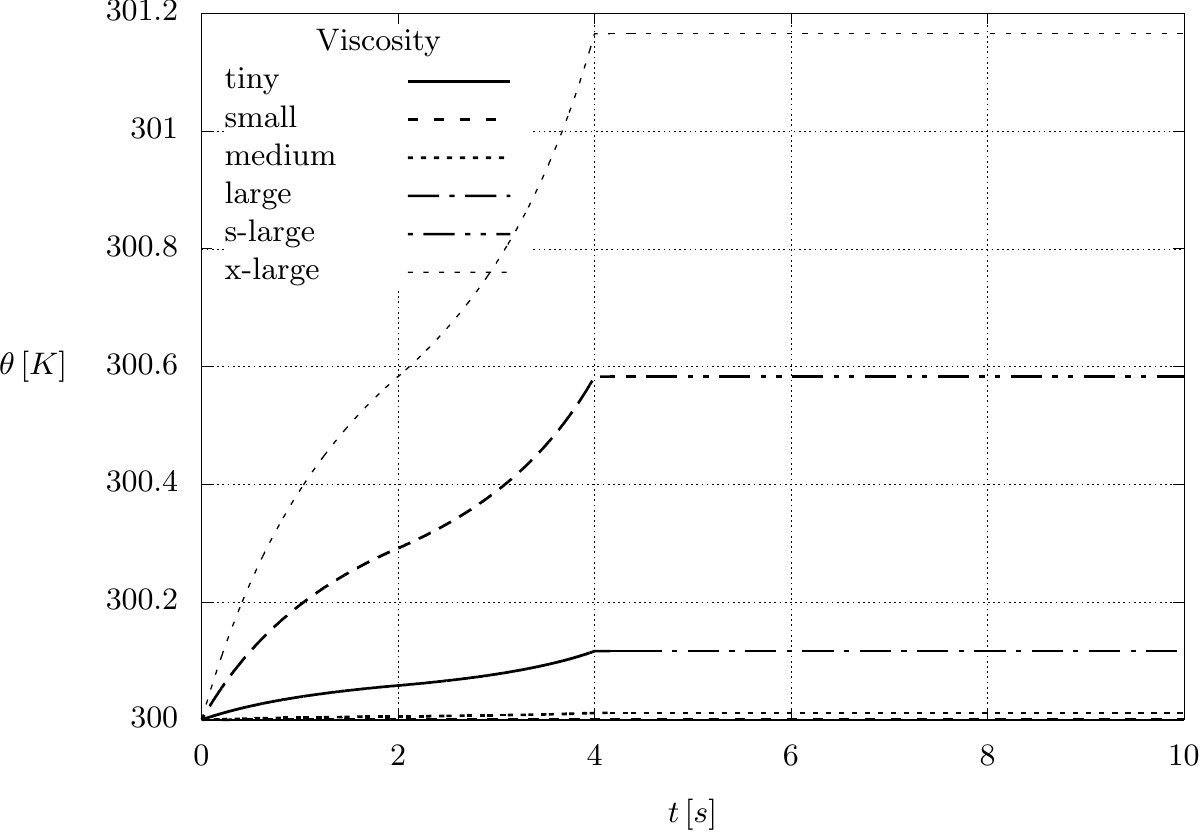}}
  \quad
  \subfloat[Measurement site B.]{\includegraphics[width=0.32\textwidth]{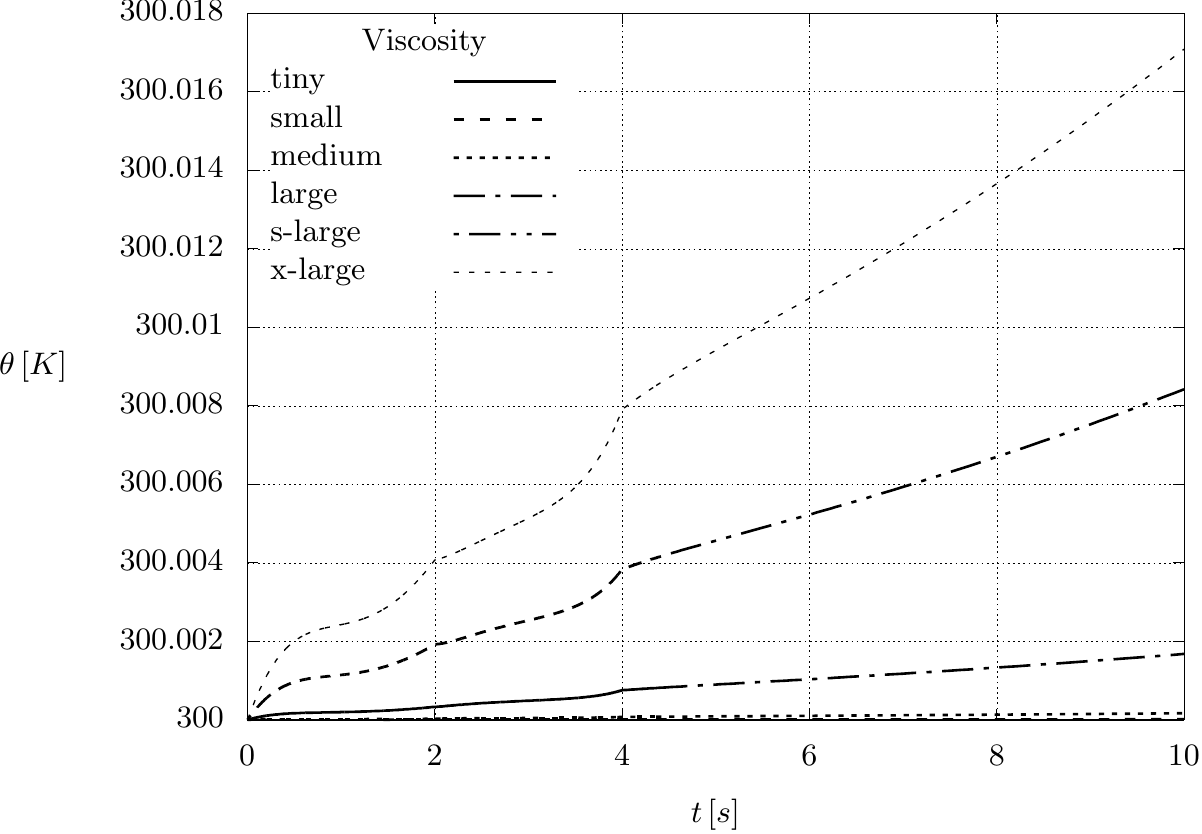}}
  \quad
  \subfloat[Measurement site C.]{\includegraphics[width=0.32\textwidth]{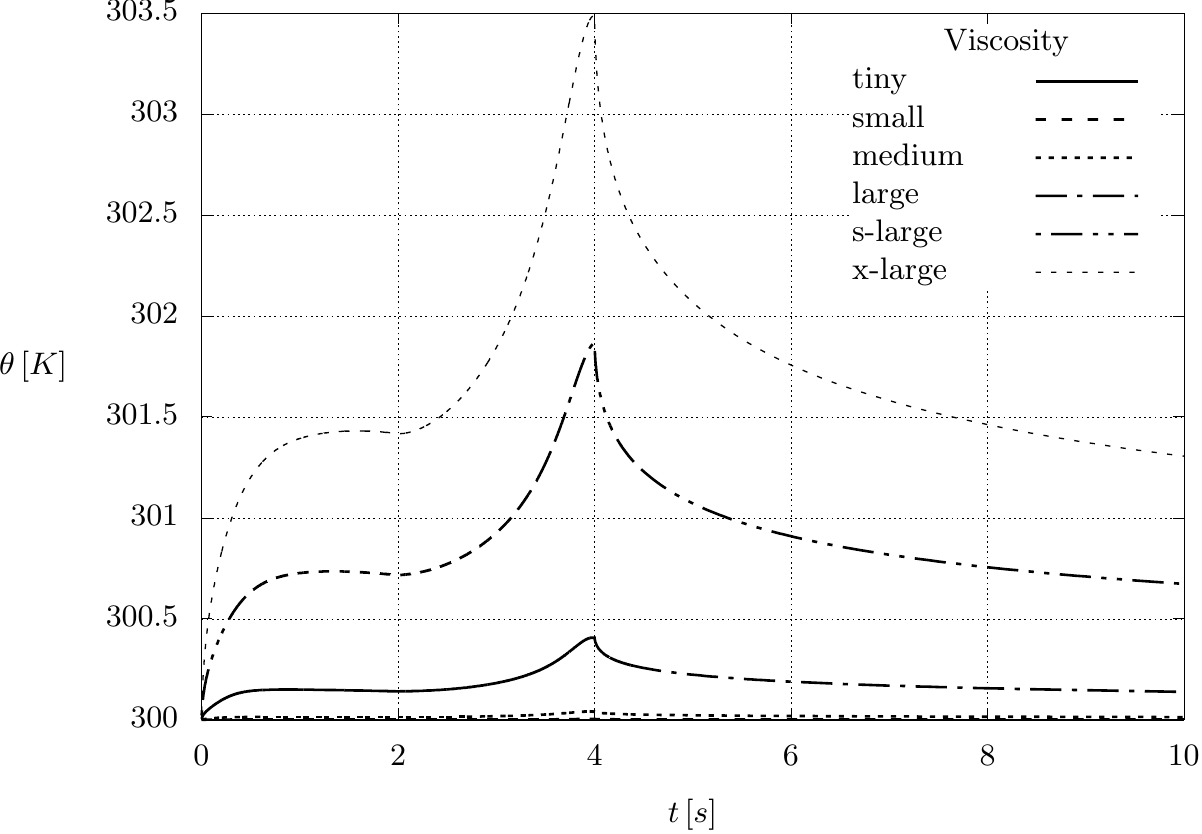}}
  \caption{Temperature values at given measurement sites. Models with \emph{constant} elastic moduli, comparison of different viscosity values. Nomenclature for viscosity values is described in Table~\ref{tab:parameter-values-viscosity}, remaining material parameters are given in Table~\ref{tab:parameter-values} and Table~\ref{tab:parameter-values-elastic}.}
  \label{fig:temperature-sites-constant-various-viscosities}
\end{figure}

\begin{figure}[h]
  \subfloat[Measurement site A.]{\includegraphics[width=0.32\textwidth]{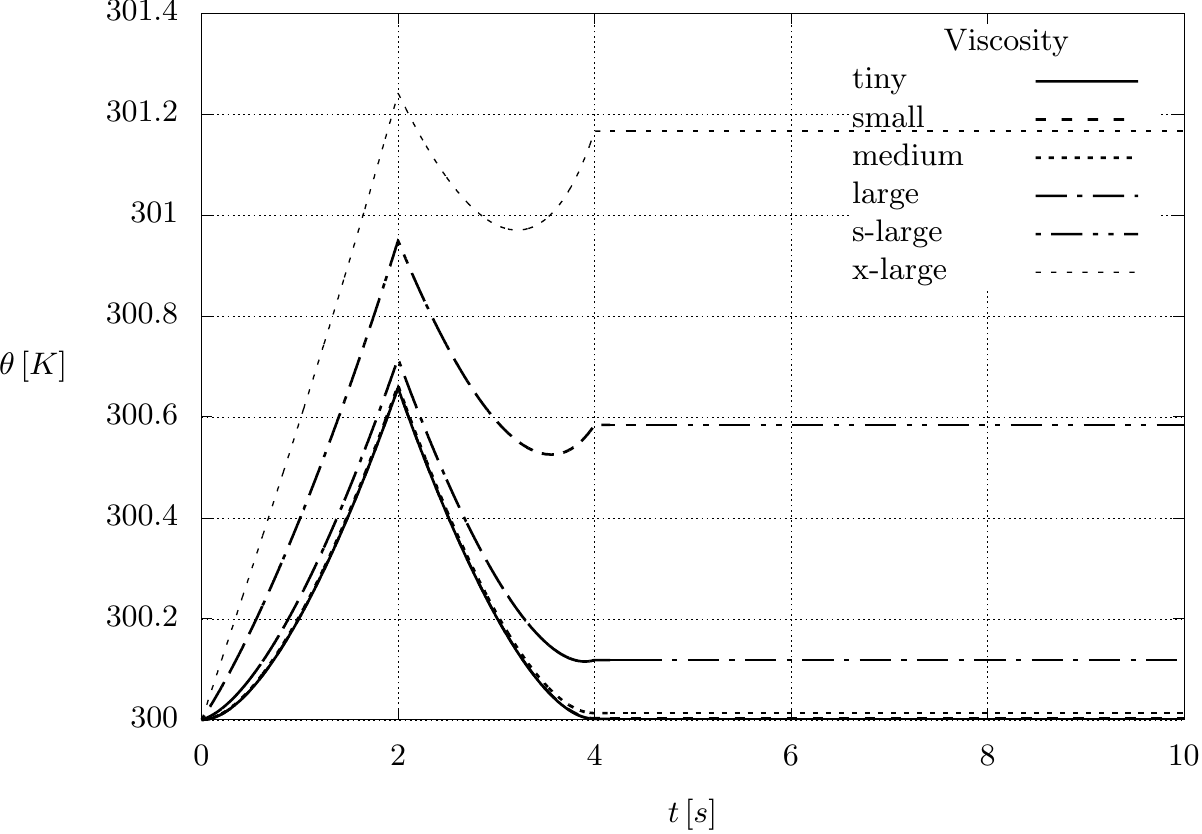}}
  \quad
  \subfloat[Measurement site B.]{\includegraphics[width=0.32\textwidth]{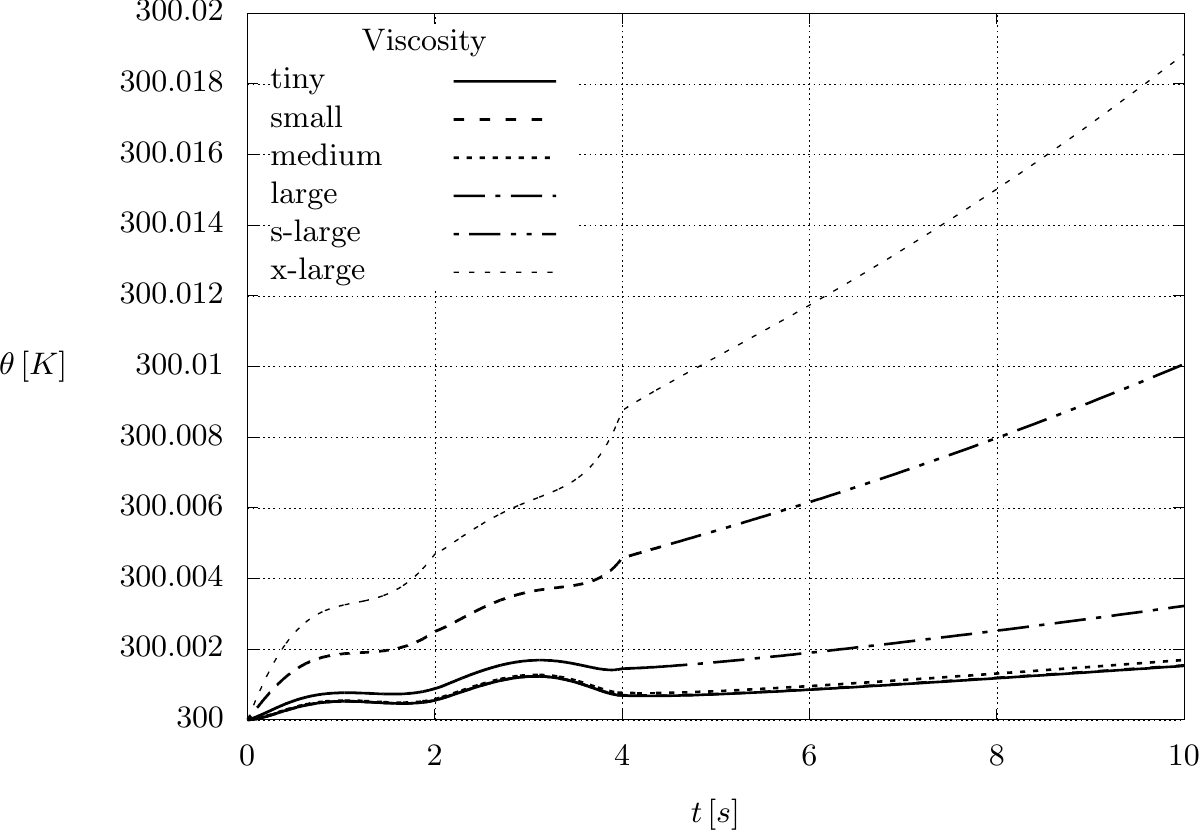}}
  \quad
  \subfloat[Measurement site C.\label{fig:temperature-sites-linear-various-viscosities-c}]{\includegraphics[width=0.32\textwidth]{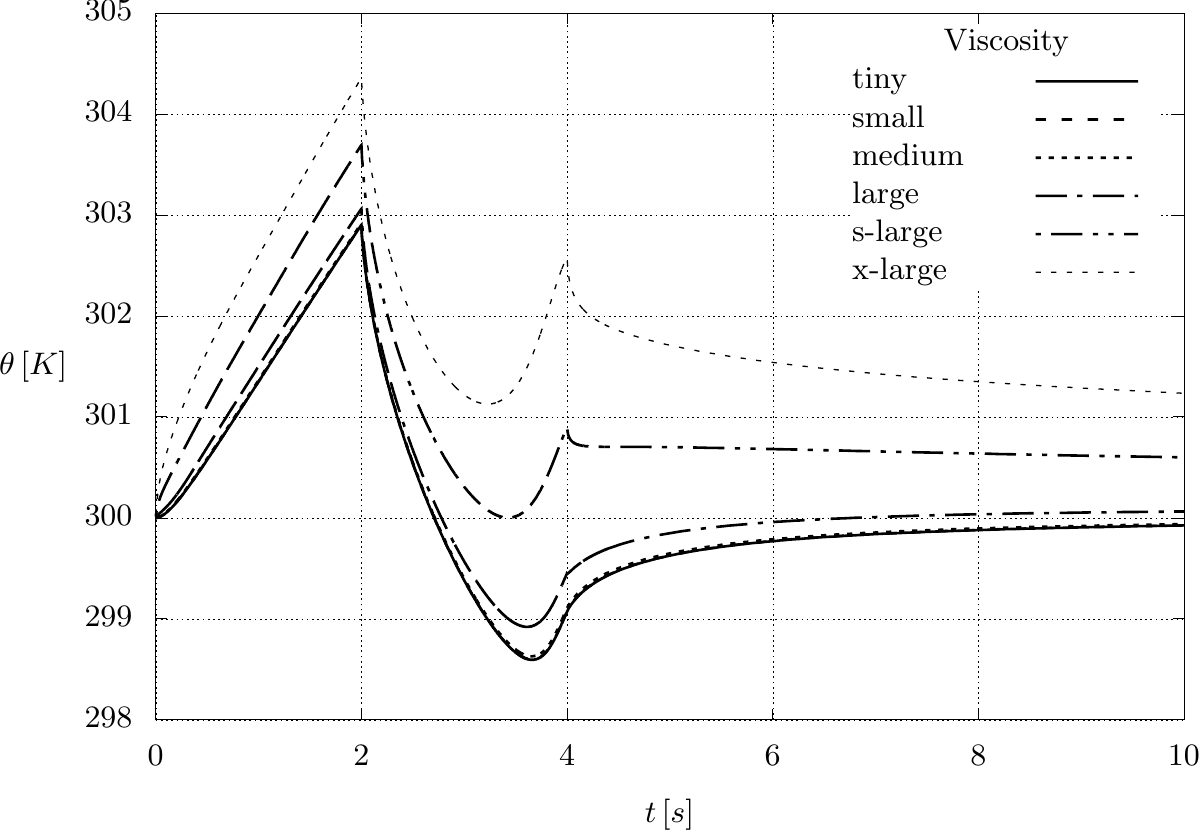}}
  \caption{Temperature values at given measurement sites. Models with \emph{temperature dependent} elastic moduli, comparison of different viscosity values. Nomenclature for viscosity values is described in Table~\ref{tab:parameter-values-viscosity}, remaining material parameters are given in Table~\ref{tab:parameter-values} and Table~\ref{tab:parameter-values-elastic}.}
  \label{fig:temperature-sites-linear-various-viscosities}
\end{figure}

The influence of viscosity on the temperature changes is also visualised in Figure~\ref{fig:temperature-sites-constant-various-viscosities} and Figure~\ref{fig:temperature-sites-linear-various-viscosities}. An inspection of these figures reveals that---in the given parameter range---the models with entropic elasticity alone, that is the models with the \emph{temperature dependent (linearly) elastic moduli} and (almost) no viscosity, predict strong temperature field changes especially in the vicinity of the cutout tip (measurement site C). Interestingly, the temperature values in the deforming material can even~\emph{drop} below the initial temperature value $\temp_{\reference} = \unit[300]{K}$, see especially Figure~\ref{fig:temperature-sites-linear-various-viscosities-c}. This phenomenon stops to occur once the viscosity is high enough, which in our setting happens for viscosities of order \emph{s-large}.

On the other hand the models with \emph{constant elastic moduli} predict monotonous increase of temperature at all measurement sites during the loading--unloading cycle, that is for $t \in [0,4]$, see Figure~\ref{fig:temperature-sites-constant-various-viscosities}. We can also again observe that for small viscosities this class of models predicts negligible temperature changes, see again Figure~\ref{fig:temperature-sites-constant-various-viscosities}.

\FloatBarrier

\subsubsection{Long time behaviour}
\label{sec:long-time-behaviour}
\begin{figure}[h]
  \subfloat[Tiny viscosity.]{\includegraphics[width=0.32\textwidth]{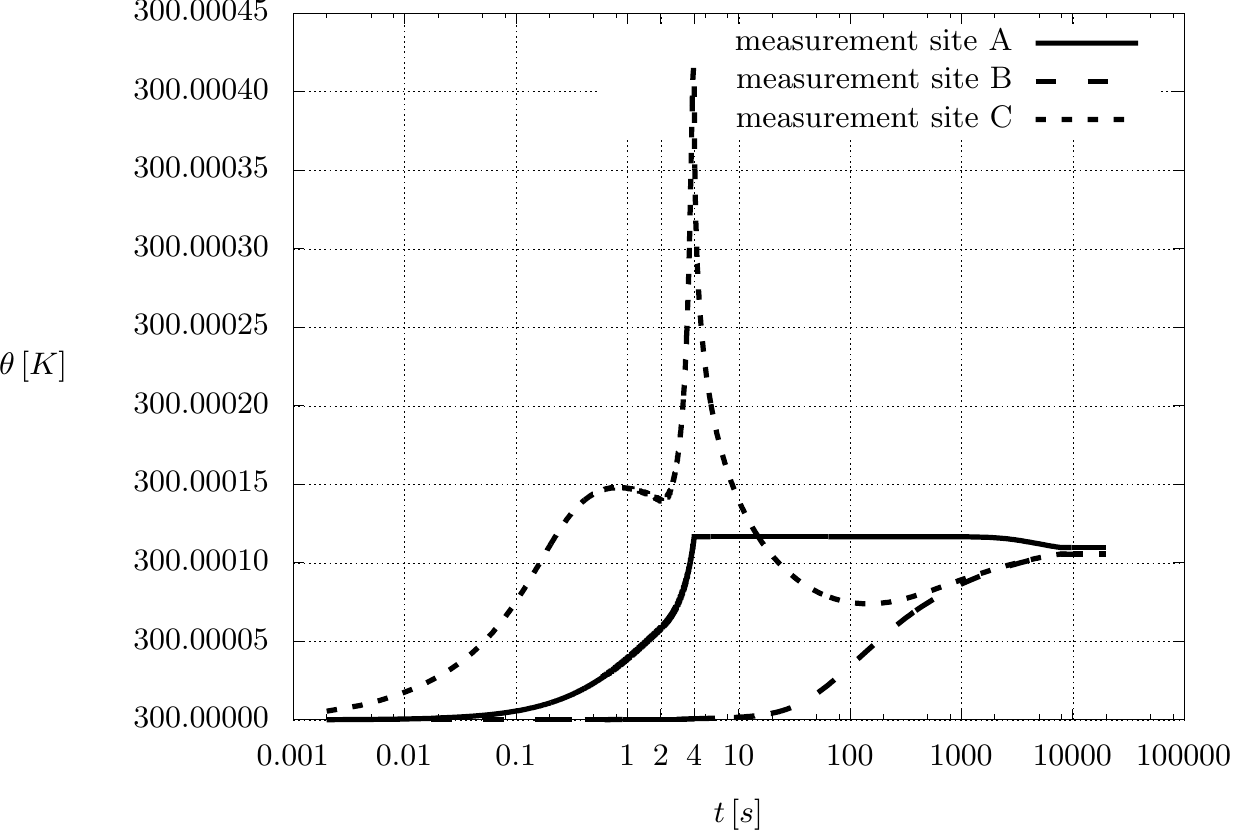}}
  \quad
  \subfloat[Small viscosity.]{\includegraphics[width=0.32\textwidth]{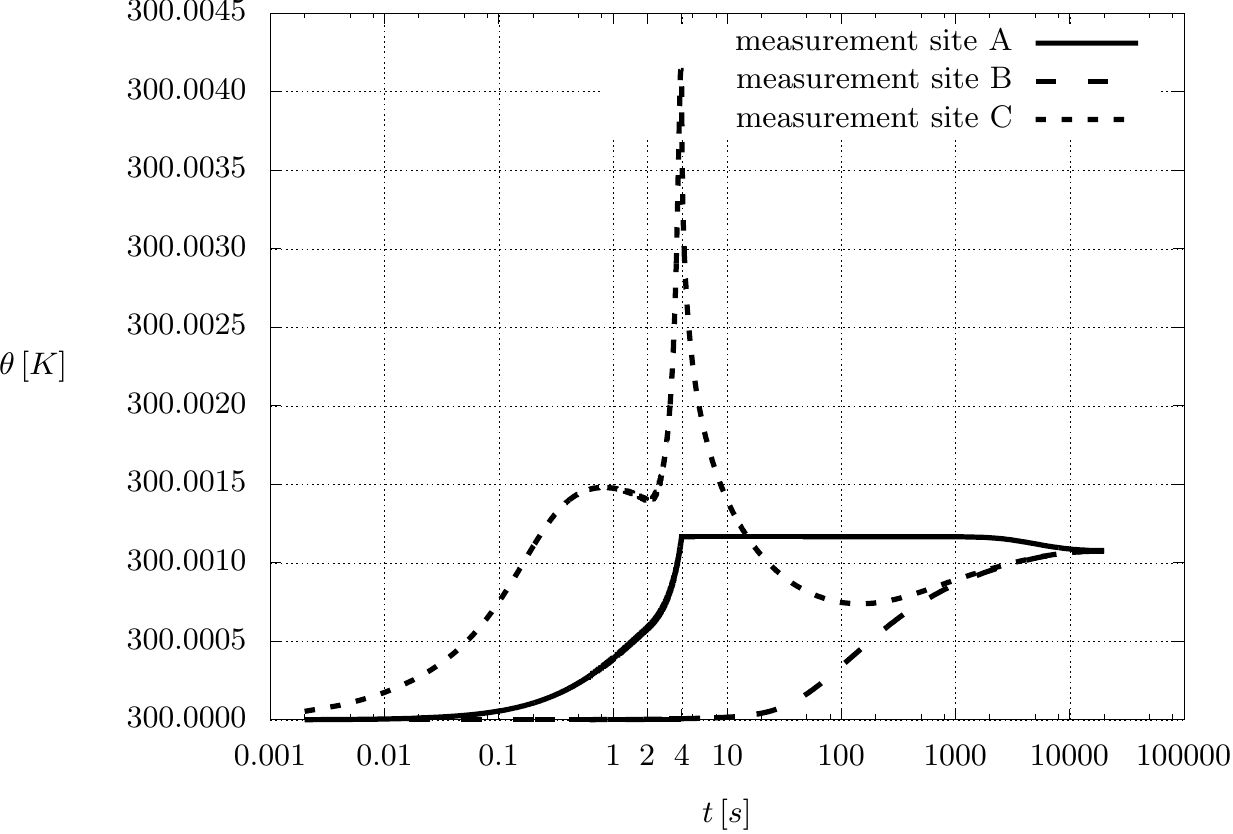}}
  \quad
  \subfloat[Medium viscosity.]{\includegraphics[width=0.32\textwidth]{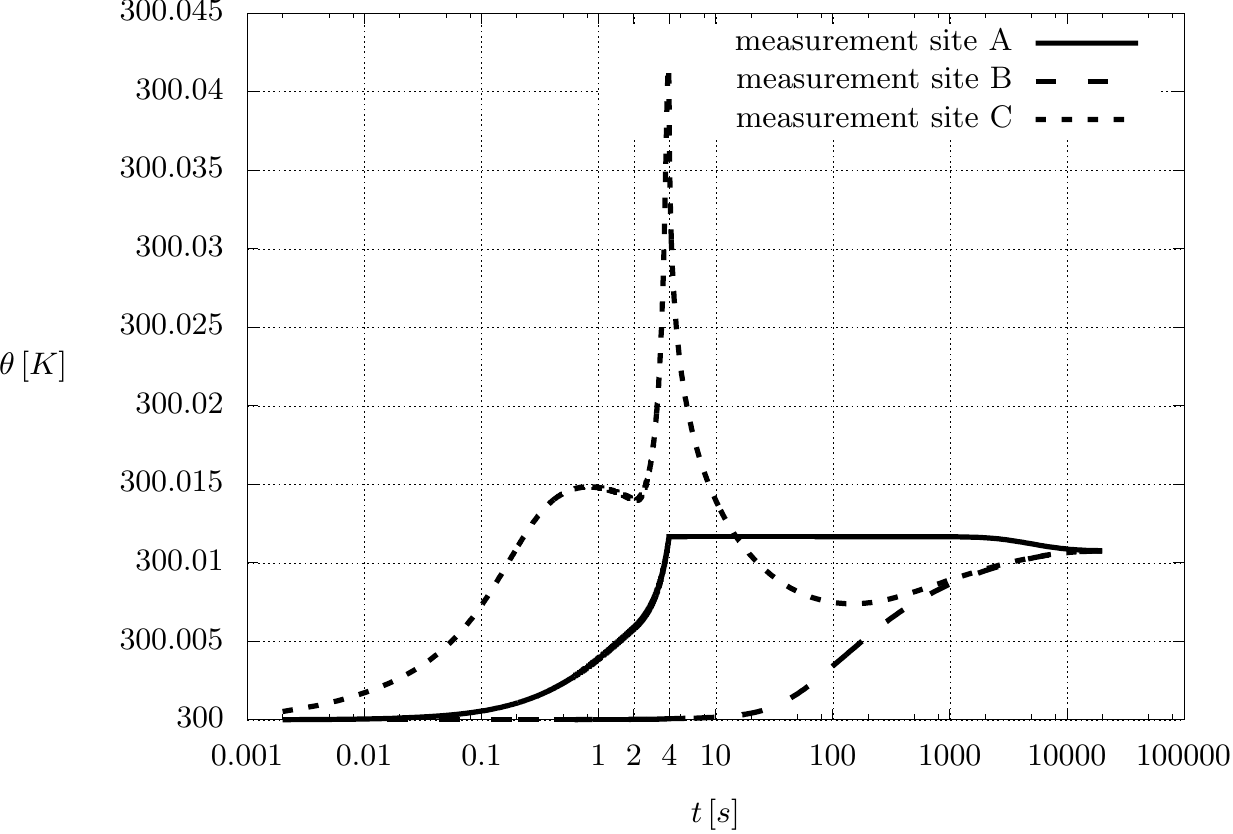}}
  \\
  \subfloat[Large viscosity.]{\includegraphics[width=0.32\textwidth]{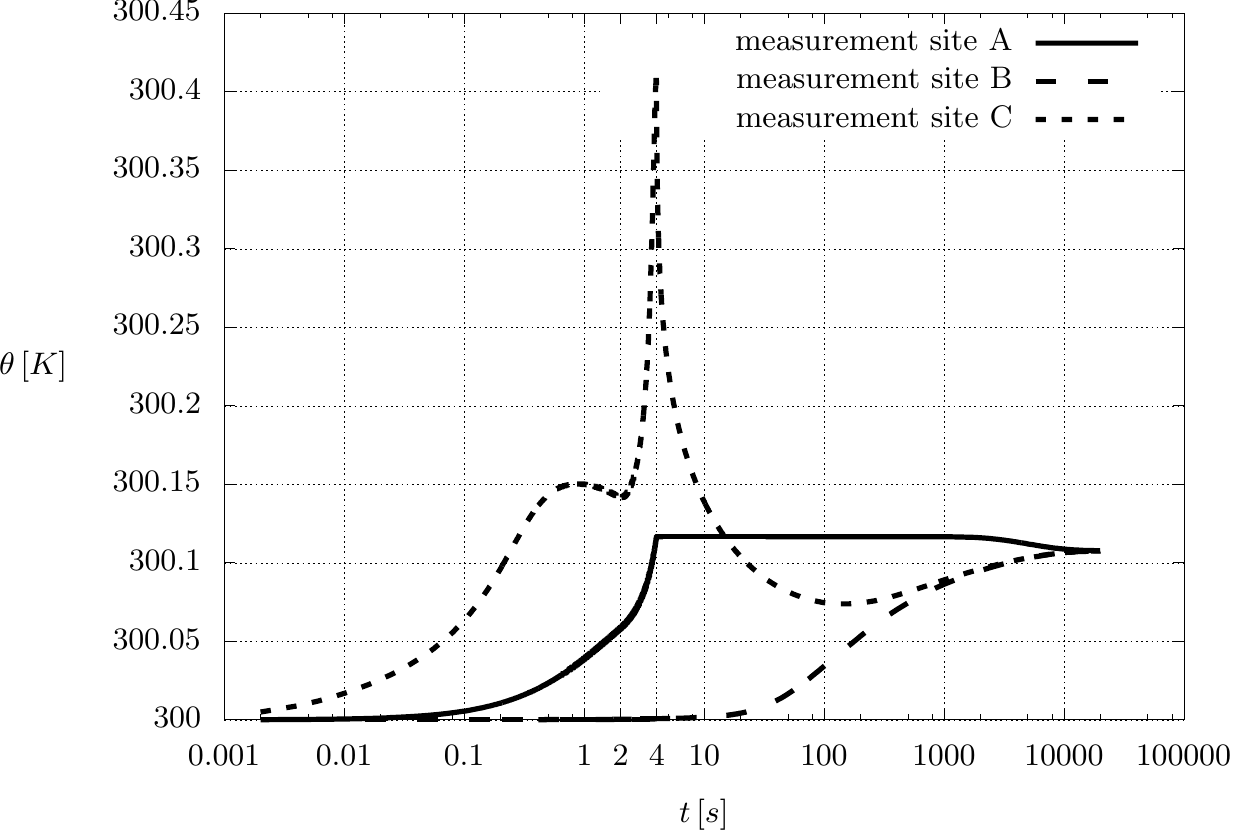}}
  \quad
  \subfloat[S-large viscosity.]{\includegraphics[width=0.32\textwidth]{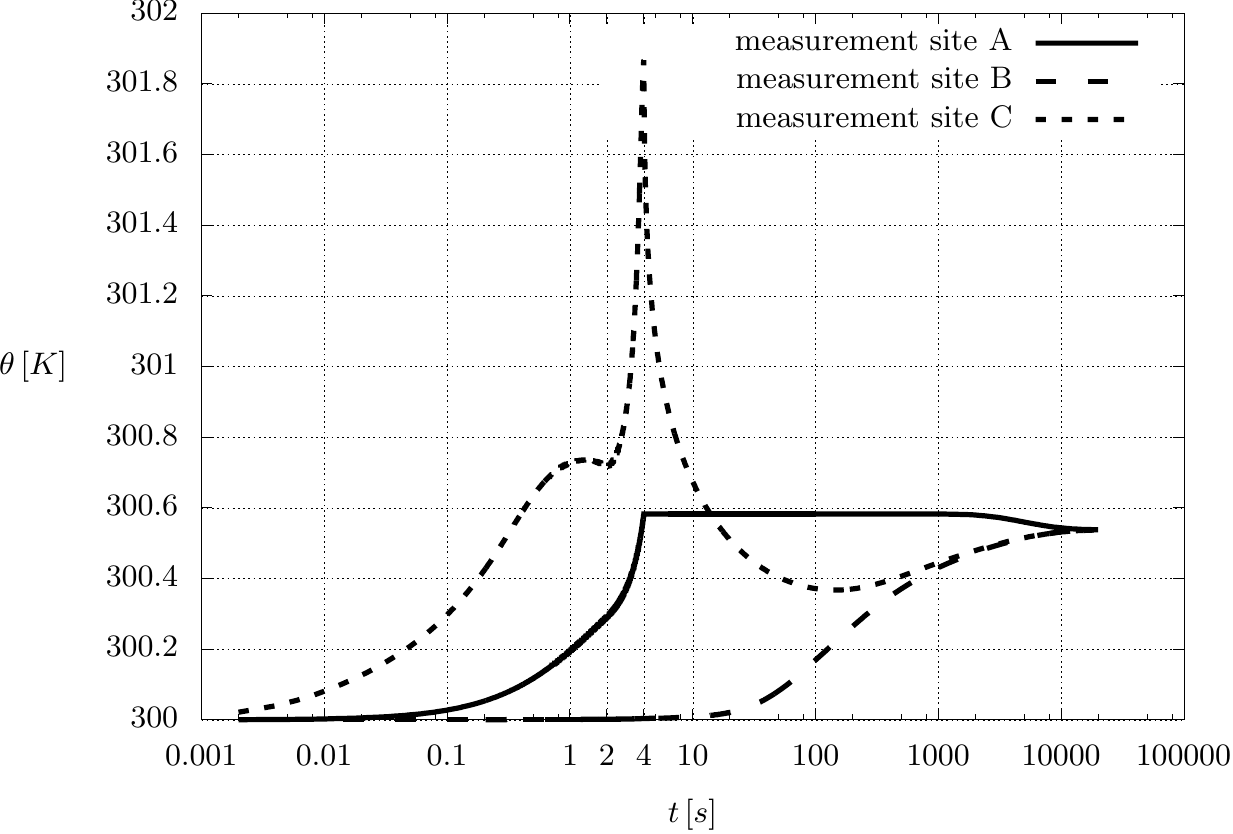}}
  \quad
  \subfloat[X-large viscosity.]{\includegraphics[width=0.32\textwidth]{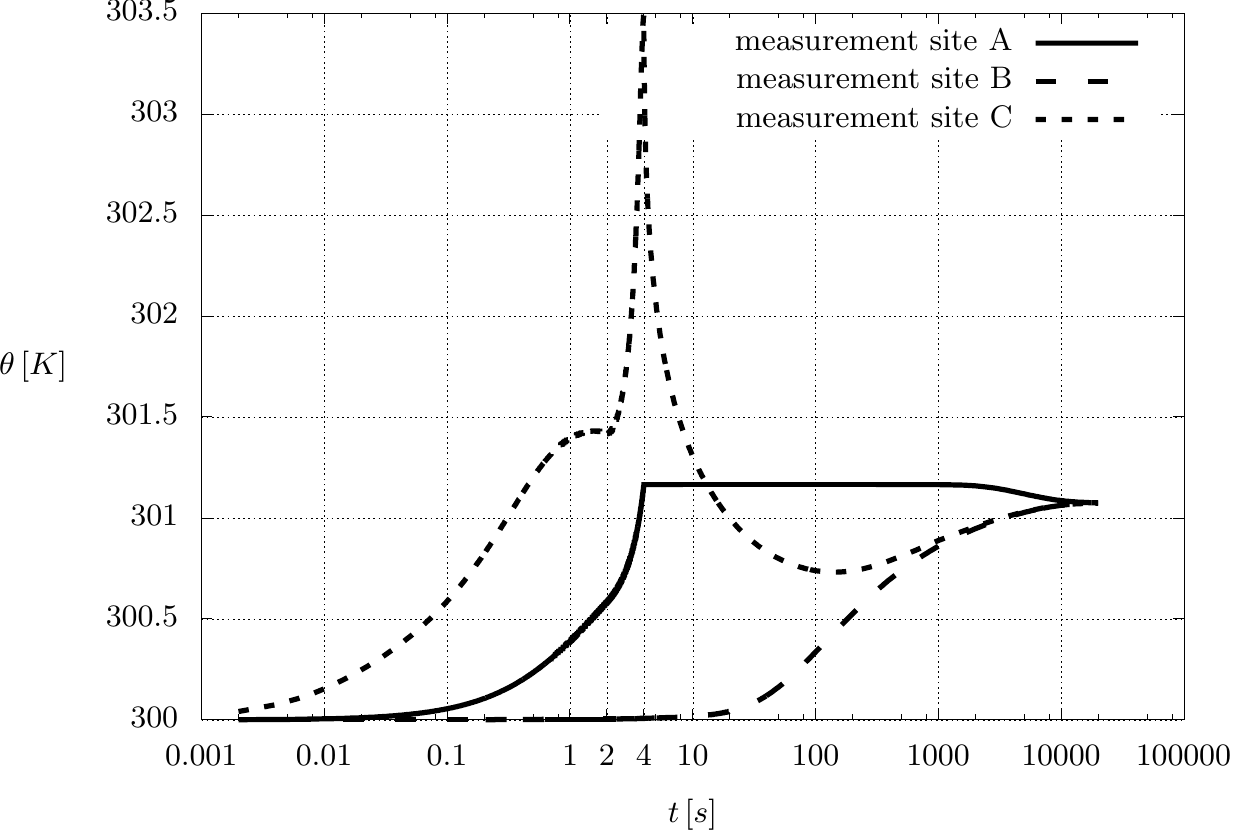}}
  \caption{Long time behaviour of temperature field, \emph{constant} elastic moduli. Nomenclature for viscosity values is described in Table~\ref{tab:parameter-values-viscosity}, remaining material parameters are given in Table~\ref{tab:parameter-values} and Table~\ref{tab:parameter-values-elastic}.}
  \label{fig:temperature-longtime-constant}
\end{figure}

Next set of numerical experiments deals with long time behaviour of the temperature field. Since the sample is thermally isolated, one can expect that once the single loading-unloading cycle is over, the temperature field would tend to a spatially homogeneous temperature field. This indeed happens. In particular, in Figure~\ref{fig:temperature-longtime-constant} we take models with particular sets of parameter values, and we observe that the temperature values at measurement sites A, B and C indeed converge to a common limit. (The ``convergence'' must be of course understood in the sense that there exists reasonable numerical evidence for the convergence to a common limit. The numerical simulation does not by any means constitute a rigorous proof of this statement.) The equilibration of the temperature field, however, happens on a quite long time scale. This numerical experiment also (partially) documents that the numerical solver for the corresponding governing equations has been correctly implemented.

\begin{figure}[h]
  \subfloat[Tiny viscosity.]{\includegraphics[width=0.32\textwidth]{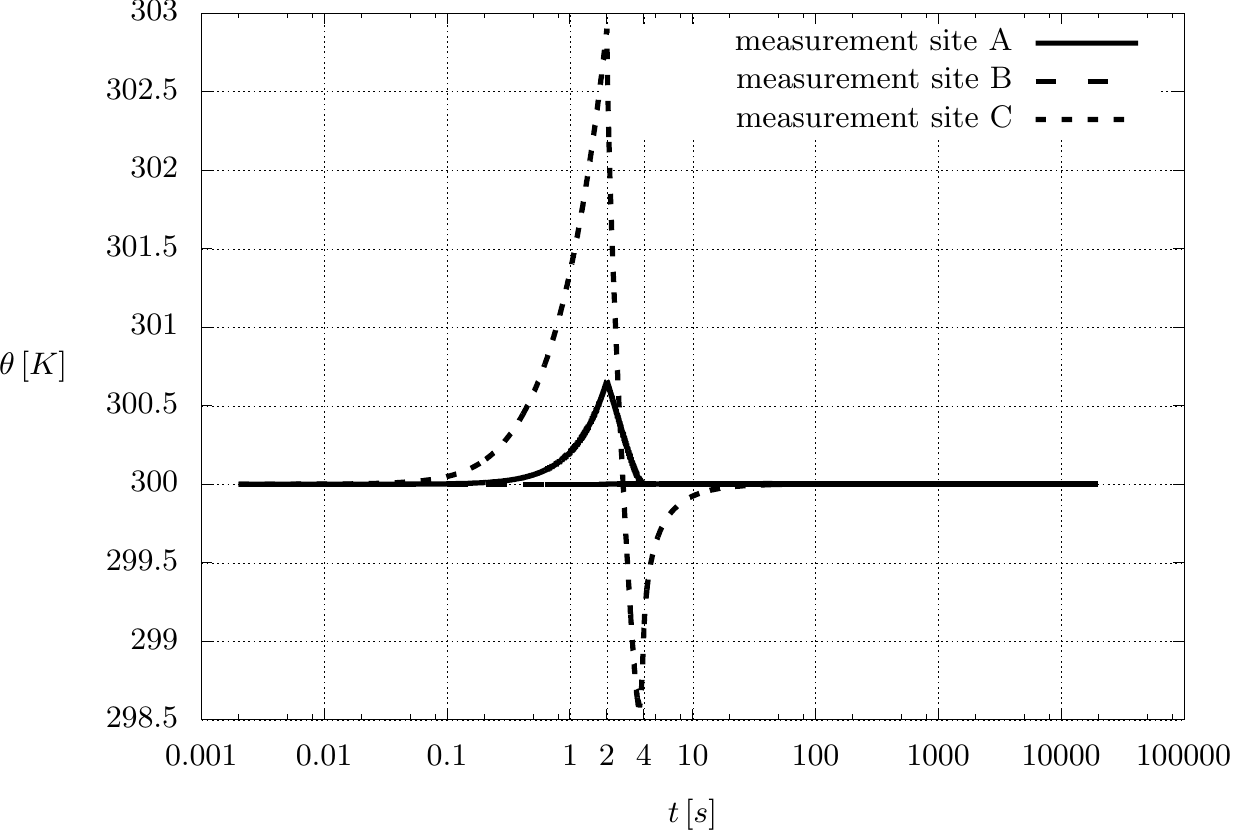}}
  \quad
  \subfloat[Small viscosity.]{\includegraphics[width=0.32\textwidth]{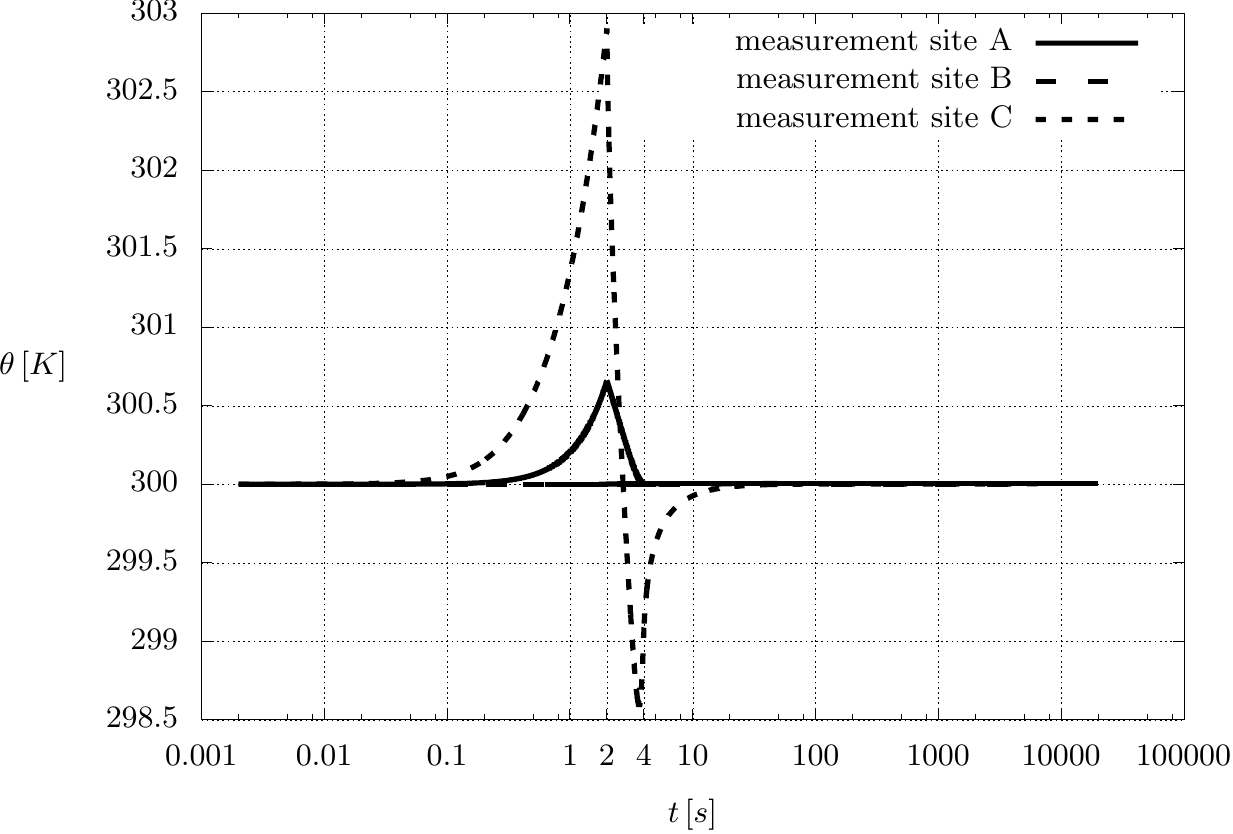}}
  \quad
  \subfloat[Medium viscosity.]{\includegraphics[width=0.32\textwidth]{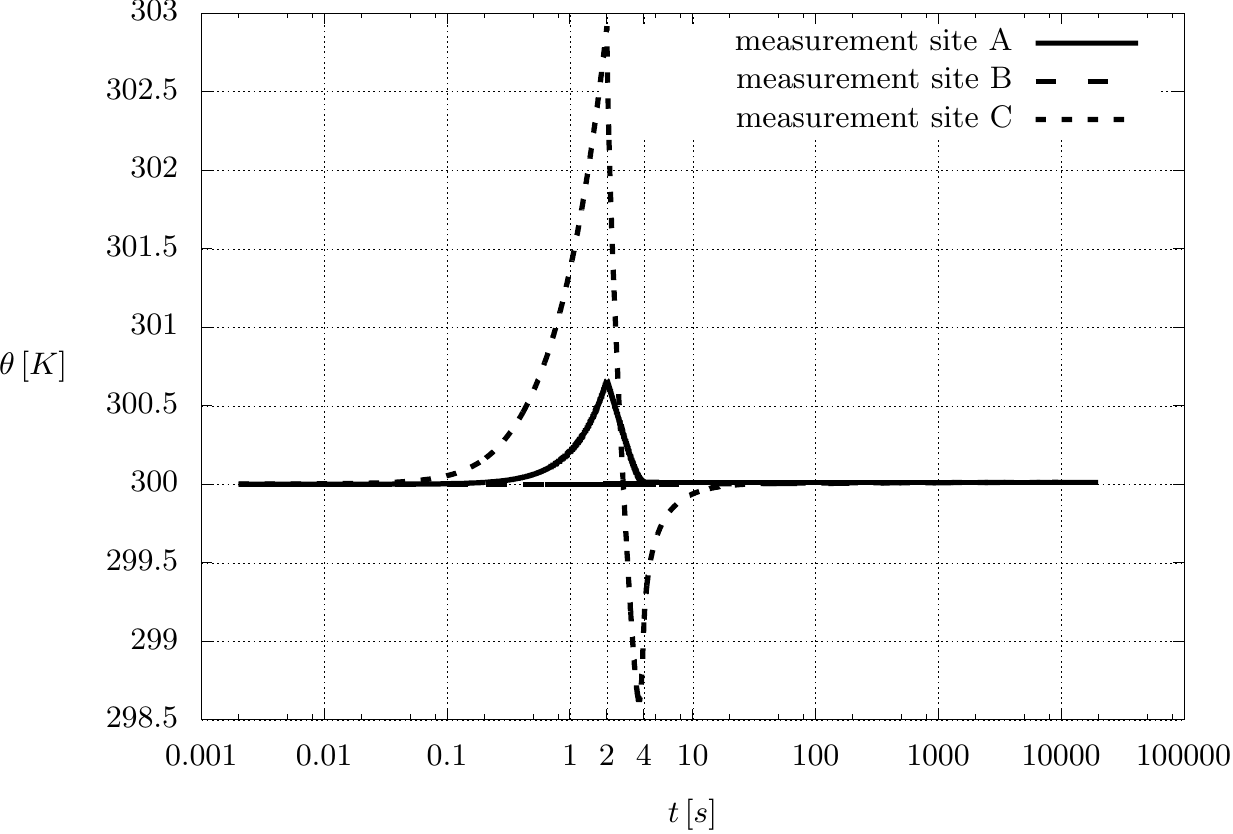}}
  \\
  \subfloat[Large viscosity.]{\includegraphics[width=0.32\textwidth]{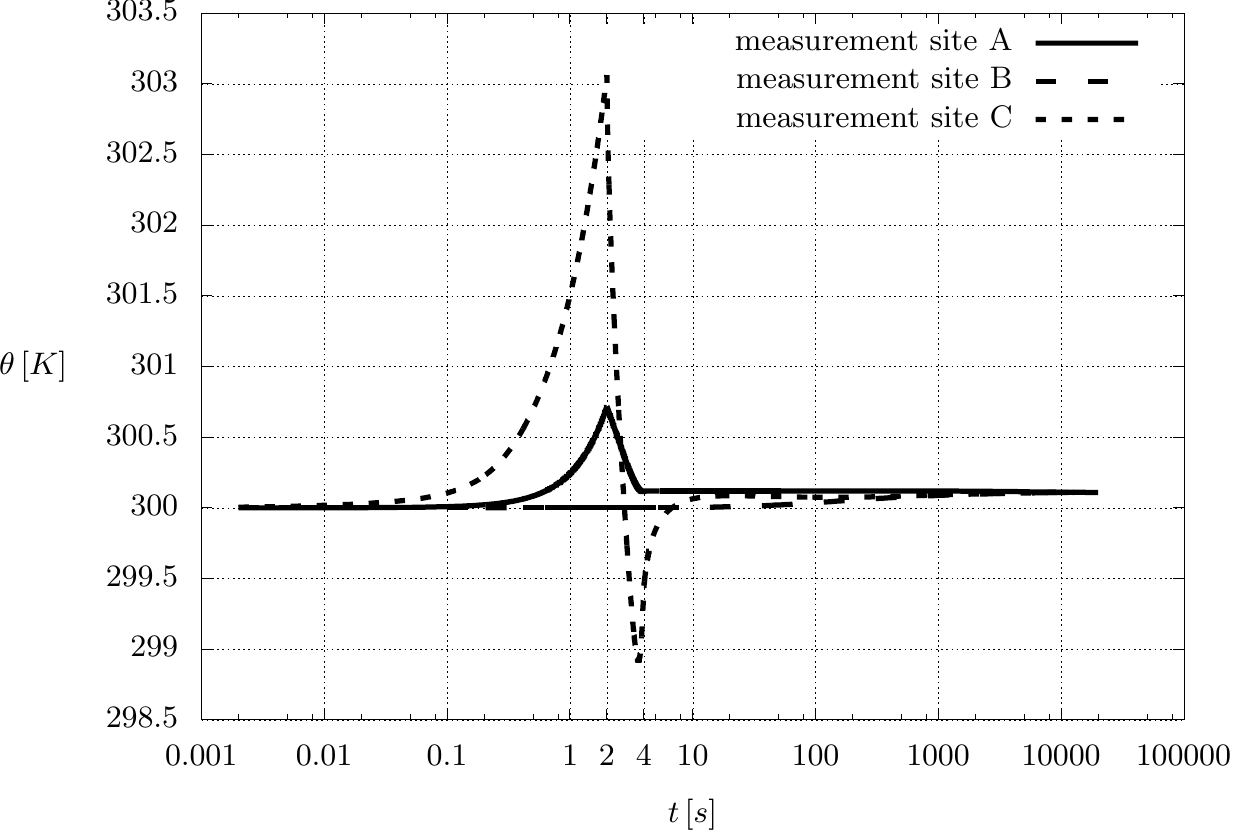}}
  \quad
  \subfloat[S-large viscosity.]{\includegraphics[width=0.32\textwidth]{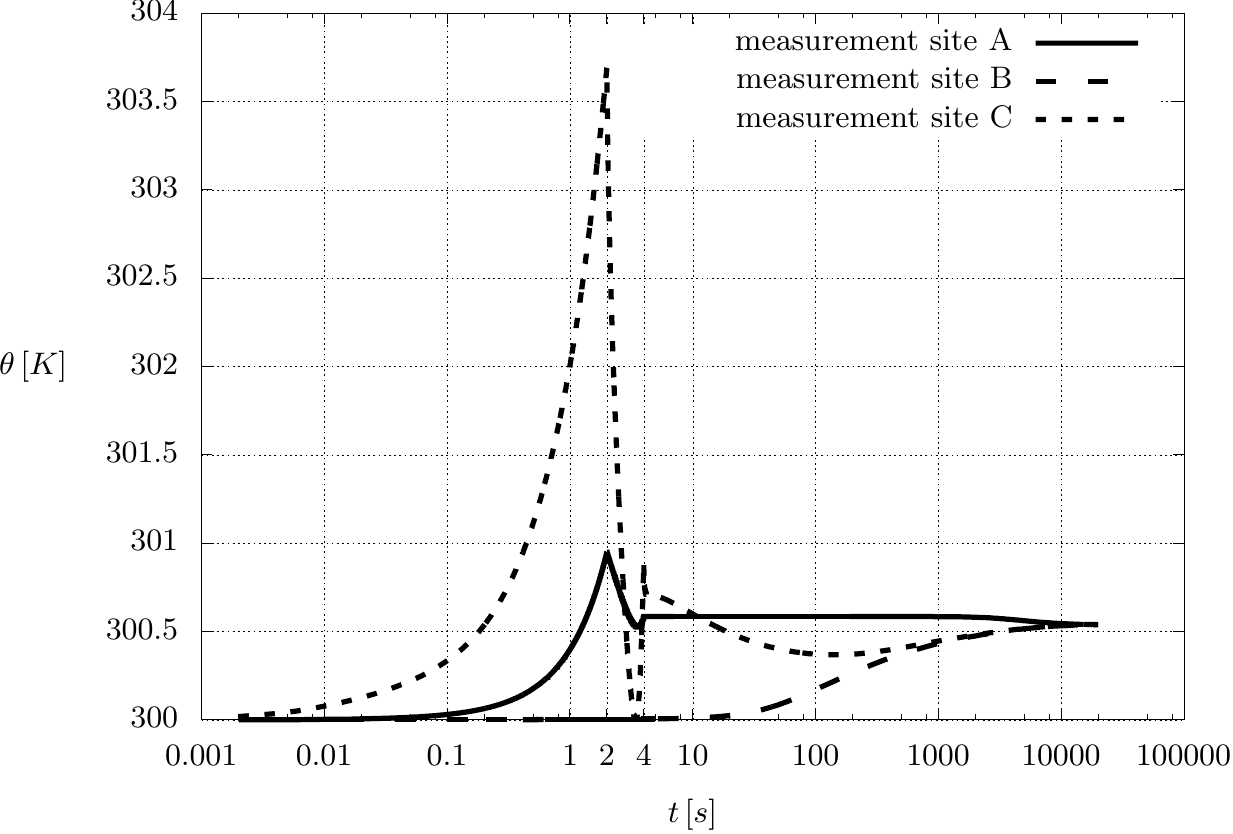}}
  \quad
  \subfloat[X-large viscosity.]{\includegraphics[width=0.32\textwidth]{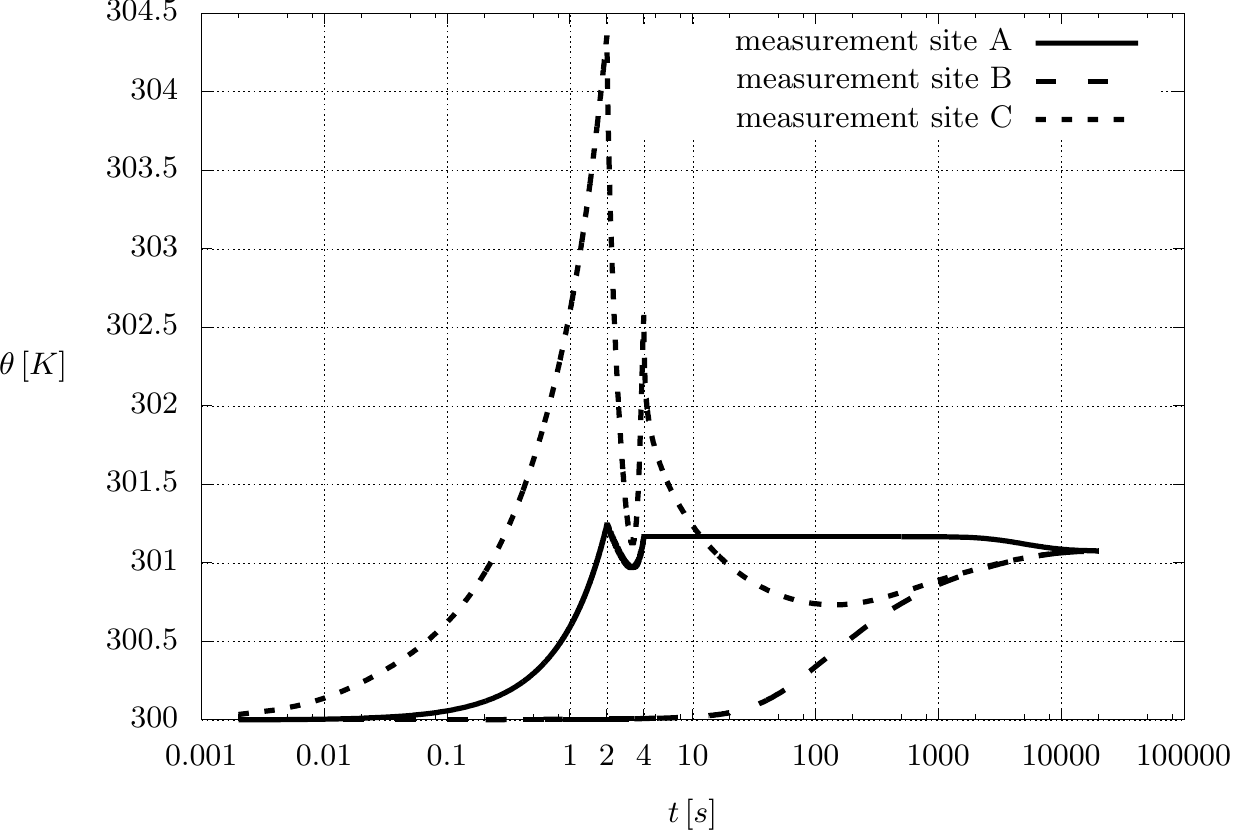}}
  \caption{Long time behaviour of temperature field, \emph{temperature dependent} elastic moduli. Nomenclature for viscosity values is described in Table~\ref{tab:parameter-values-viscosity}, remaining material parameters are given in Table~\ref{tab:parameter-values} and Table~\ref{tab:parameter-values-elastic}.}
  \label{fig:temperature-longtime-temperature-dependent}
\end{figure}

\FloatBarrier

\subsubsection{Temperature field in the vicinity of the cutout tip}
\label{sec:temp-field-evol}
\begin{figure}[h]
  \subfloat[$t=1.0$]{\includegraphics[width=0.35\textwidth]{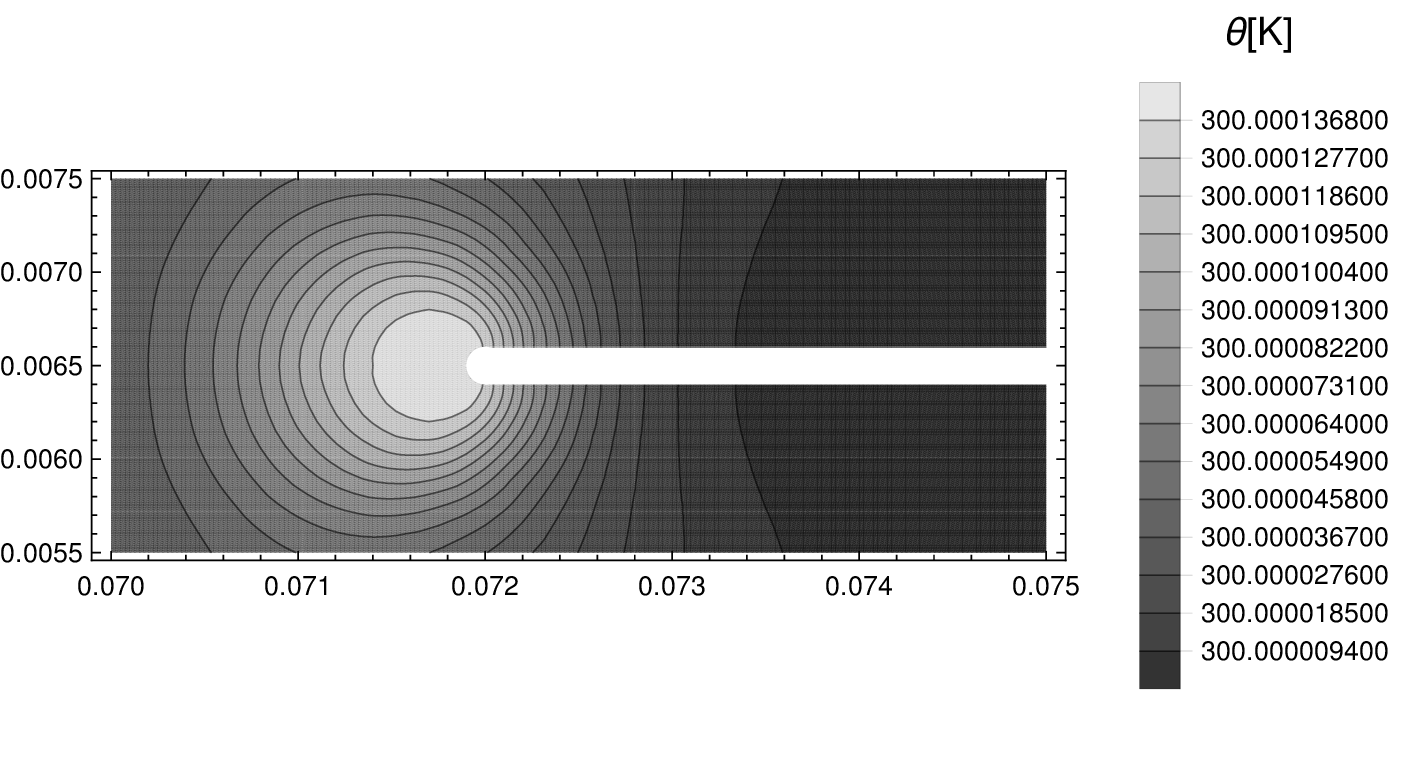}}
  \qquad
  \subfloat[$t=2.0$]{\includegraphics[width=0.35\textwidth]{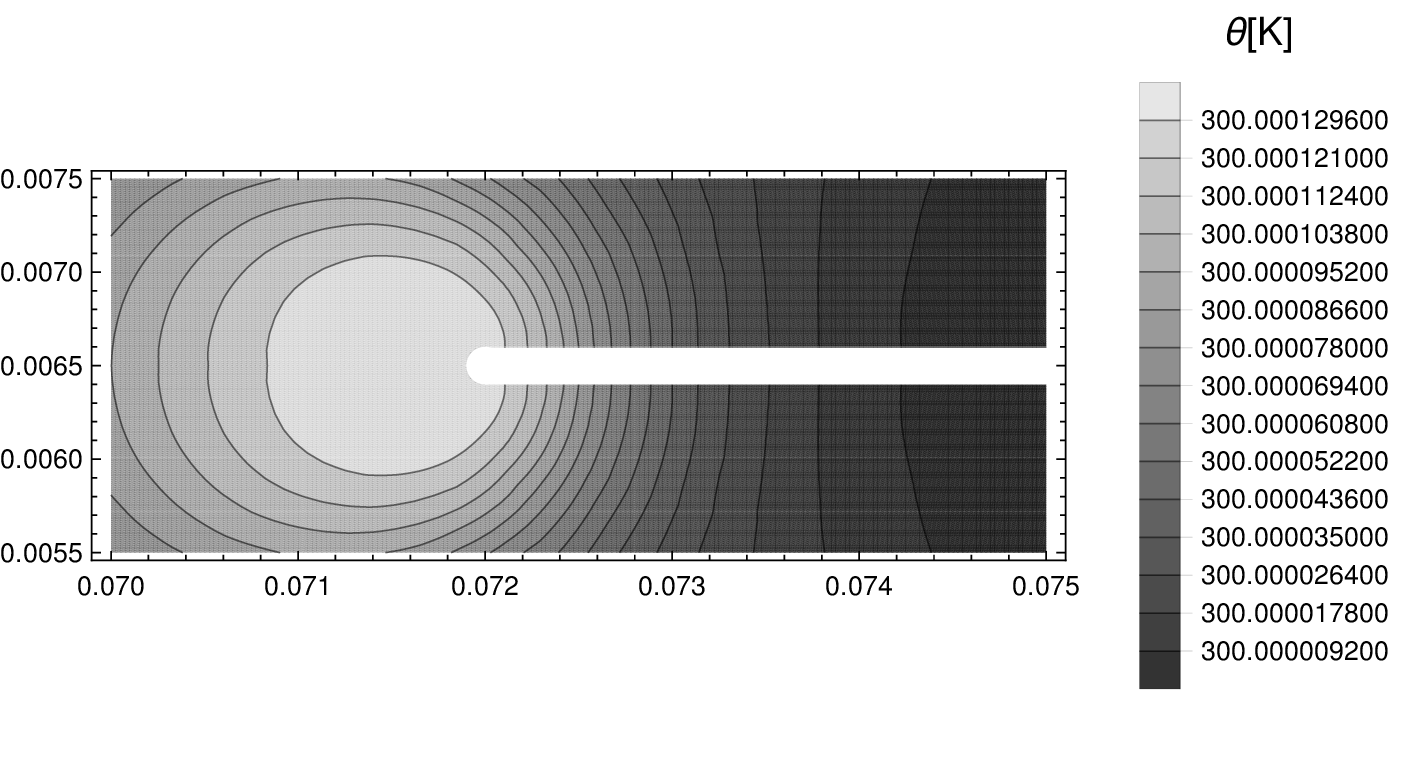}}
  \\
  \subfloat[$t=2.5$\label{fig:temperature-field-whole-specimen-constant-parameters-small-viscosity-c}]{\includegraphics[width=0.35\textwidth]{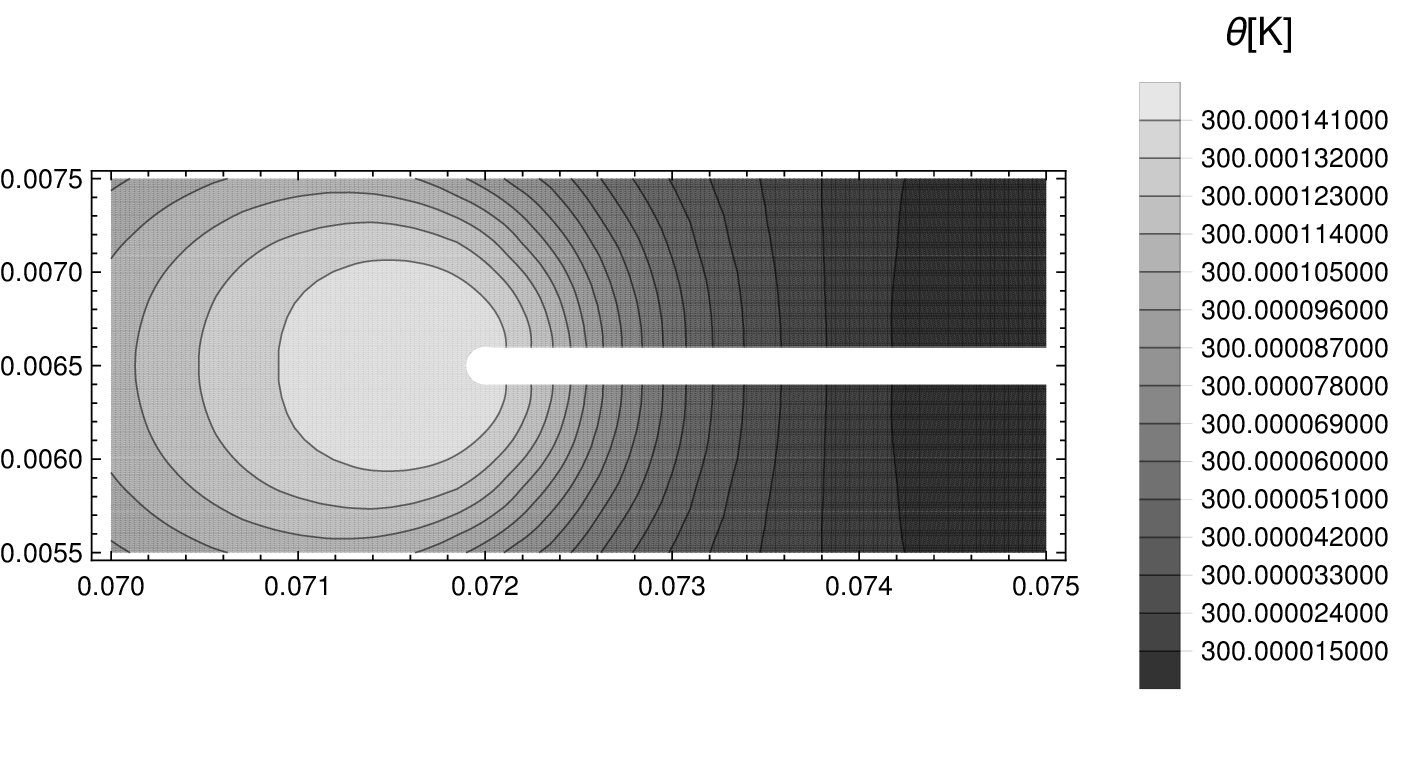}}
  \qquad
  \subfloat[$t=10$]{\includegraphics[width=0.35\textwidth]{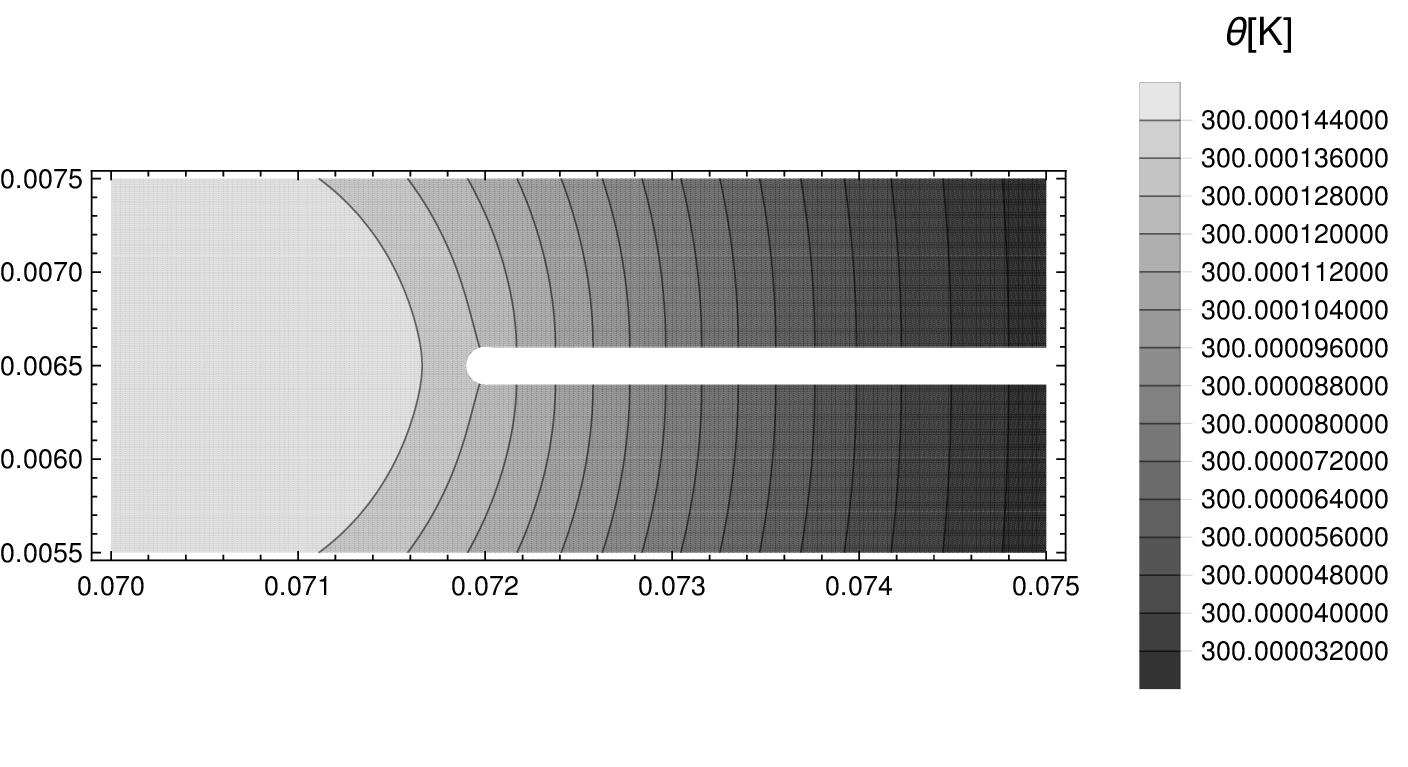}}
  \caption{Temperature field close to the cutout tip, \emph{constant} elastic moduli, \emph{tiny viscosity} value. Nomenclature for viscosity values is described in Table~\ref{tab:parameter-values-viscosity}, remaining material parameters are given in Table~\ref{tab:parameter-values} and Table~\ref{tab:parameter-values-elastic}.}
  \label{fig:temperature-field-whole-specimen-constant-parameters-small-viscosity}
\end{figure}

\begin{figure}[h]
  \subfloat[$t=1.0$]{\includegraphics[width=0.35\textwidth]{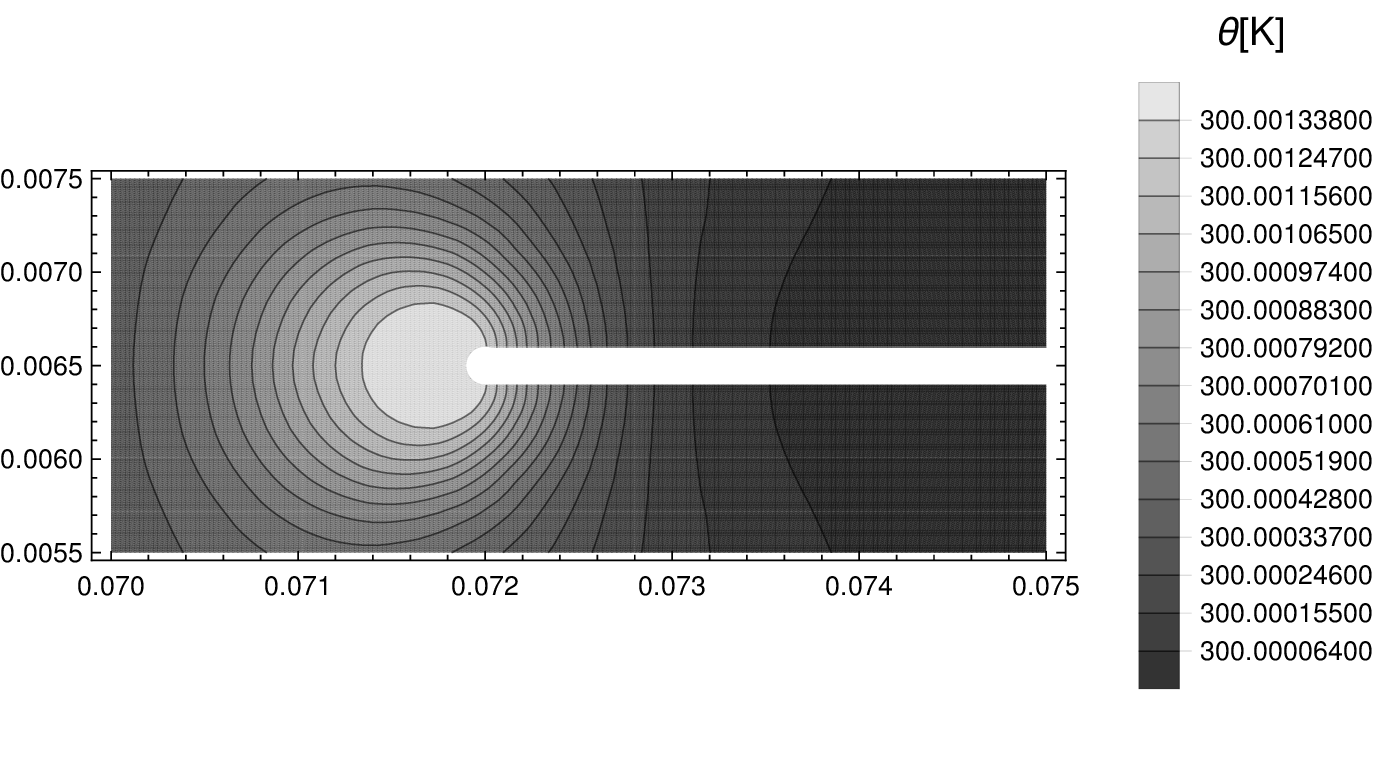}}
  \qquad
  \subfloat[$t=2.0$]{\includegraphics[width=0.35\textwidth]{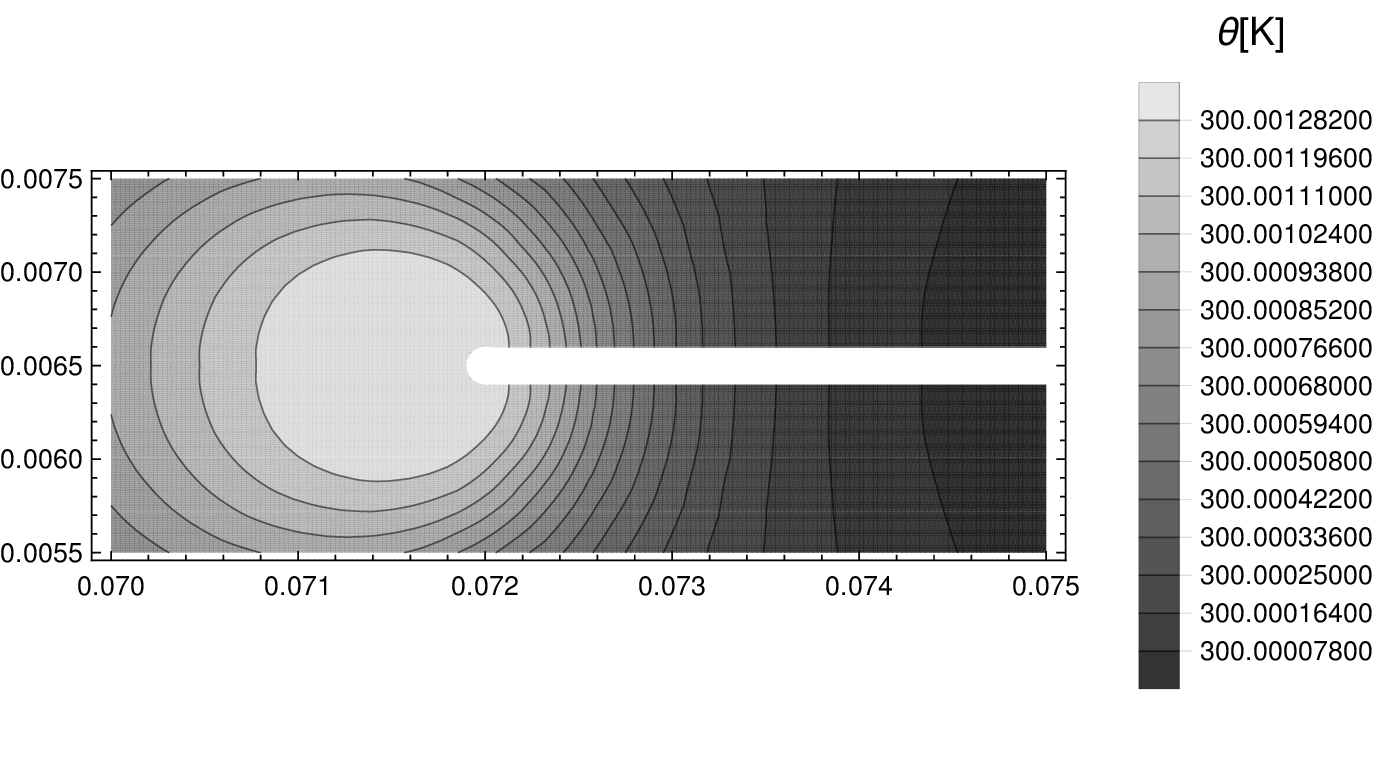}}
  \\
  \subfloat[$t=2.5$]{\includegraphics[width=0.35\textwidth]{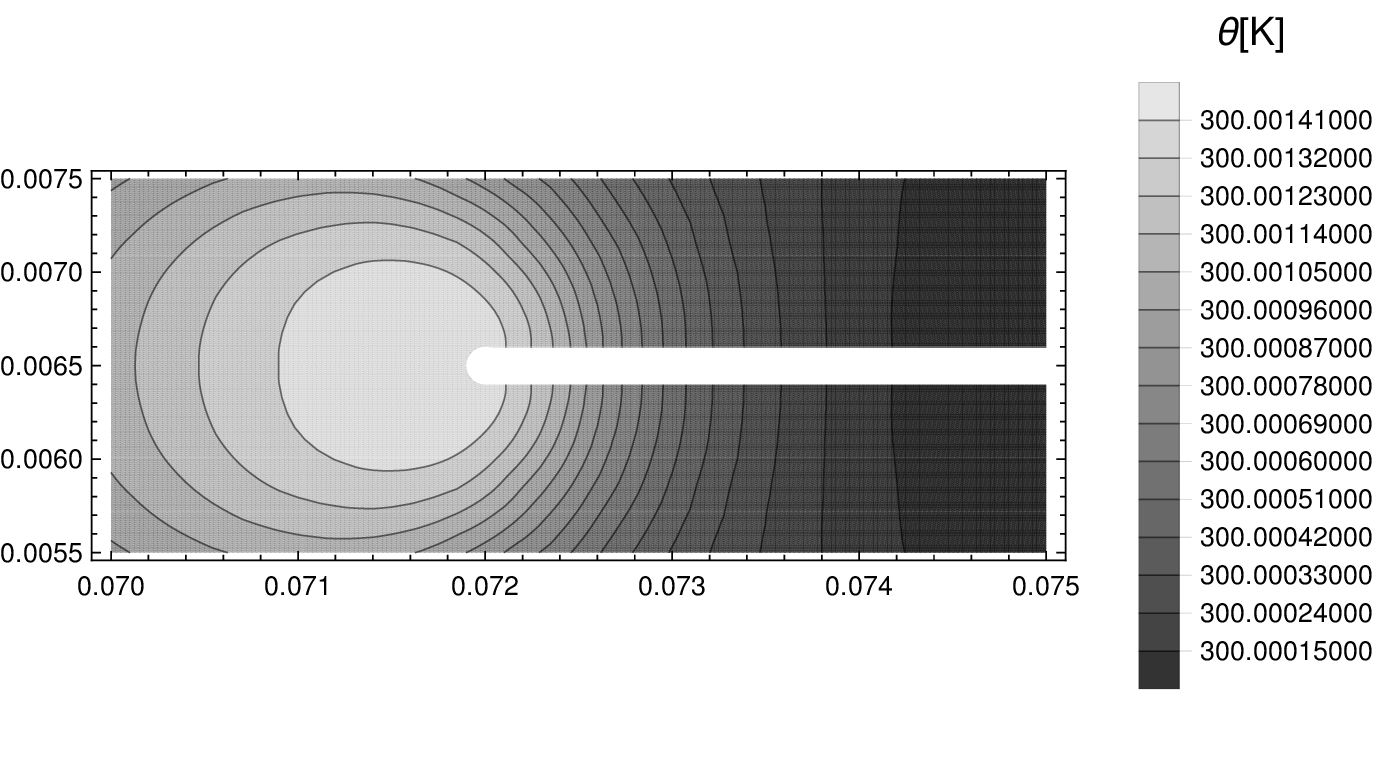}}
  \qquad
  \subfloat[$t=10$]{\includegraphics[width=0.35\textwidth]{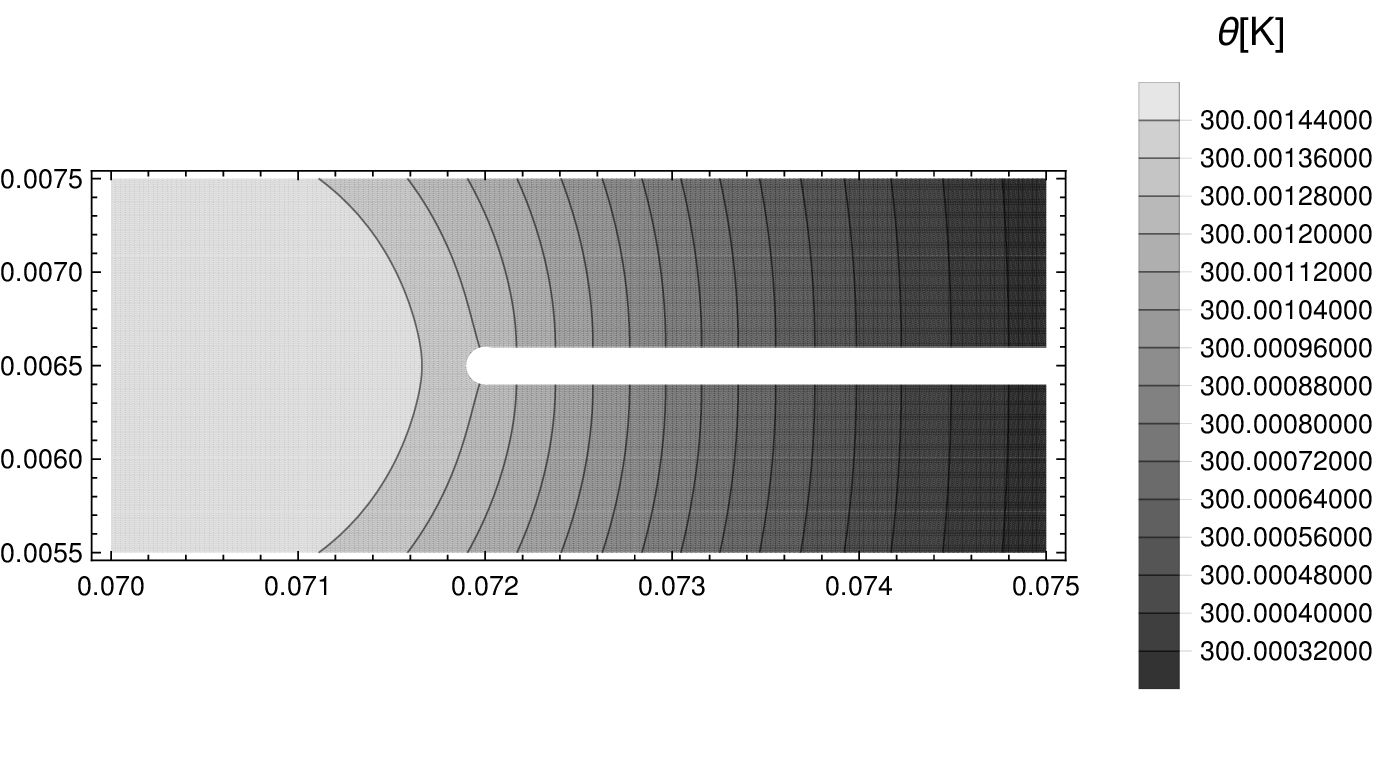}}
  \caption{Temperature field close to the cutout tip, \emph{constant} elastic moduli, \emph{small viscosity} value. Nomenclature for viscosity values is described in Table~\ref{tab:parameter-values-viscosity}, remaining material parameters are given in Table~\ref{tab:parameter-values} and Table~\ref{tab:parameter-values-elastic}.}
  \label{fig:temperature-field-whole-specimen-constant-parameters-normal-viscosity}
\end{figure}

\begin{figure}[h]
  \subfloat[$t=1.0$]{\includegraphics[width=0.35\textwidth]{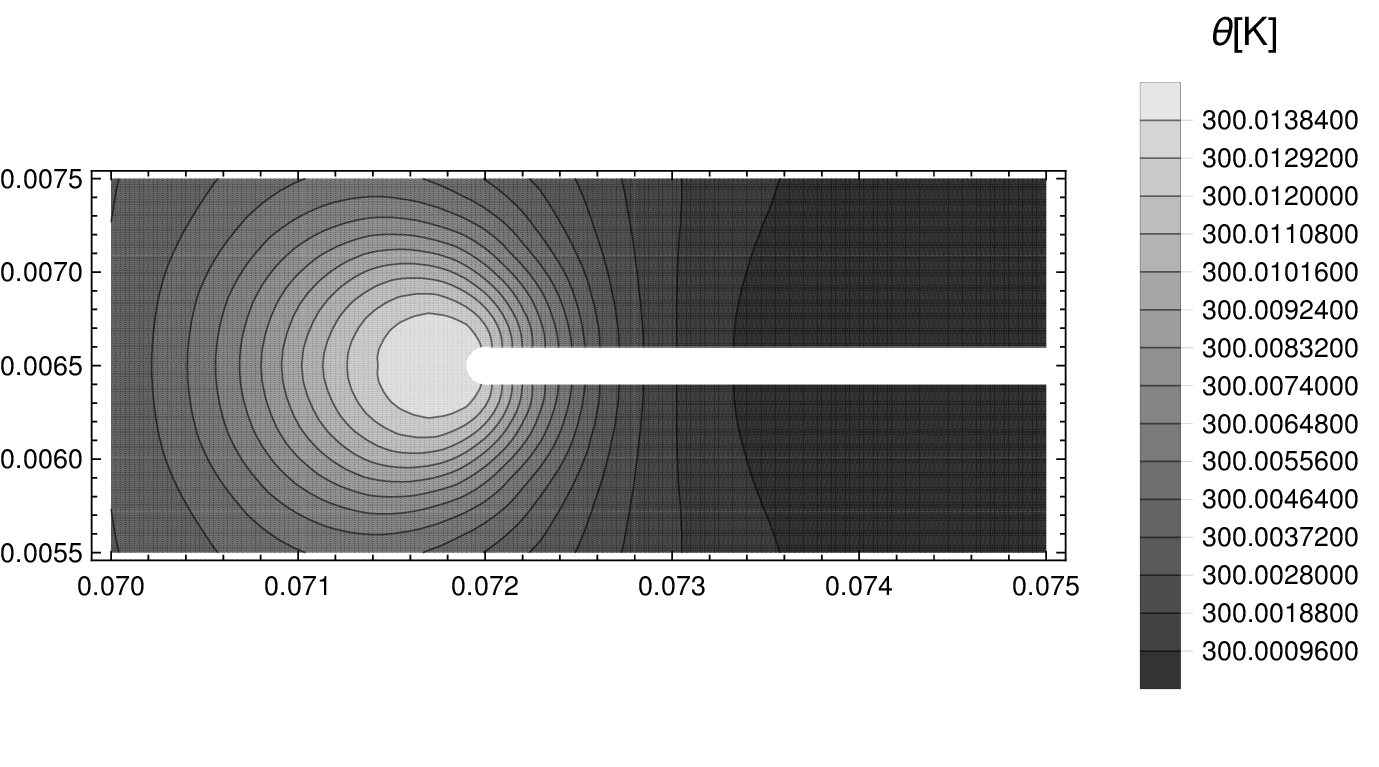}}
  \qquad
  \subfloat[$t=2.0$]{\includegraphics[width=0.35\textwidth]{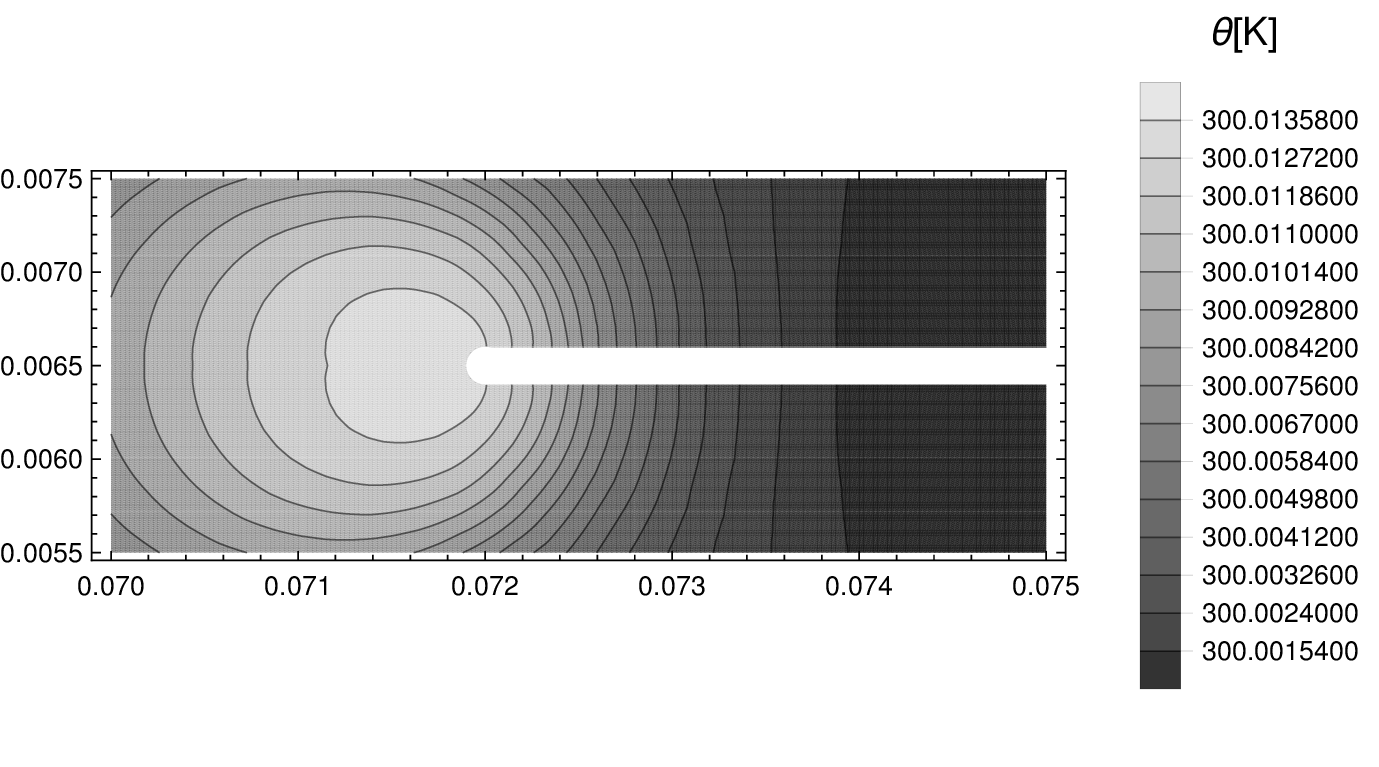}}
  \\
  \subfloat[$t=2.5$]{\includegraphics[width=0.35\textwidth]{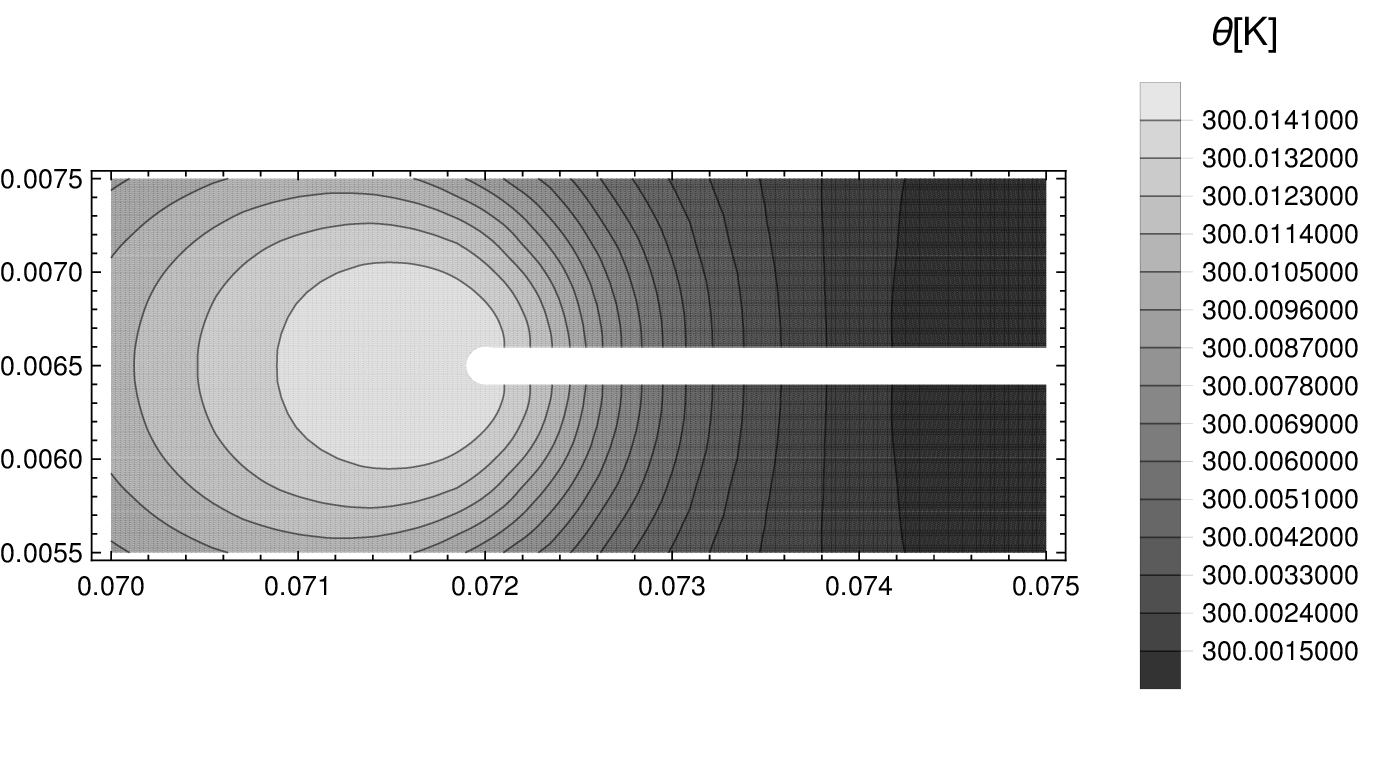}}
  \qquad
  \subfloat[$t=10$]{\includegraphics[width=0.35\textwidth]{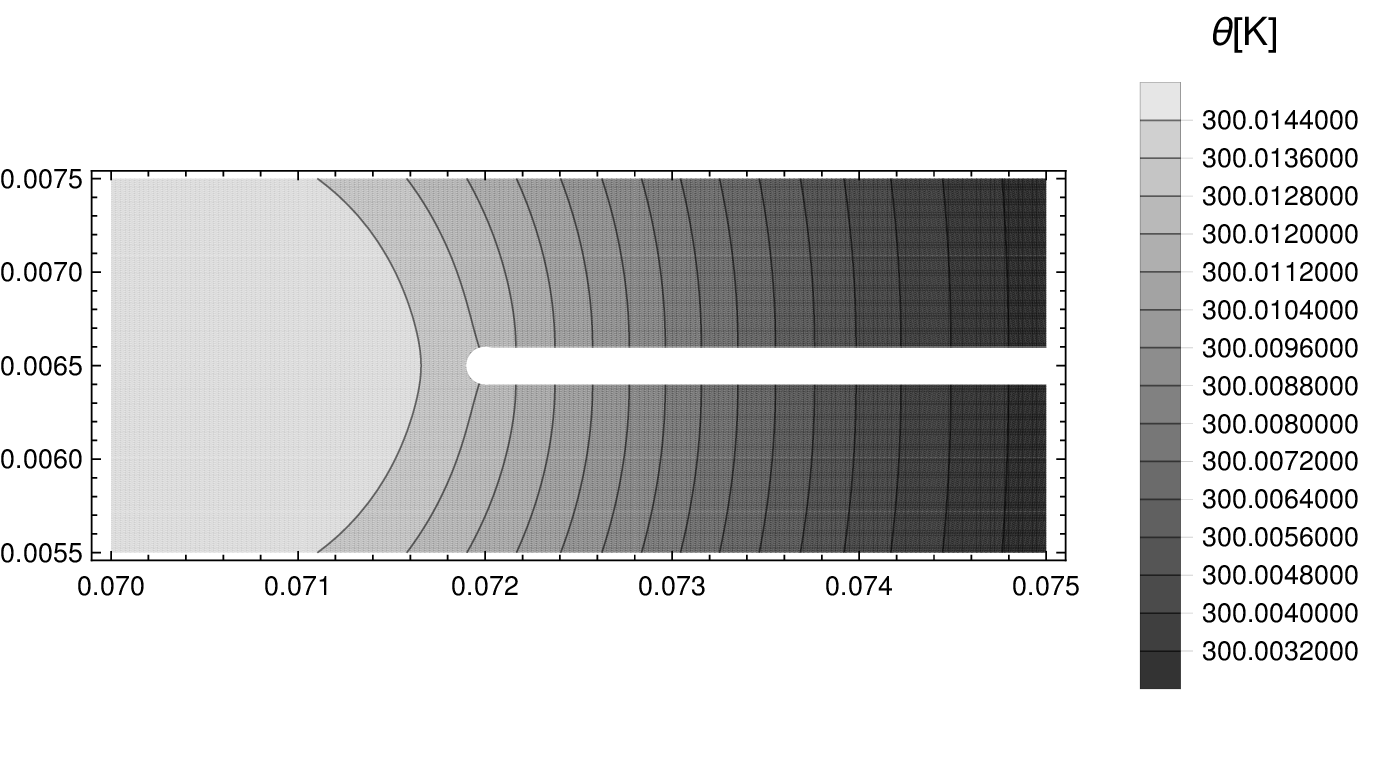}}
  \caption{Temperature field close to the cutout tip, \emph{constant} elastic moduli, \emph{medium viscosity} value. Nomenclature for viscosity values is described in Table~\ref{tab:parameter-values-viscosity}, remaining material parameters are given in Table~\ref{tab:parameter-values} and Table~\ref{tab:parameter-values-elastic}.}
  \label{fig:temperature-field-whole-specimen-constant-parameters-medium-viscosity}
\end{figure}

\begin{figure}[h]
  \subfloat[$t=1.0$]{\includegraphics[width=0.35\textwidth]{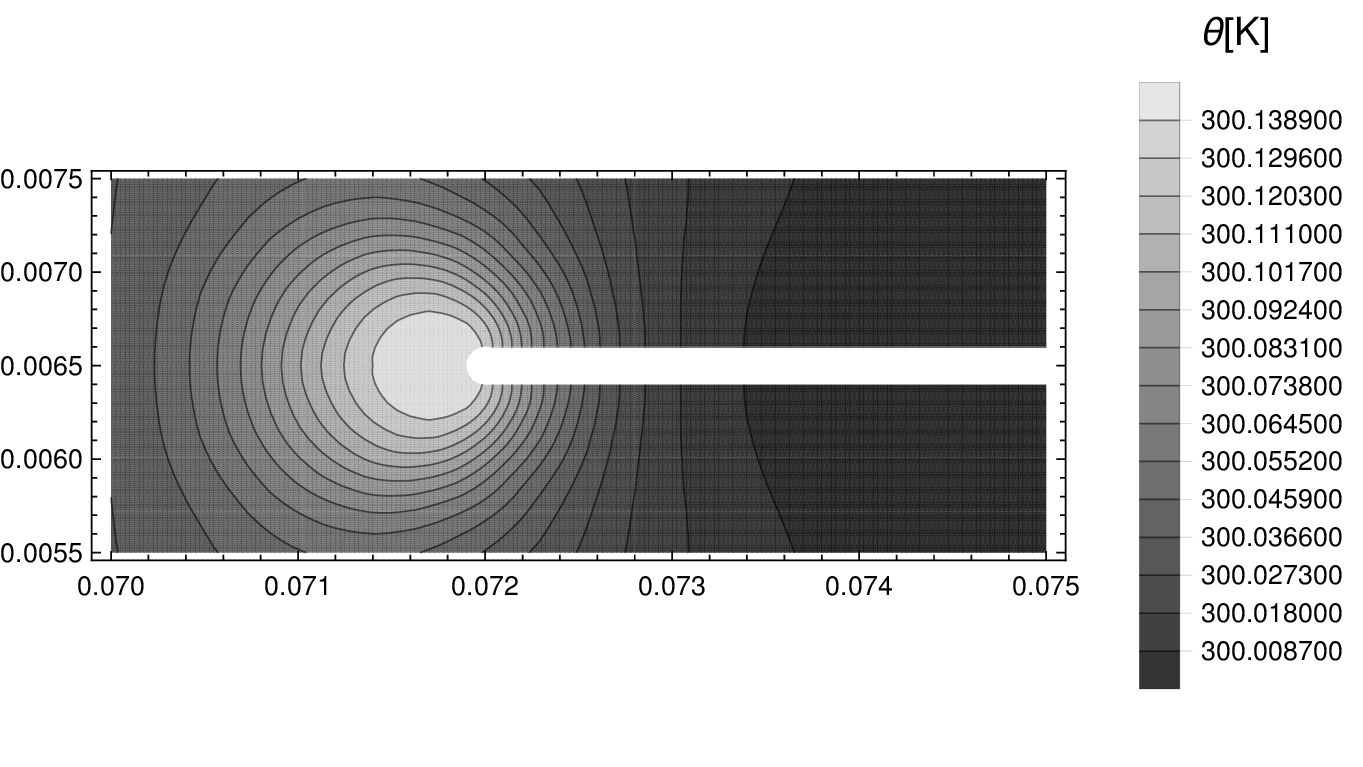}}
  \qquad
  \subfloat[$t=2.0$]{\includegraphics[width=0.35\textwidth]{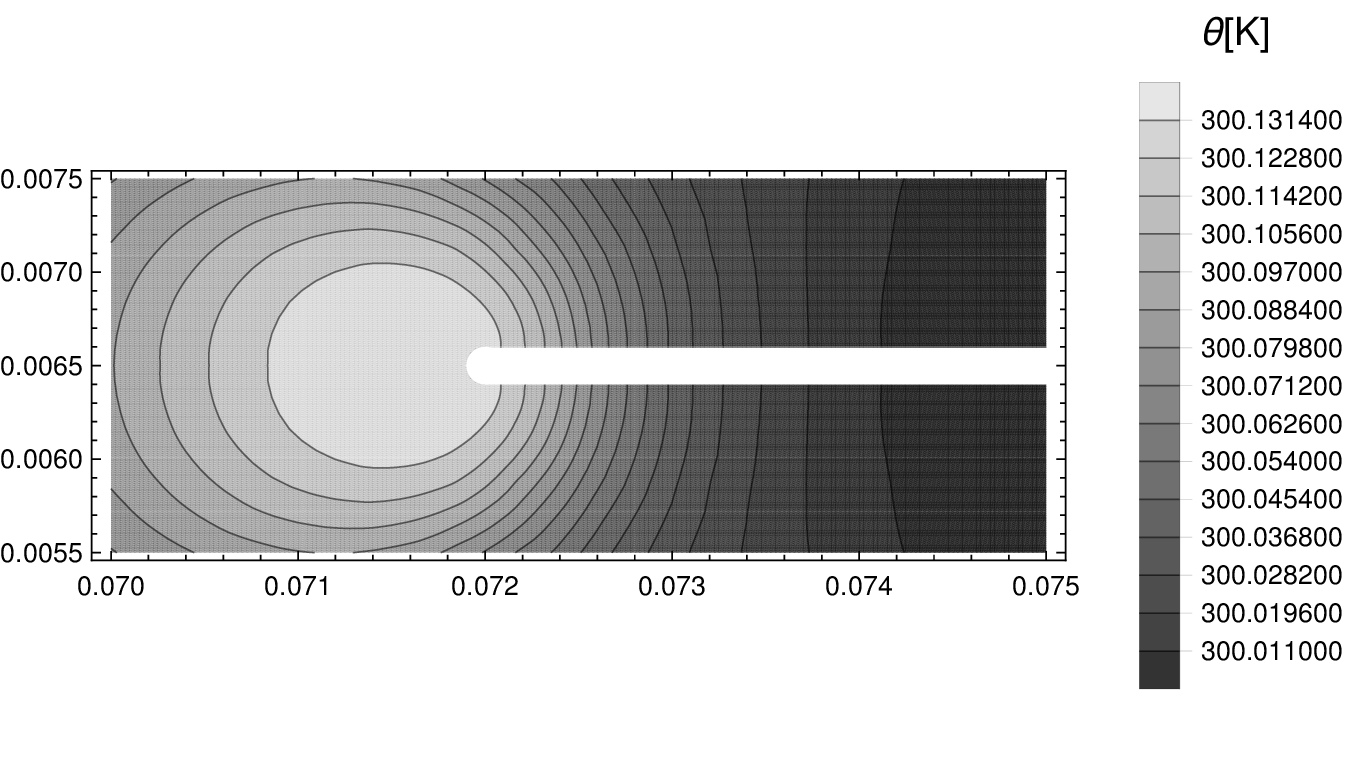}}
  \\
  \subfloat[$t=2.5$]{\includegraphics[width=0.35\textwidth]{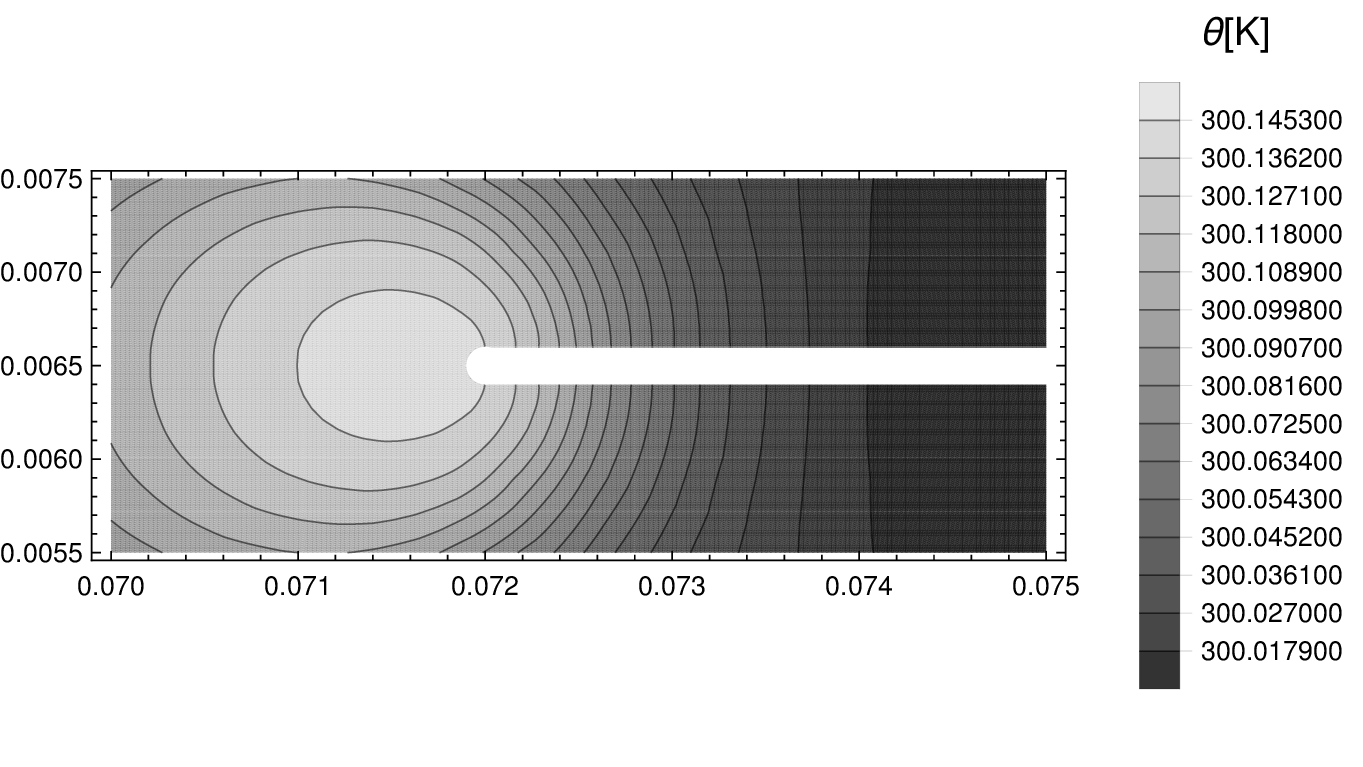}}
  \qquad
  \subfloat[$t=10$]{\includegraphics[width=0.35\textwidth]{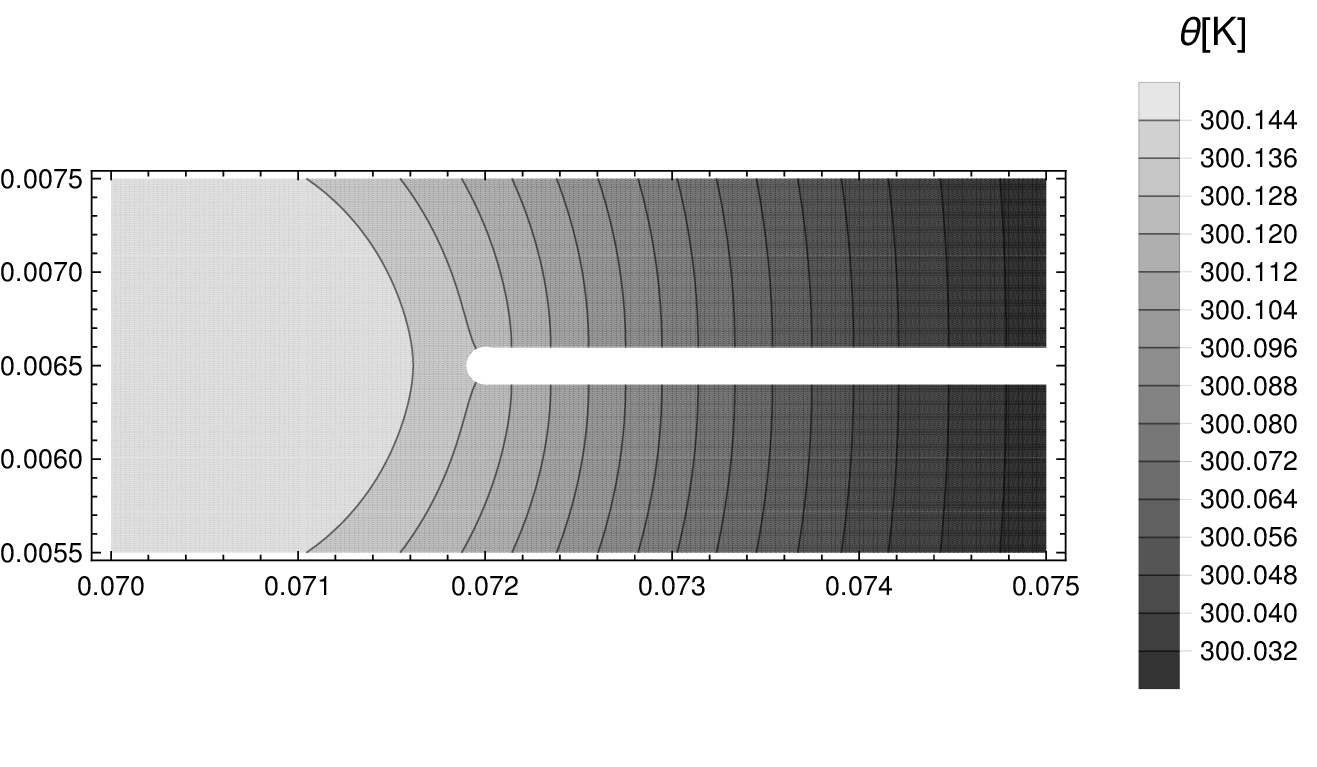}}
  \caption{Temperature field close to the cutout tip, \emph{constant} elastic moduli, \emph{large viscosity} value. Nomenclature for viscosity values is described in Table~\ref{tab:parameter-values-viscosity}, remaining material parameters are given in Table~\ref{tab:parameter-values} and Table~\ref{tab:parameter-values-elastic}.}
  \label{fig:temperature-field-whole-specimen-constant-parameters-large-viscosity}
\end{figure}

\begin{figure}[h]
  \subfloat[$t=1.0$]{\includegraphics[width=0.35\textwidth]{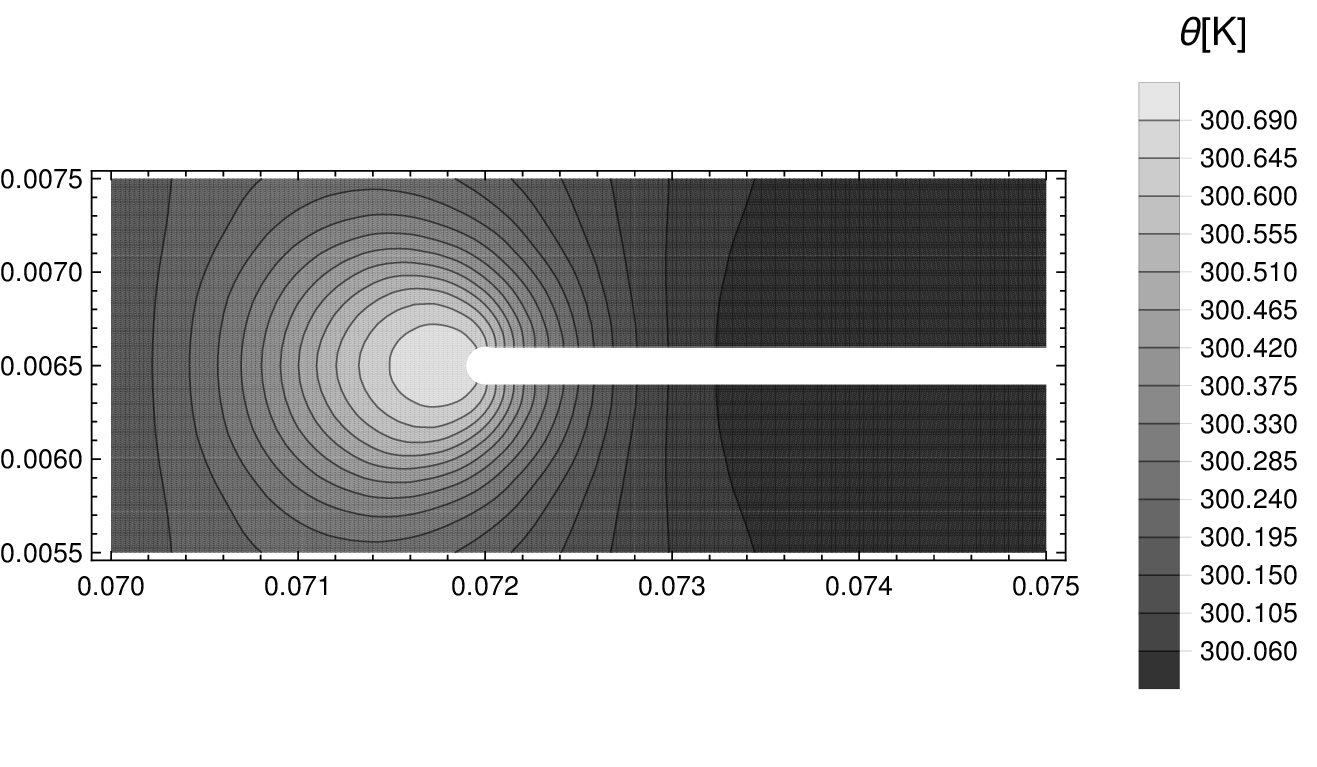}}
  \qquad
  \subfloat[$t=2.0$]{\includegraphics[width=0.35\textwidth]{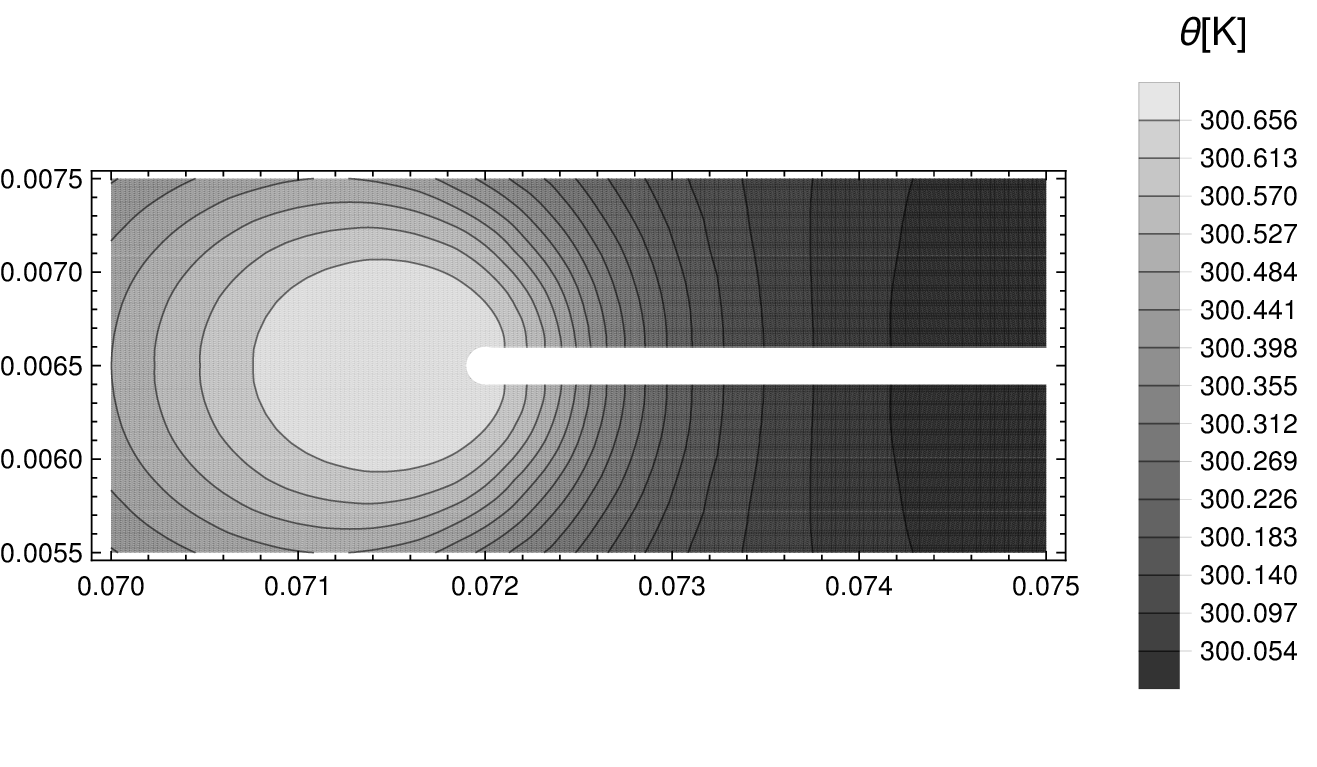}}
  \\
  \subfloat[$t=2.5$]{\includegraphics[width=0.35\textwidth]{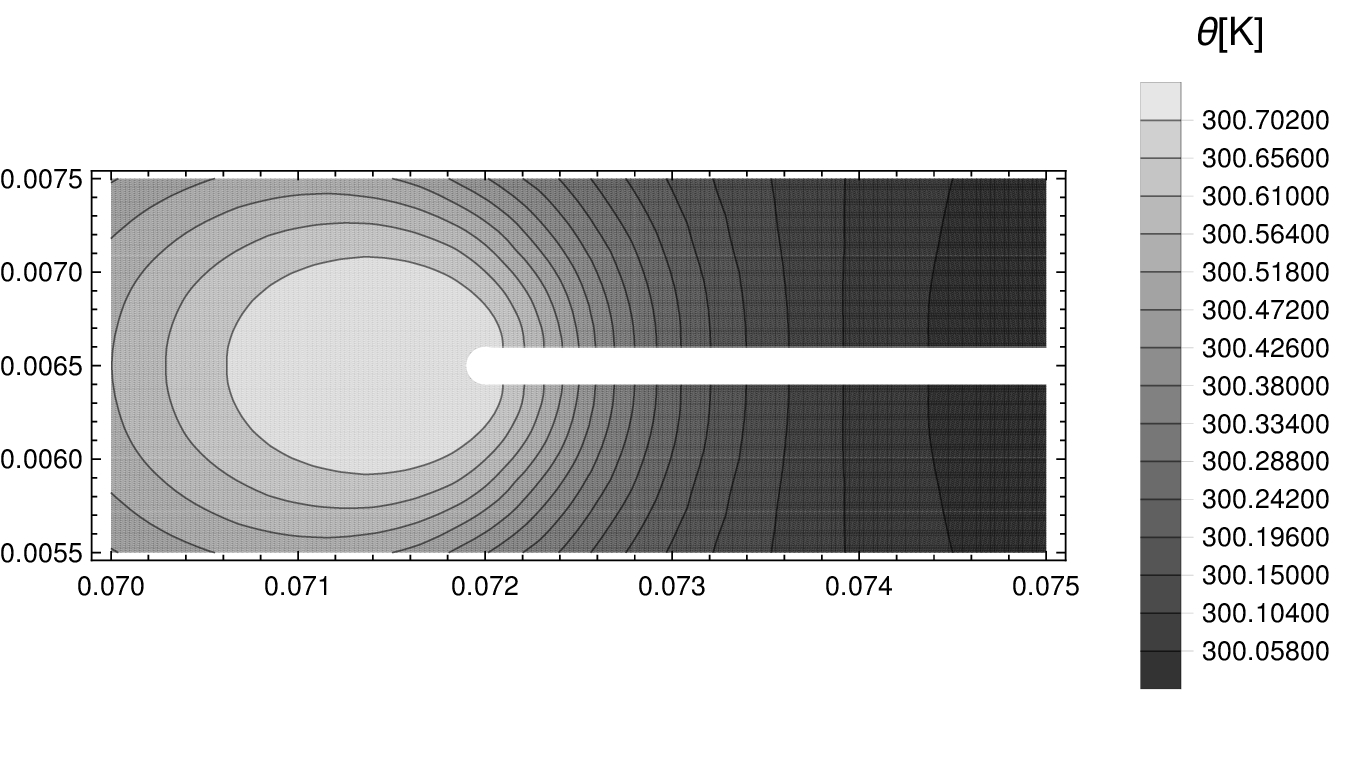}}
  \qquad
  \subfloat[$t=10$]{\includegraphics[width=0.35\textwidth]{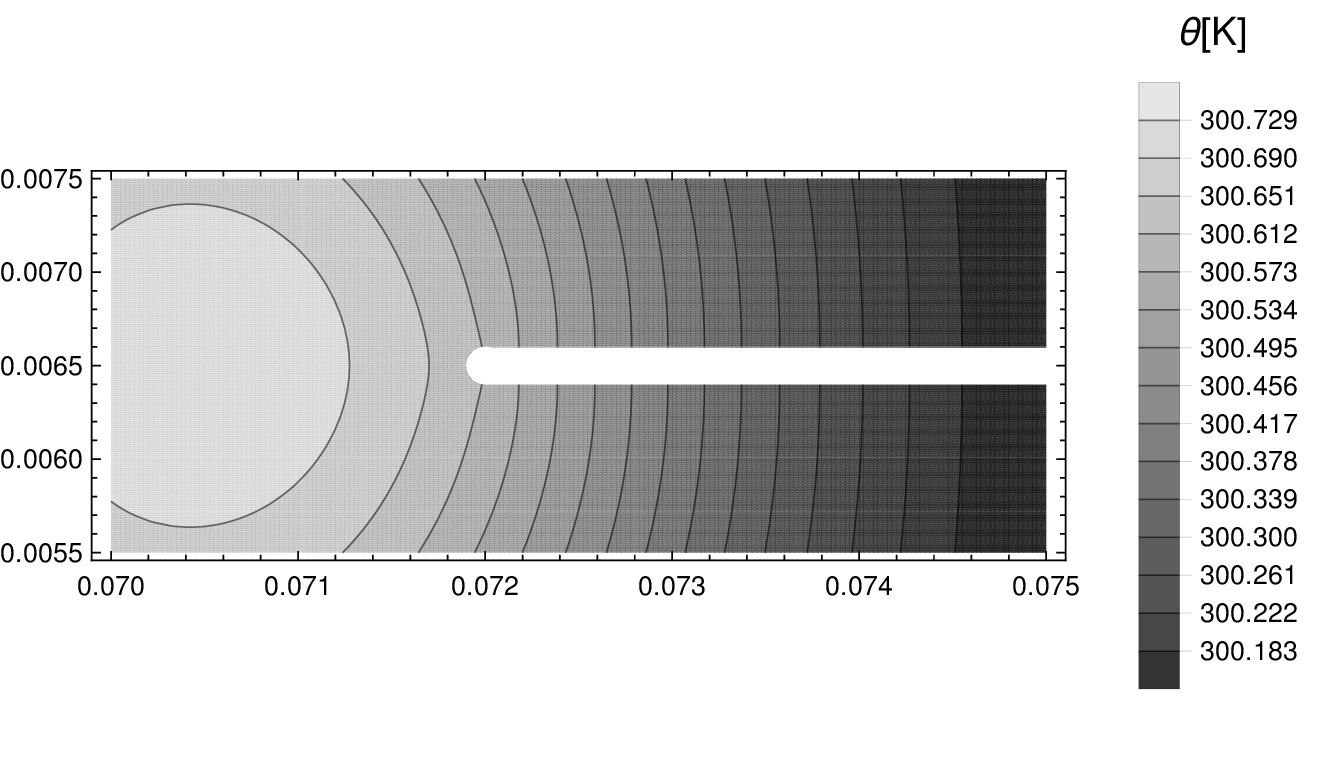}}
  \caption{Temperature field close to the cutout tip, \emph{constant} elastic moduli, \emph{s-large viscosity} value. Nomenclature for viscosity values is described in Table~\ref{tab:parameter-values-viscosity}, remaining material parameters are given in Table~\ref{tab:parameter-values} and Table~\ref{tab:parameter-values-elastic}.}
  \label{fig:temperature-field-whole-specimen-constant-parameters-s-large-viscosity}
\end{figure}

\begin{figure}[h]
  \subfloat[$t=1.0$]{\includegraphics[width=0.35\textwidth]{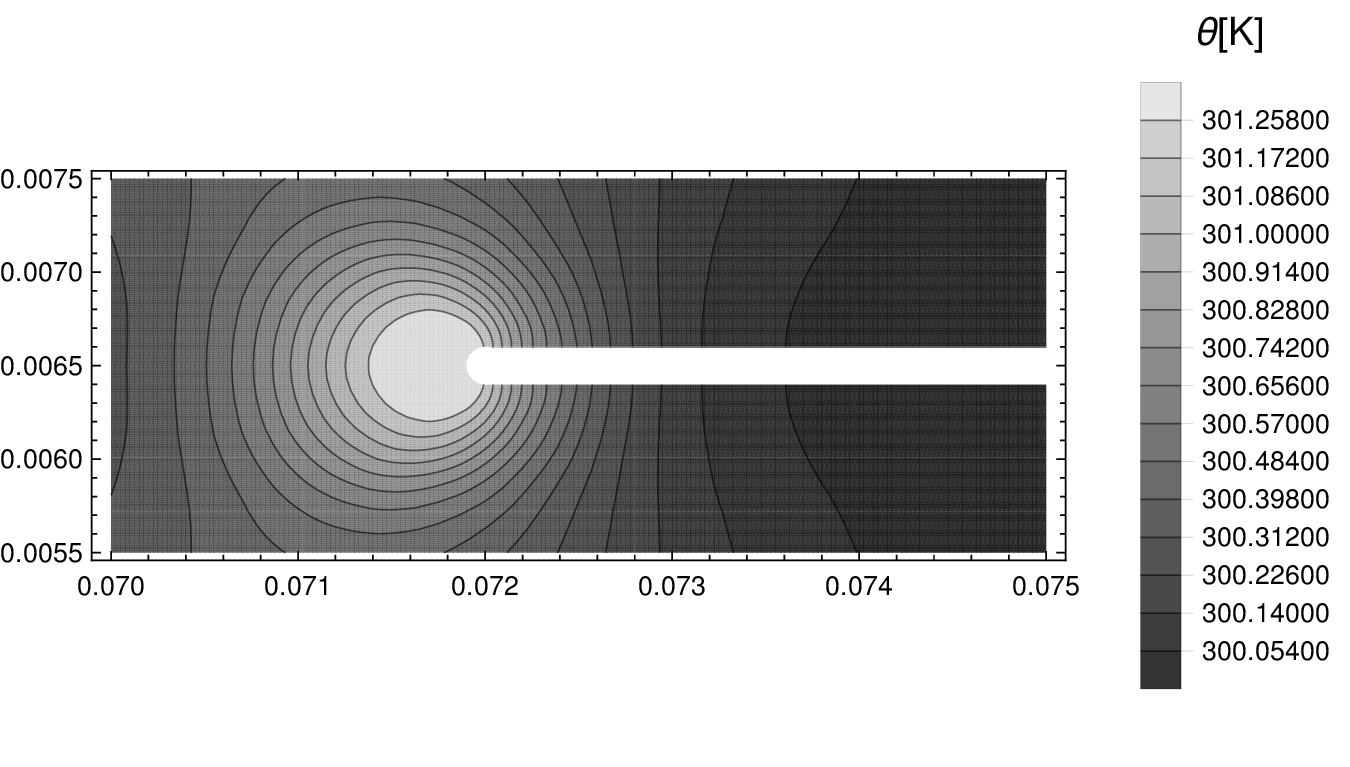}}
  \qquad
  \subfloat[$t=2.0$]{\includegraphics[width=0.35\textwidth]{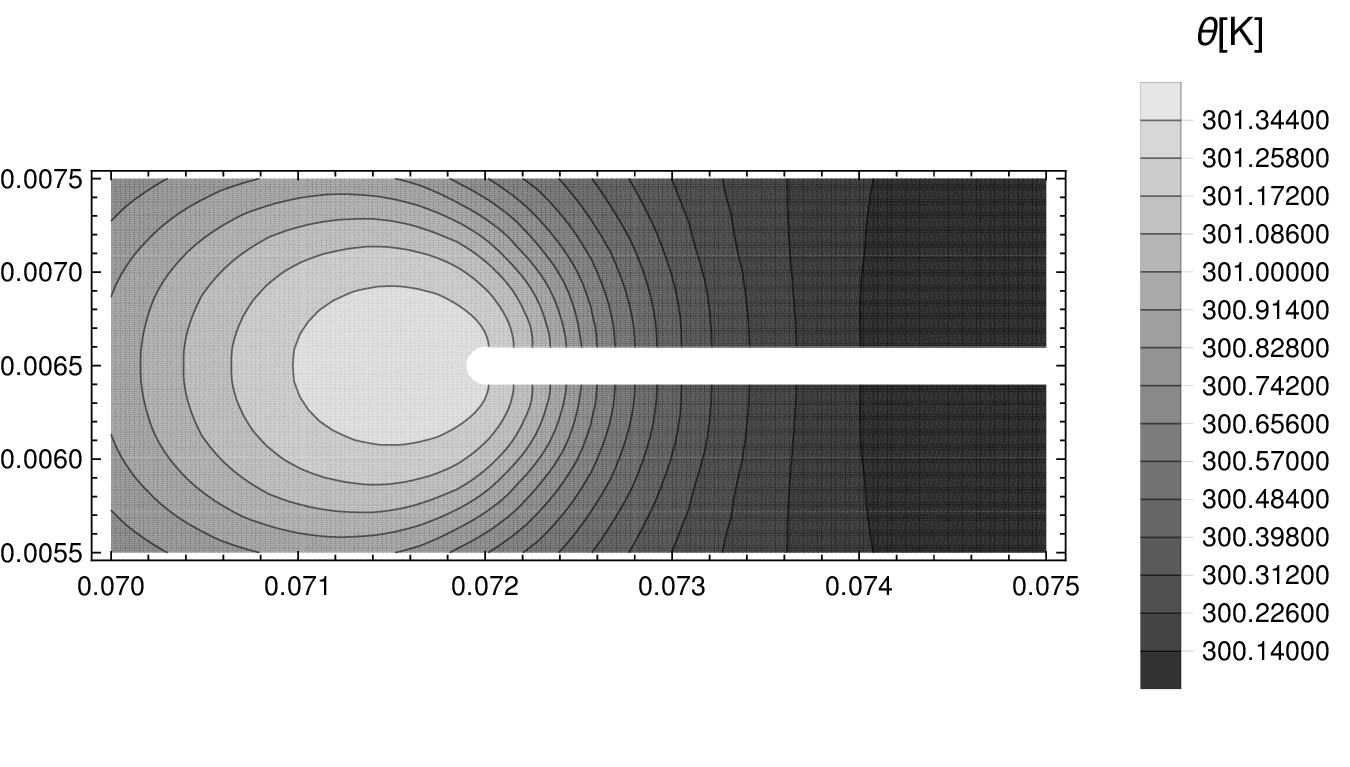}}
  \\
  \subfloat[$t=2.5$]{\includegraphics[width=0.35\textwidth]{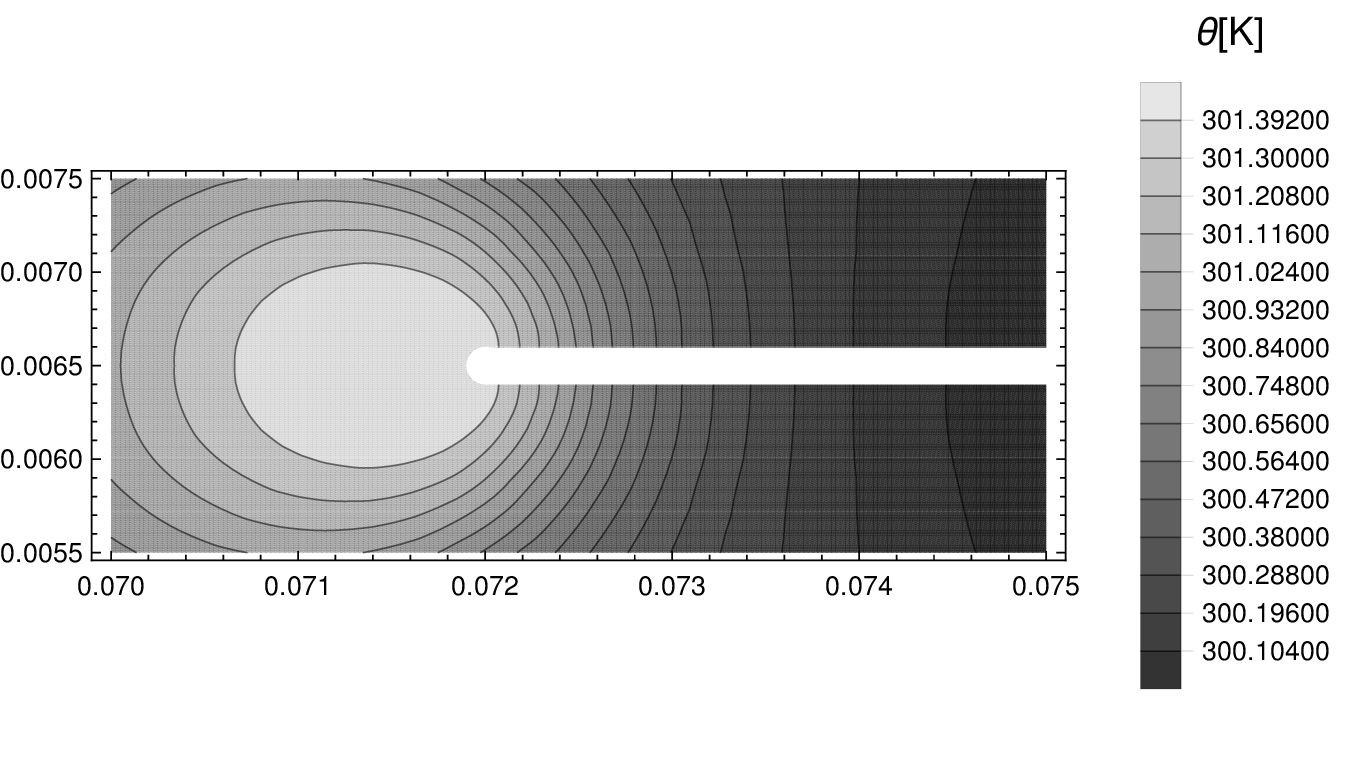}}
  \qquad
  \subfloat[$t=10$]{\includegraphics[width=0.35\textwidth]{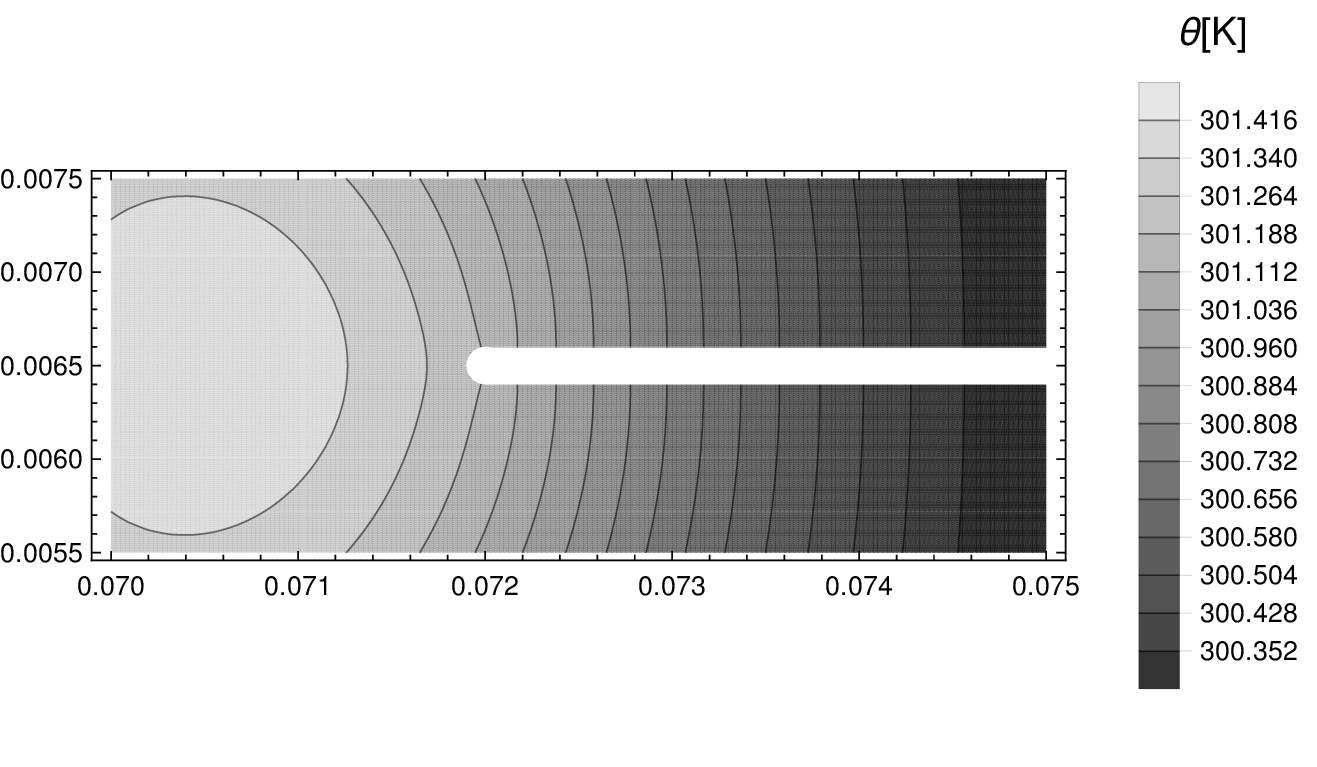}}
  \caption{Temperature field close to the cutout tip, \emph{constant} elastic moduli, \emph{x-large viscosity} value. Nomenclature for viscosity values is described in Table~\ref{tab:parameter-values-viscosity}, remaining material parameters are given in Table~\ref{tab:parameter-values} and Table~\ref{tab:parameter-values-elastic}.}
  \label{fig:temperature-field-whole-specimen-constant-parameters-x-large-viscosity}
\end{figure}

\begin{figure}[h]
  \subfloat[$t=1.0$]{\includegraphics[width=0.35\textwidth]{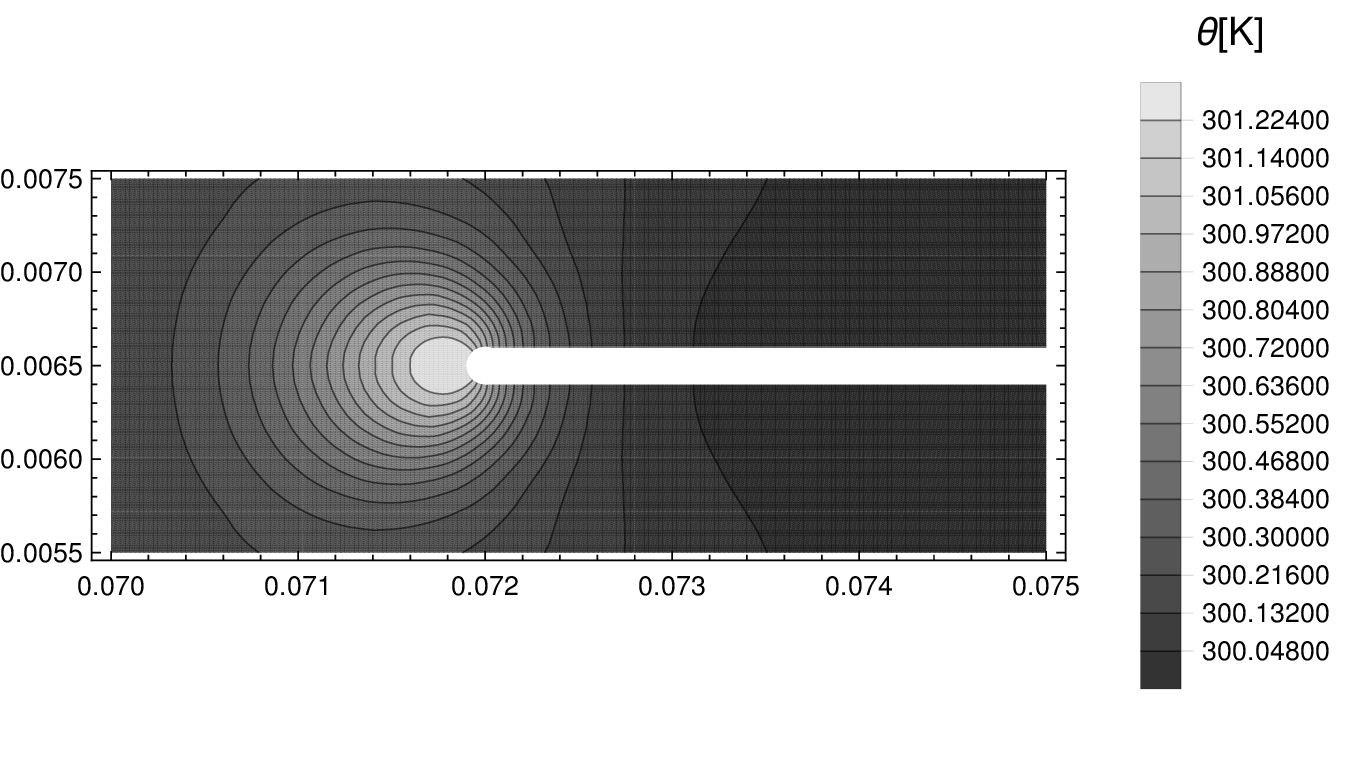}}
  \qquad
  \subfloat[$t=2.0$]{\includegraphics[width=0.35\textwidth]{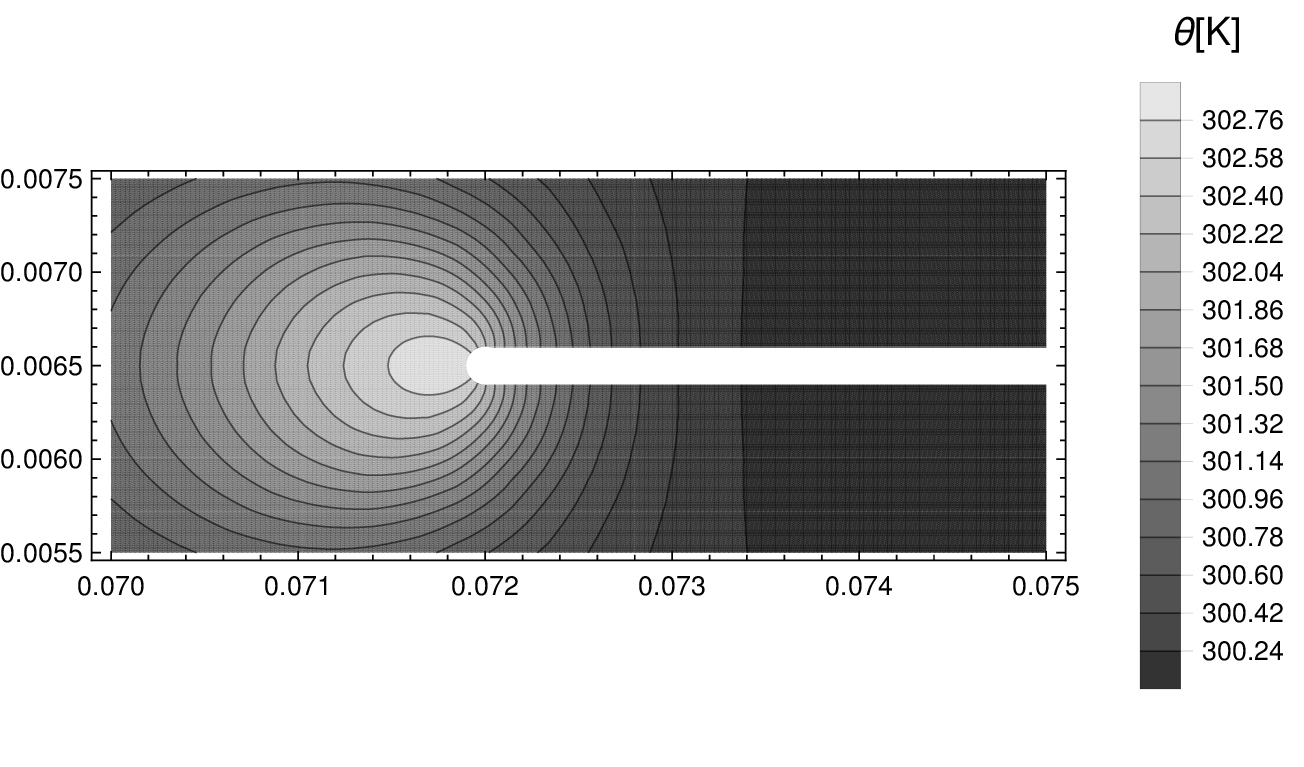}}
  \\
  \subfloat[$t=2.5$\label{fig:temperature-field-whole-specimen-linear-parameters-small-viscosity-c}]{\includegraphics[width=0.35\textwidth]{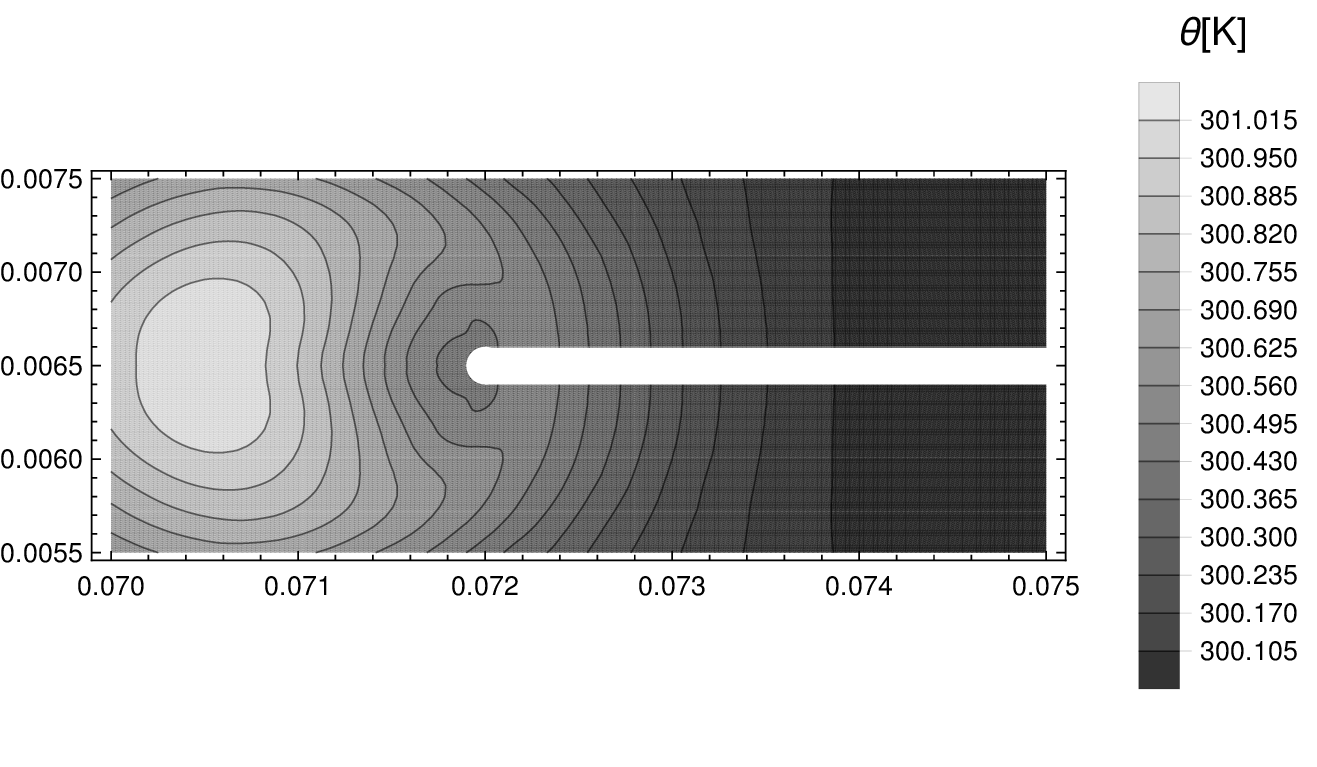}}
  \qquad
  \subfloat[$t=10$]{\includegraphics[width=0.35\textwidth]{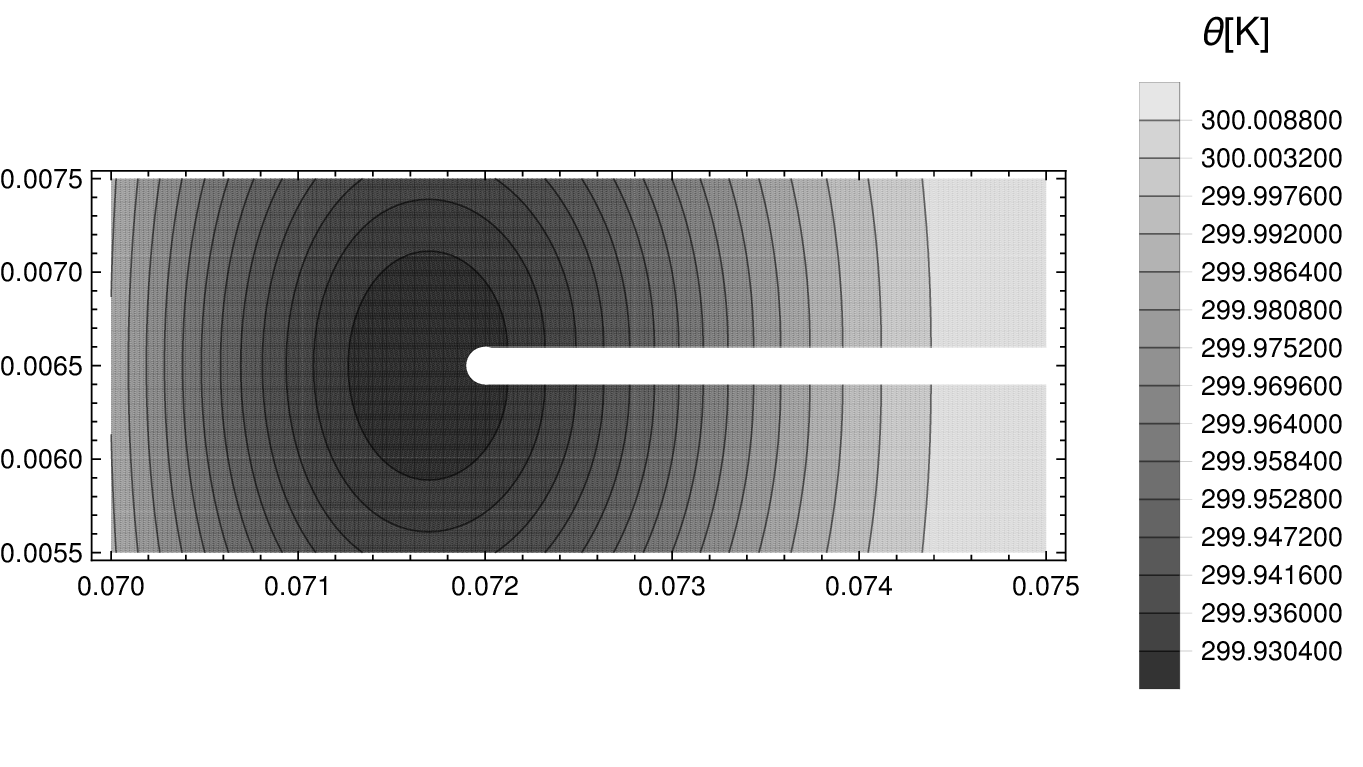}}
  \caption{Temperature field close to the cutout tip, \emph{temperature dependent} elastic moduli, \emph{tiny viscosity} value. Nomenclature for viscosity values is described in Table~\ref{tab:parameter-values-viscosity}, remaining material parameters are given in Table~\ref{tab:parameter-values} and Table~\ref{tab:parameter-values-elastic}.}
  \label{fig:temperature-field-whole-specimen-linear-parameters-small-viscosity}
\end{figure}

\begin{figure}[h]
  \subfloat[$t=1.0$]{\includegraphics[width=0.35\textwidth]{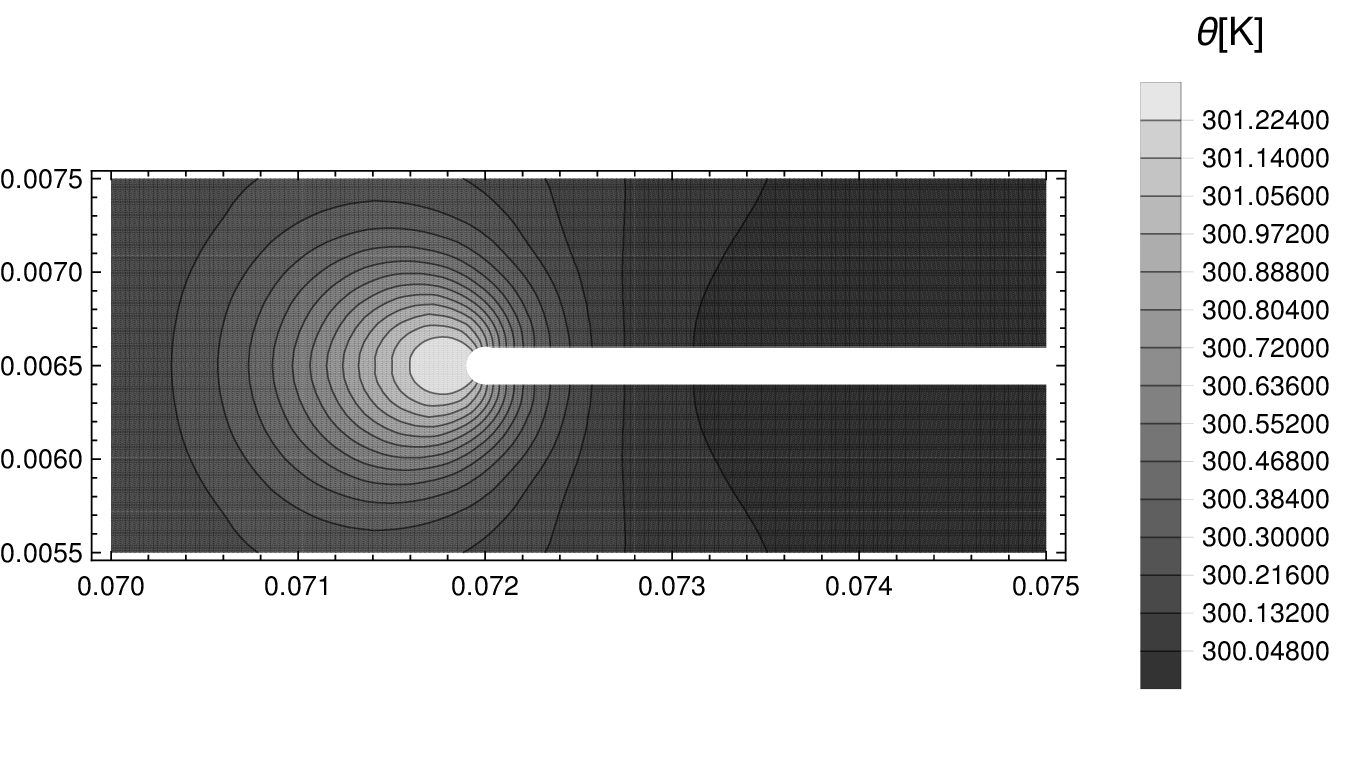}}
  \qquad
  \subfloat[$t=2.0$]{\includegraphics[width=0.35\textwidth]{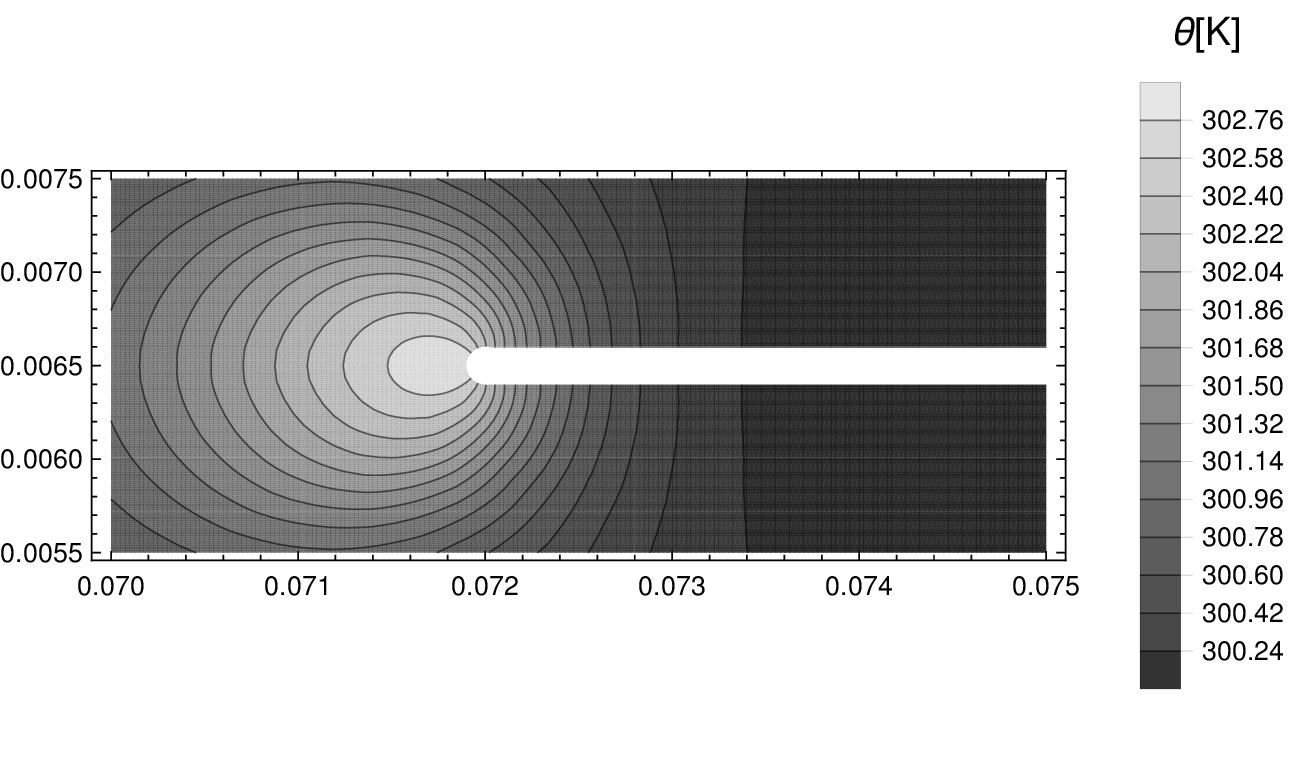}}
  \\
  \subfloat[$t=2.5$]{\includegraphics[width=0.35\textwidth]{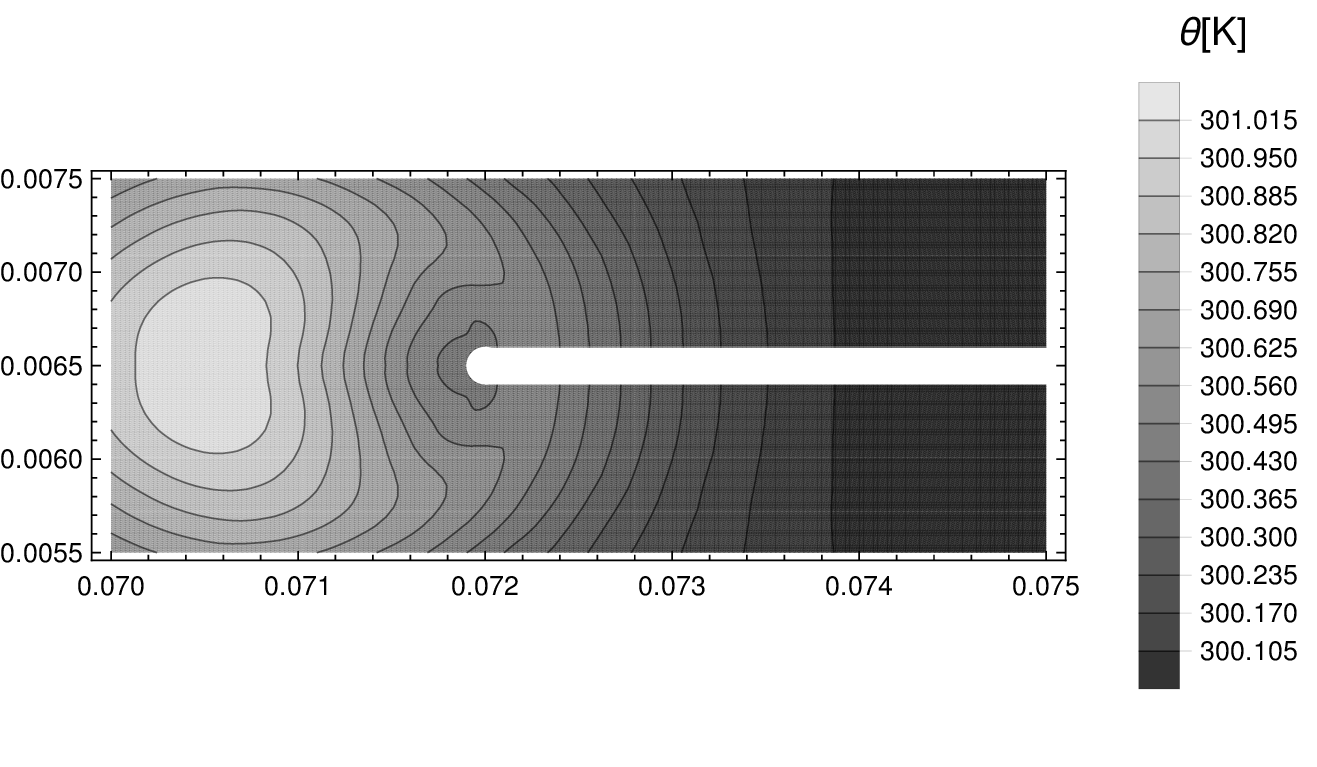}}
  \qquad
  \subfloat[$t=10$]{\includegraphics[width=0.35\textwidth]{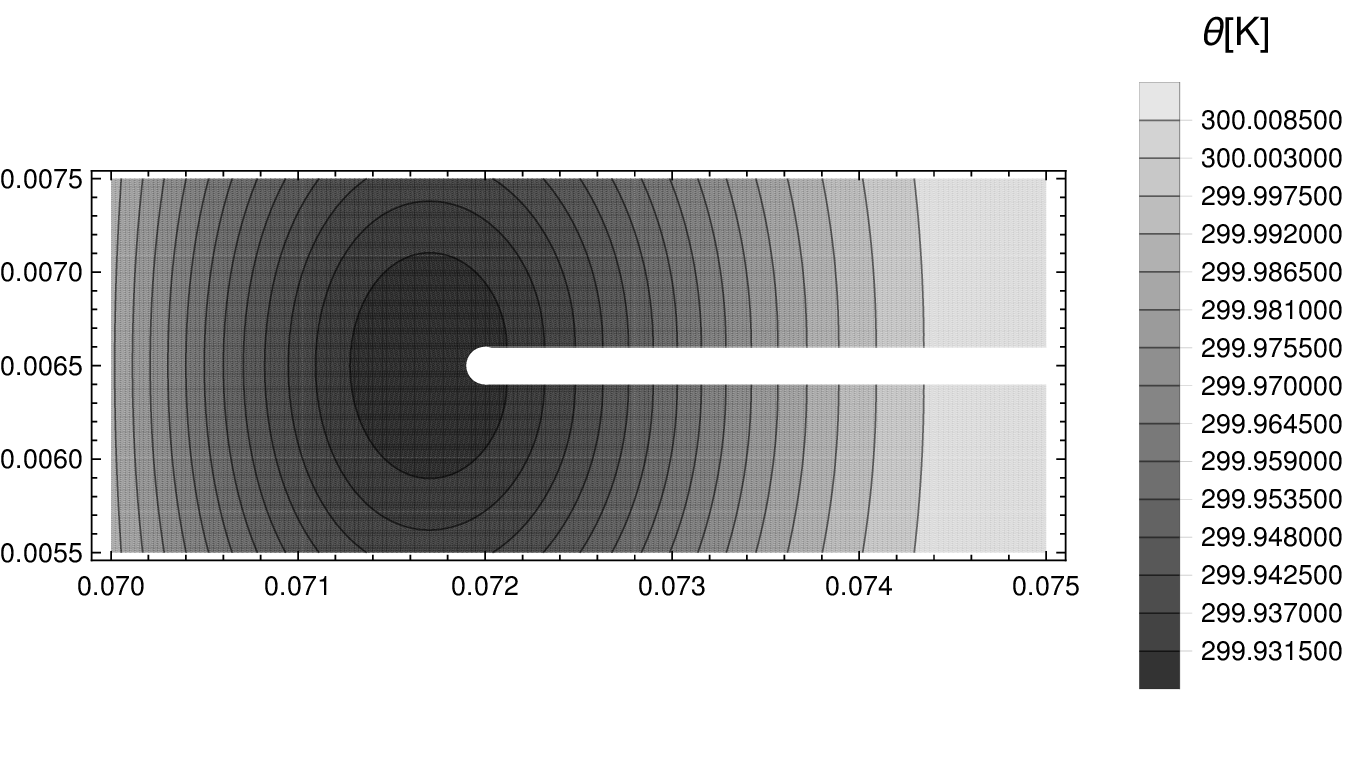}}
  \caption{Temperature field close to the cutout tip, \emph{temperature dependent} elastic moduli, \emph{small viscosity} value. Nomenclature for viscosity values is described in Table~\ref{tab:parameter-values-viscosity}, remaining material parameters are given in Table~\ref{tab:parameter-values} and Table~\ref{tab:parameter-values-elastic}.}
  \label{fig:temperature-field-whole-specimen-linear-parameters-normal-viscosity}
\end{figure}

\begin{figure}[h]
  \subfloat[$t=1.0$]{\includegraphics[width=0.35\textwidth]{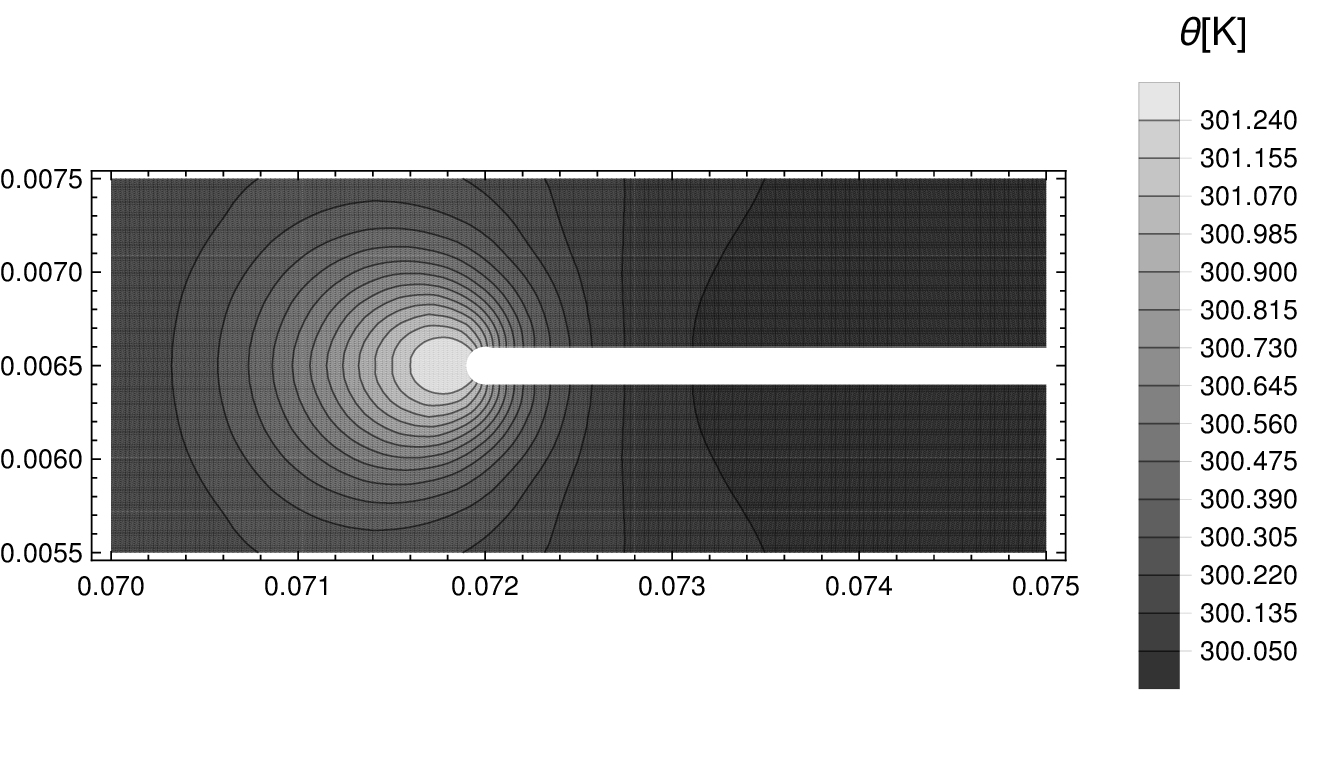}}
  \qquad
  \subfloat[$t=2.0$]{\includegraphics[width=0.35\textwidth]{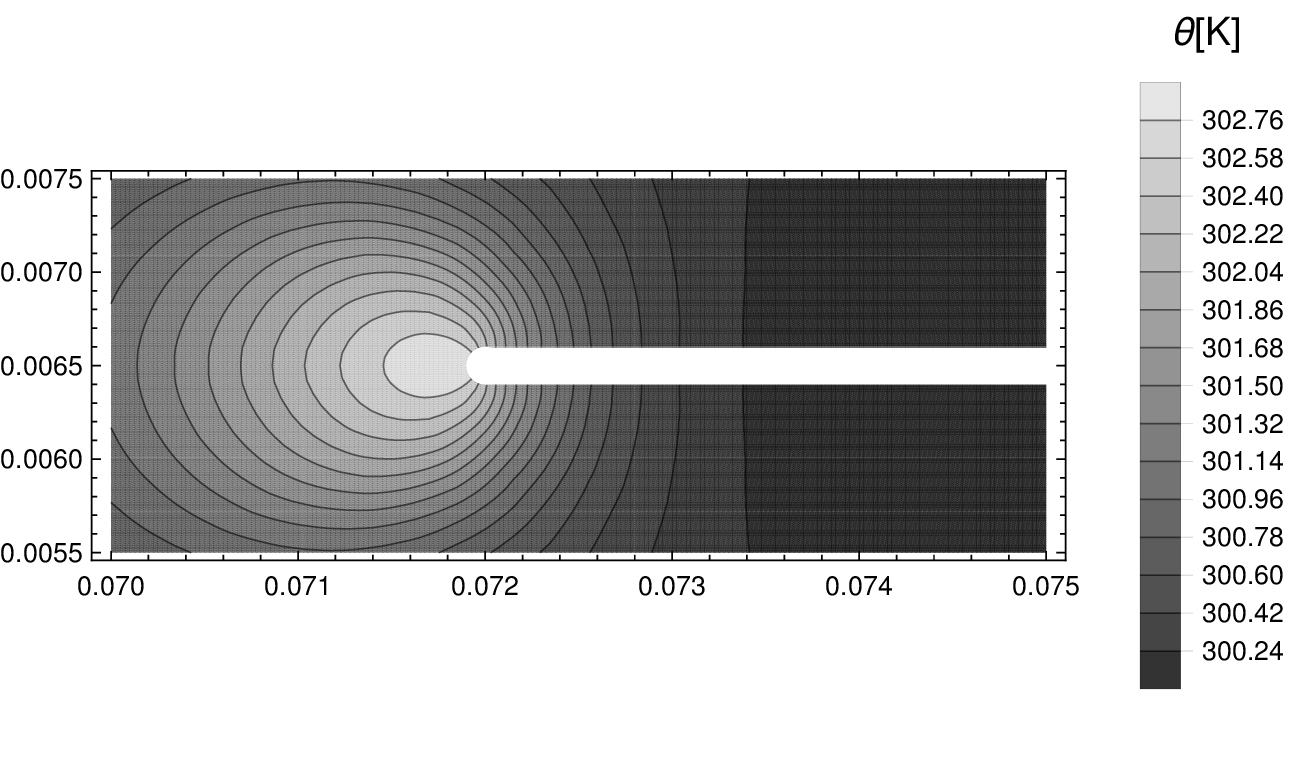}}
  \\
  \subfloat[$t=2.5$]{\includegraphics[width=0.35\textwidth]{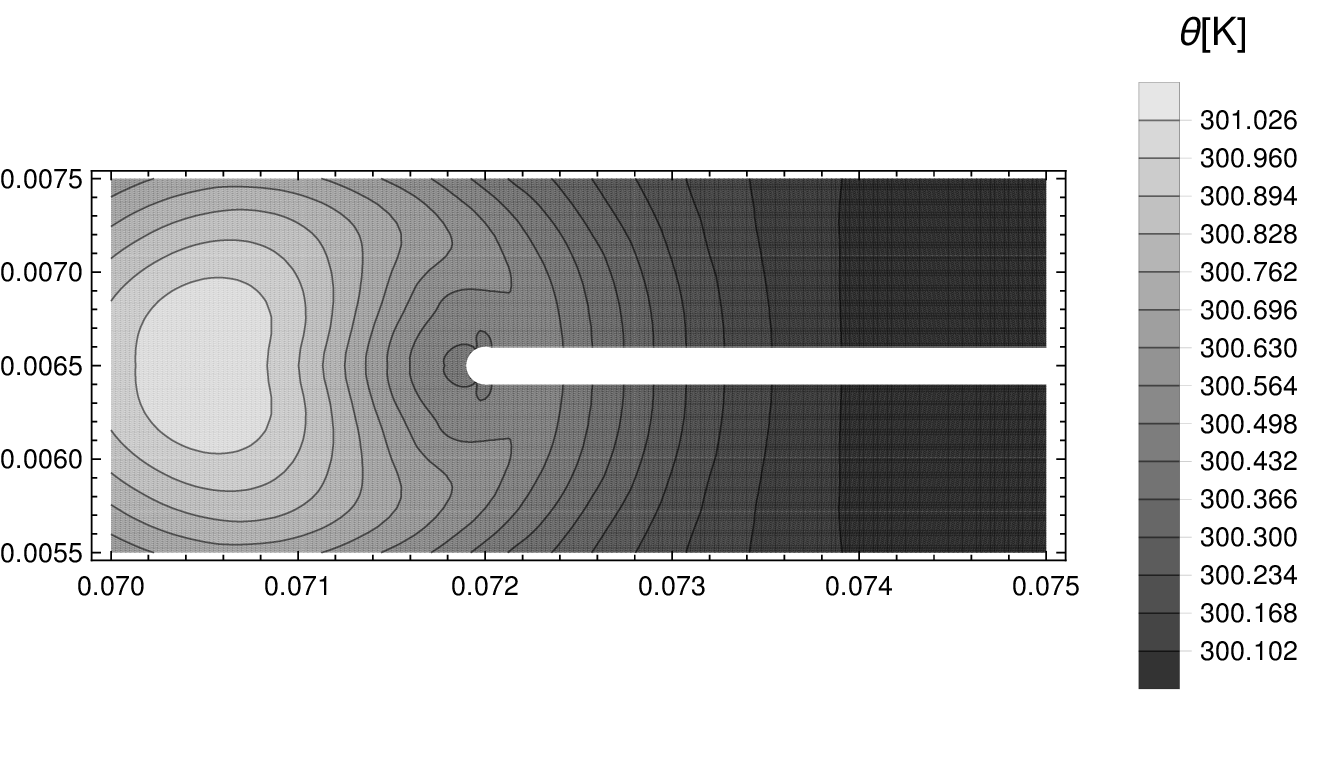}}
  \qquad
  \subfloat[$t=10$]{\includegraphics[width=0.35\textwidth]{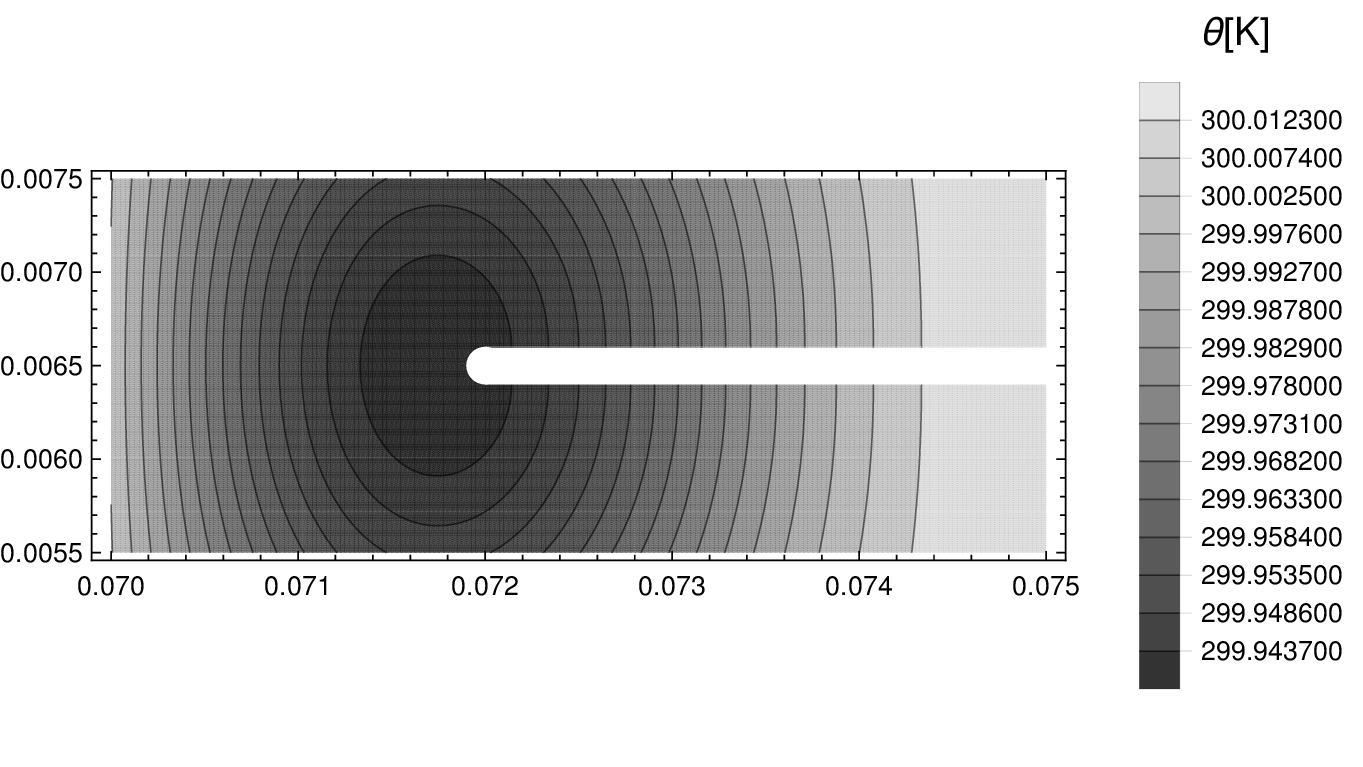}}
  \caption{Temperature field close to the cutout tip, \emph{temperature dependent} elastic moduli, \emph{medium viscosity} value. Nomenclature for viscosity values is described in Table~\ref{tab:parameter-values-viscosity}, remaining material parameters are given in Table~\ref{tab:parameter-values} and Table~\ref{tab:parameter-values-elastic}.}
  \label{fig:temperature-field-whole-specimen-linear-parameters-medium-viscosity}
\end{figure}

\begin{figure}[h]
  \subfloat[$t=1.0$]{\includegraphics[width=0.35\textwidth]{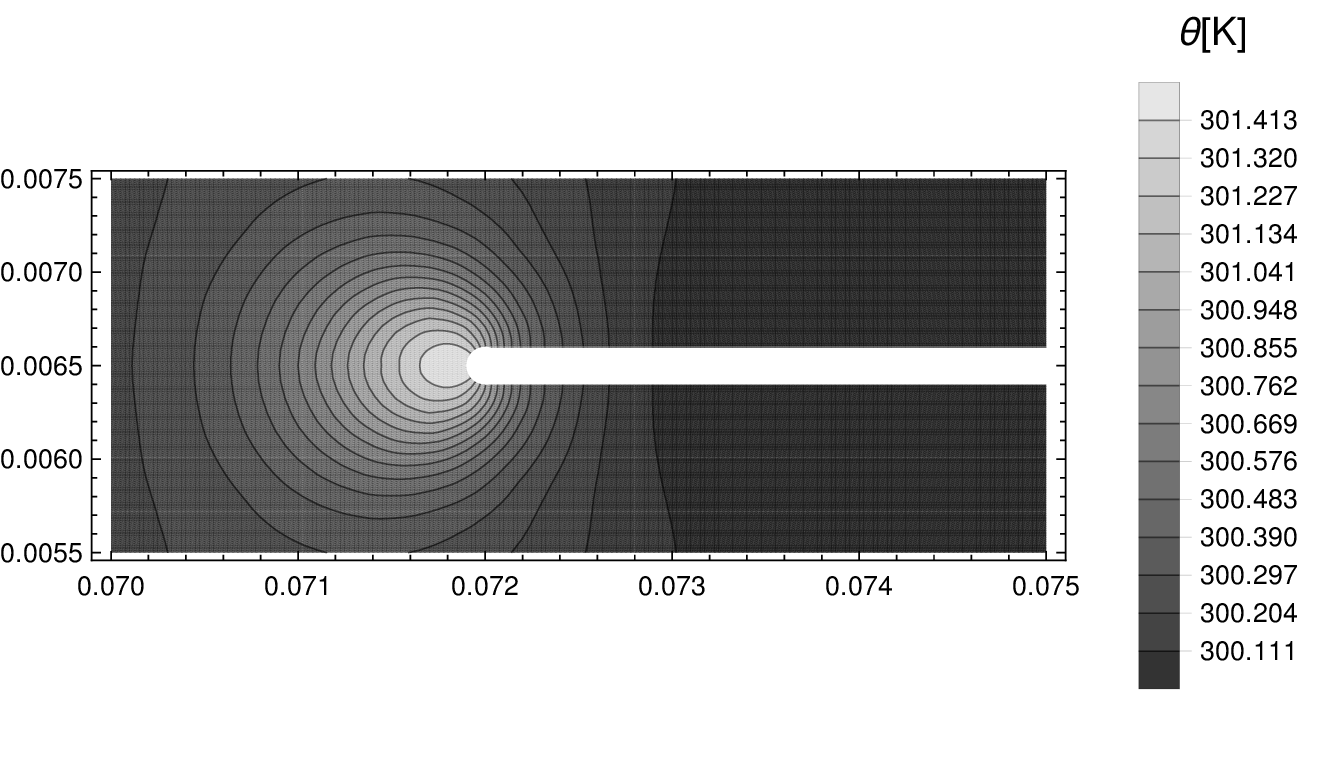}}
  \qquad
  \subfloat[$t=2.0$]{\includegraphics[width=0.35\textwidth]{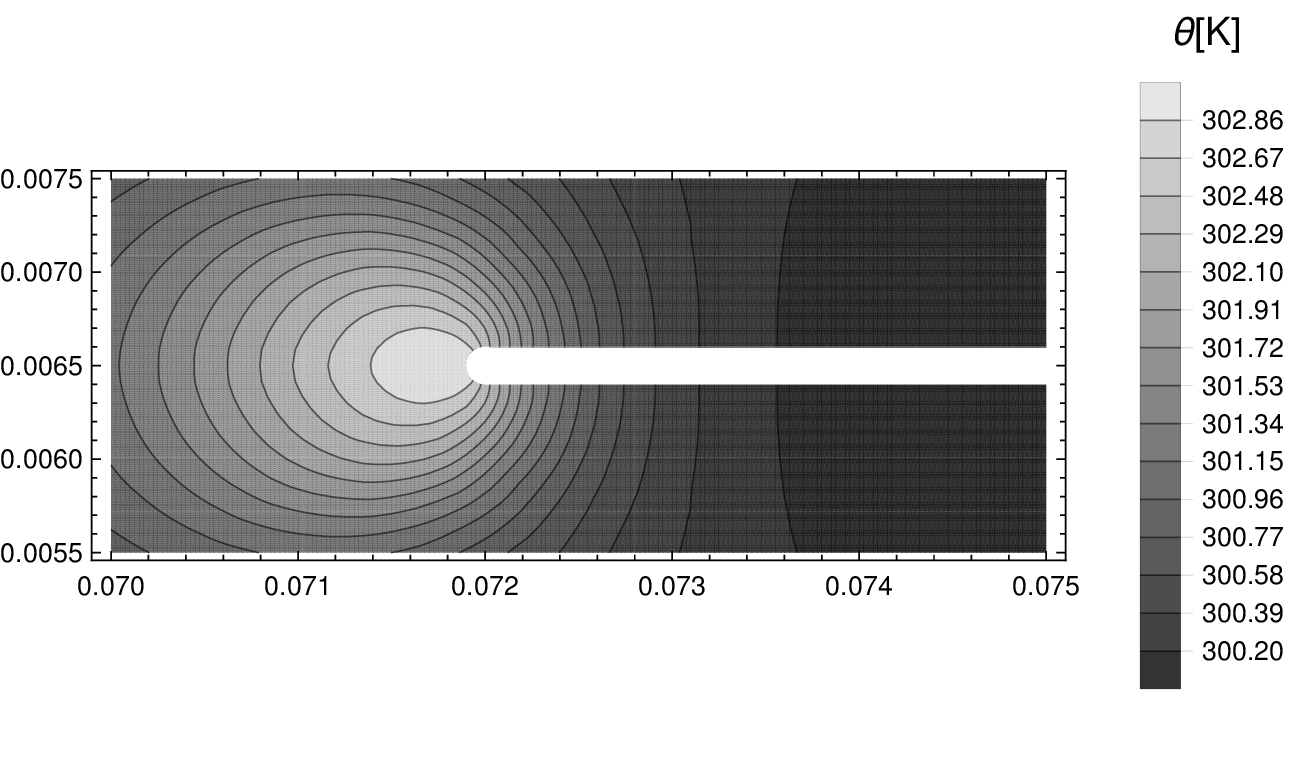}}
  \\
  \subfloat[$t=2.5$]{\includegraphics[width=0.35\textwidth]{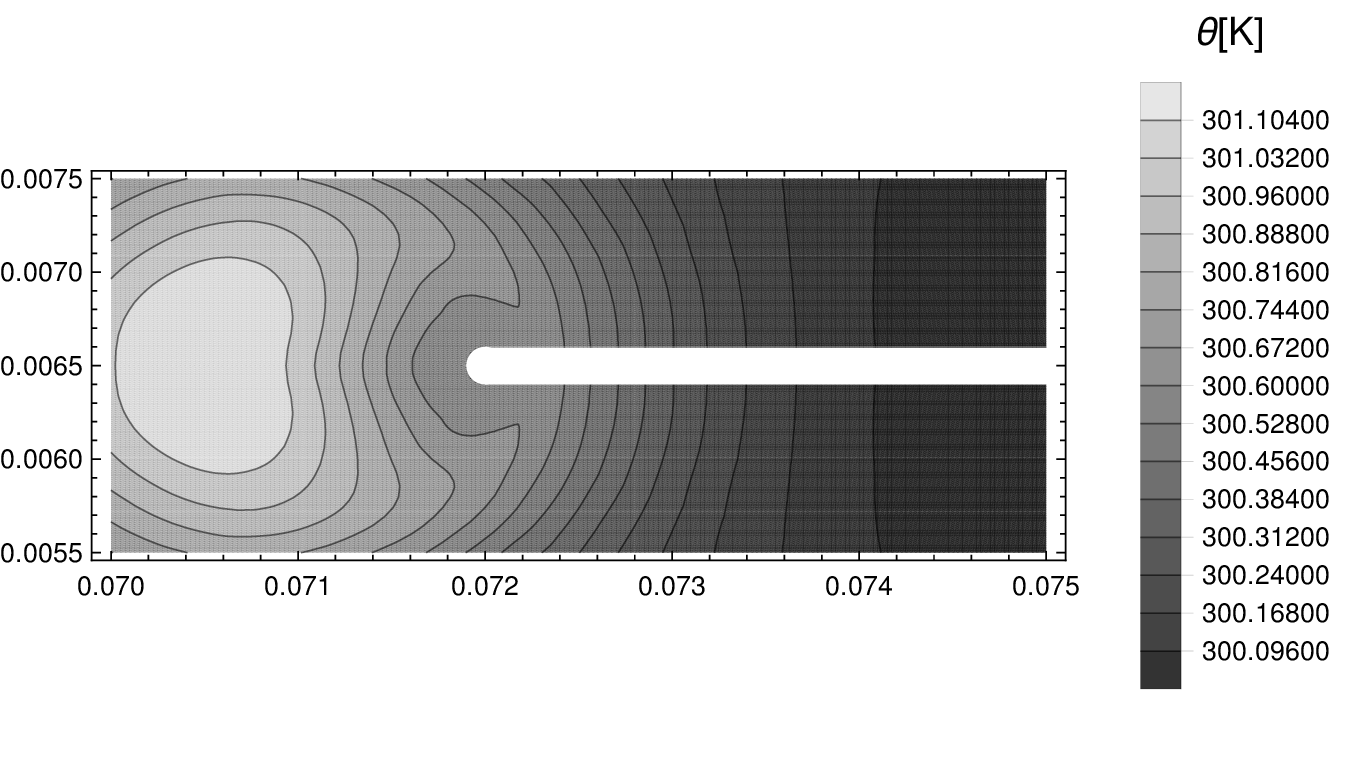}}
  \qquad
  \subfloat[$t=10$]{\includegraphics[width=0.35\textwidth]{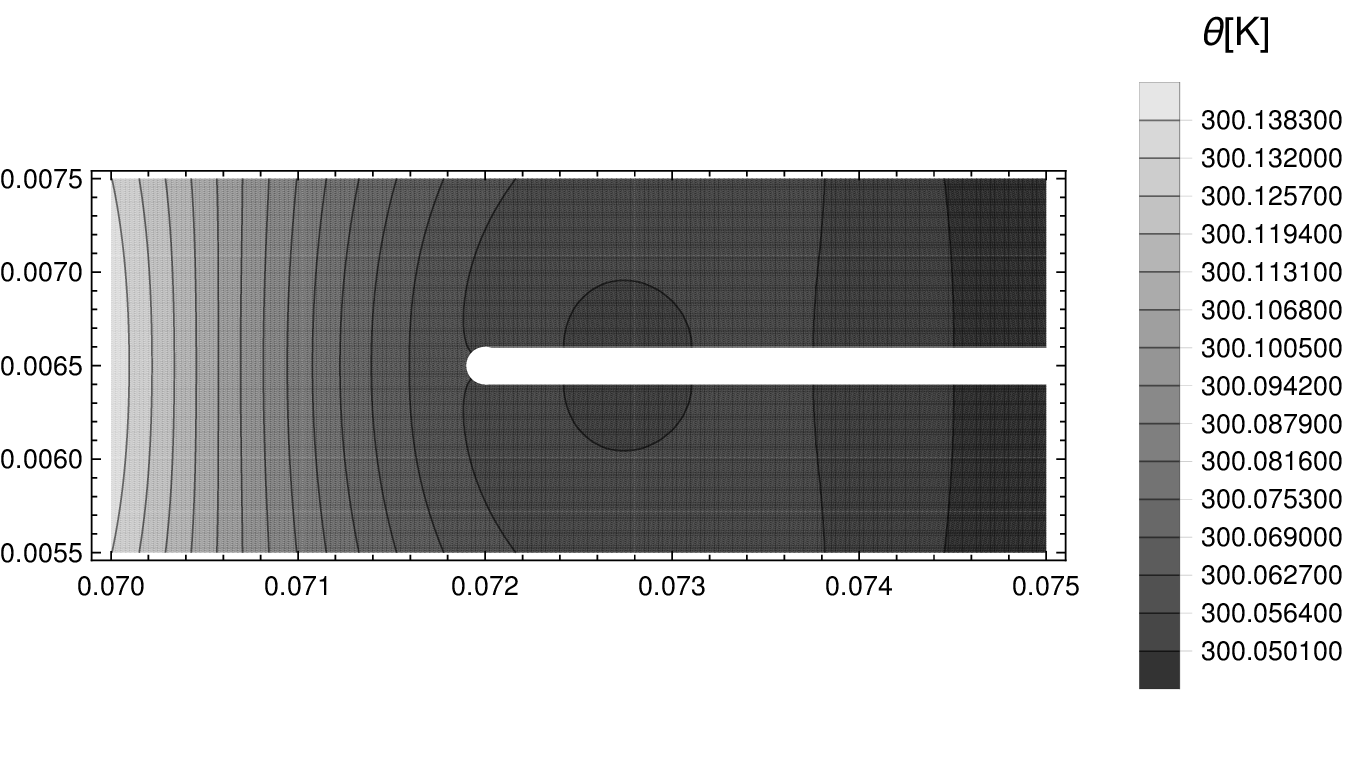}}
  \caption{Temperature field close to the cutout tip, \emph{temperature dependent} elastic moduli, \emph{large viscosity} value. Nomenclature for viscosity values is described in Table~\ref{tab:parameter-values-viscosity}, remaining material parameters are given in Table~\ref{tab:parameter-values} and Table~\ref{tab:parameter-values-elastic}.}
  \label{fig:temperature-field-whole-specimen-linear-parameters-large-viscosity}
\end{figure}

\begin{figure}[h]
  \subfloat[$t=1.0$]{\includegraphics[width=0.35\textwidth]{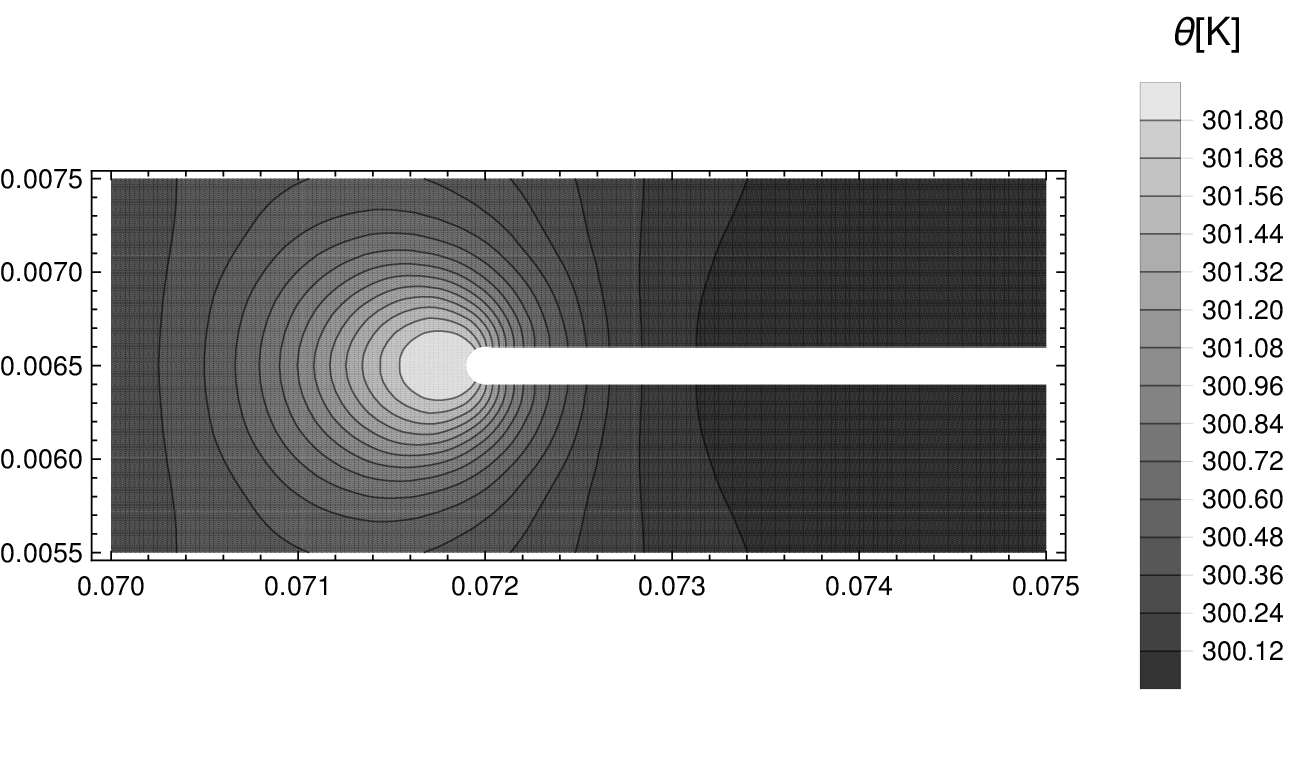}}
  \qquad
  \subfloat[$t=2.0$]{\includegraphics[width=0.35\textwidth]{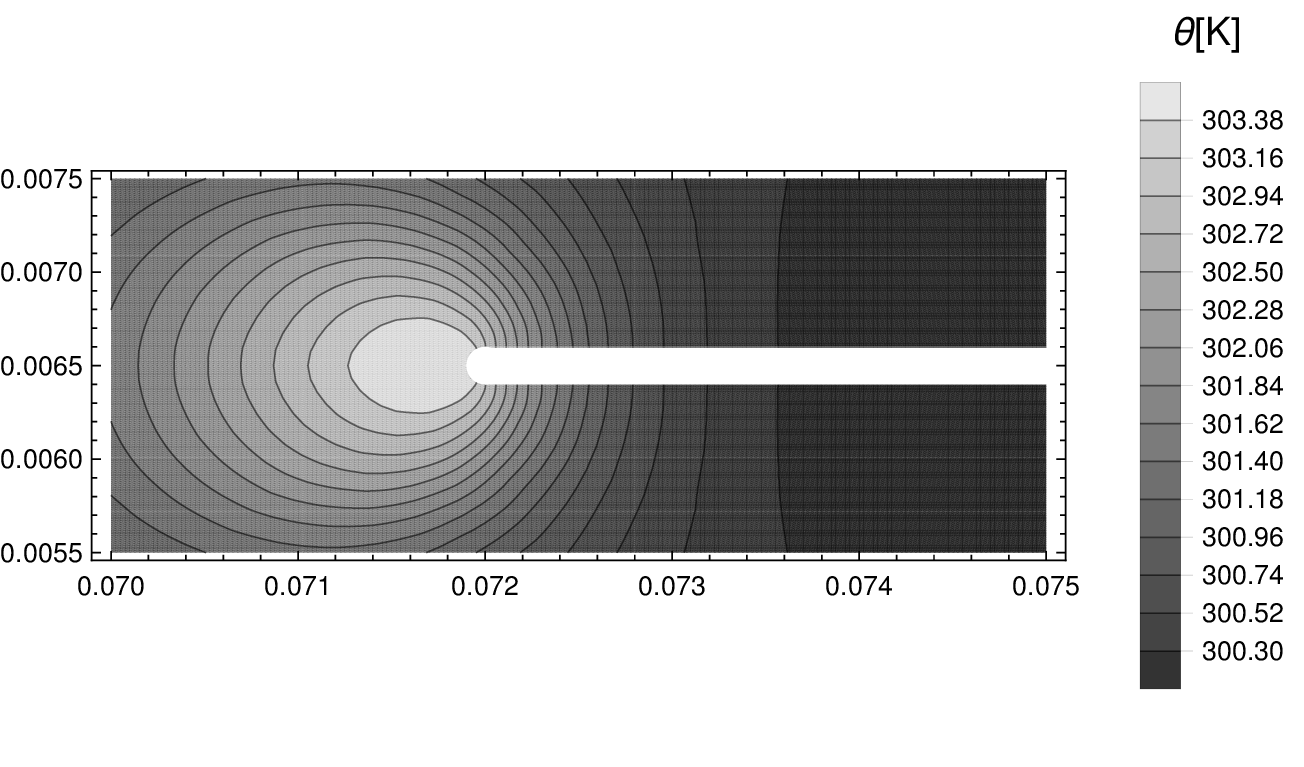}}
  \\
  \subfloat[$t=2.5$ \label{fig:temperature-field-whole-specimen-linear-parameters-s-large-viscosity-c}]{\includegraphics[width=0.35\textwidth]{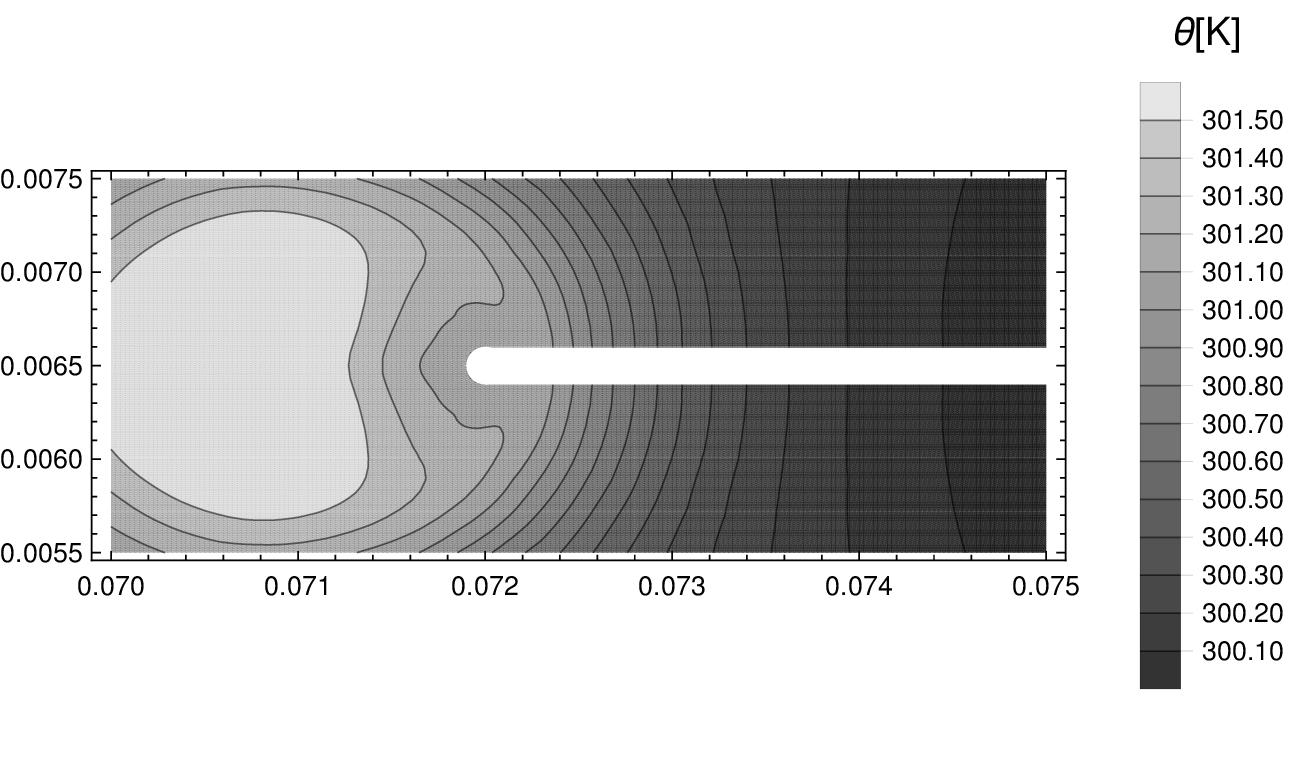}}
  \qquad
  \subfloat[$t=10$]{\includegraphics[width=0.35\textwidth]{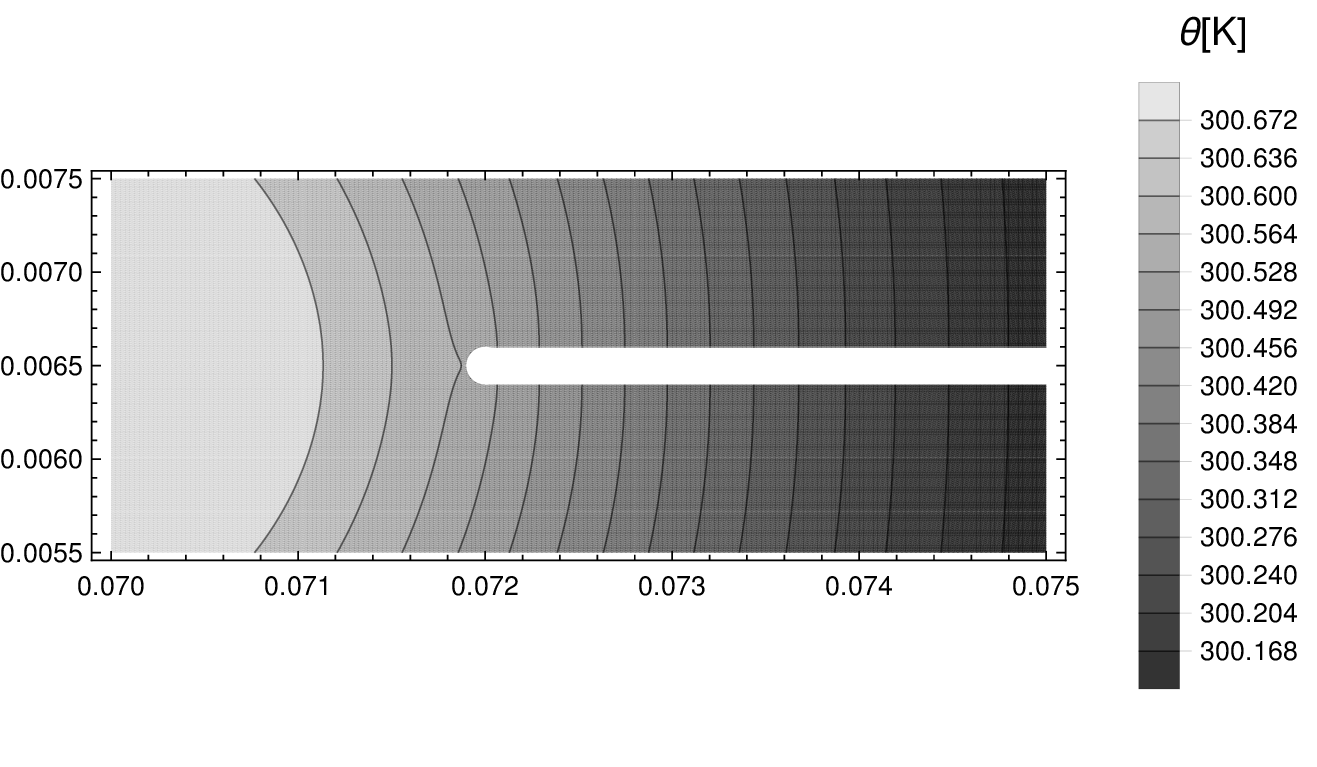}}
  \caption{Temperature field close to the cutout tip, \emph{temperature dependent} elastic moduli, \emph{s-large viscosity} value. Nomenclature for viscosity values is described in Table~\ref{tab:parameter-values-viscosity}, remaining material parameters are given in Table~\ref{tab:parameter-values} and Table~\ref{tab:parameter-values-elastic}.}
  \label{fig:temperature-field-whole-specimen-linear-parameters-s-large-viscosity}
\end{figure}

\begin{figure}[h]
  \subfloat[$t=1.0$]{\includegraphics[width=0.35\textwidth]{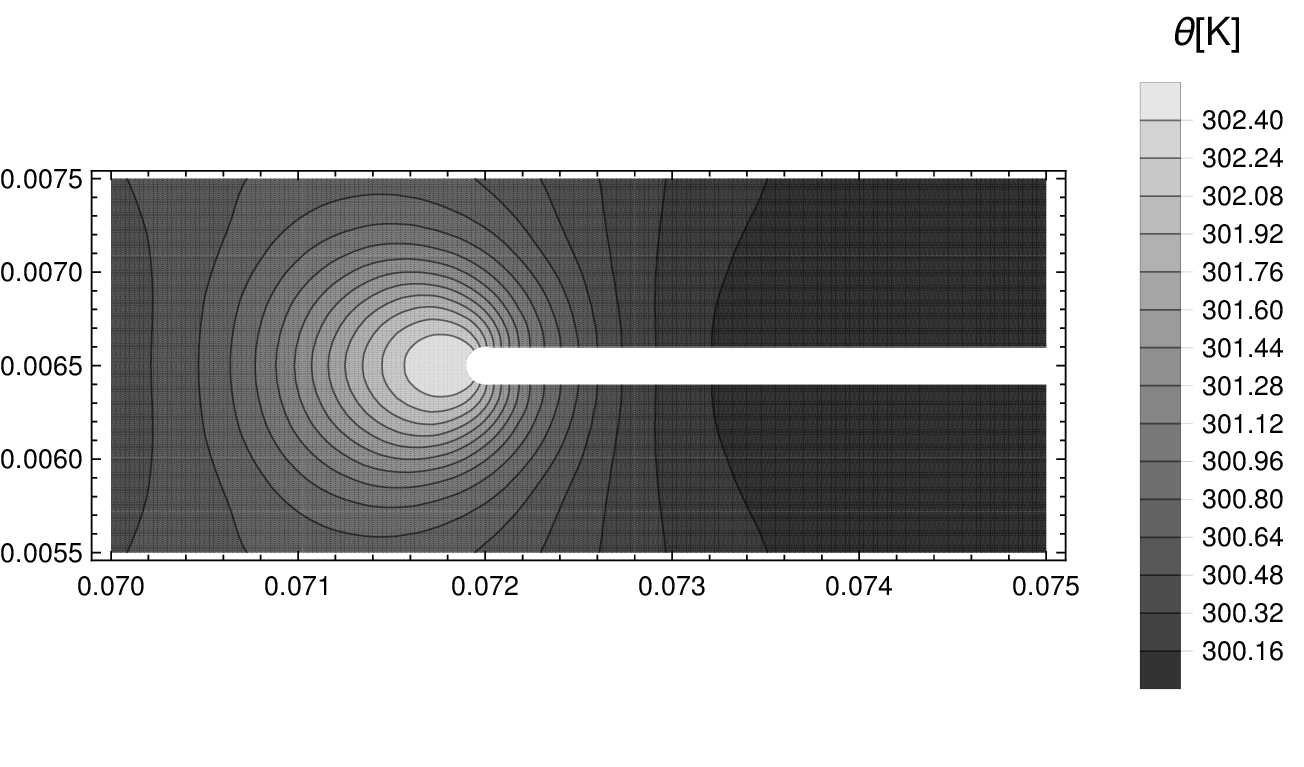}}
  \qquad
  \subfloat[$t=2.0$]{\includegraphics[width=0.35\textwidth]{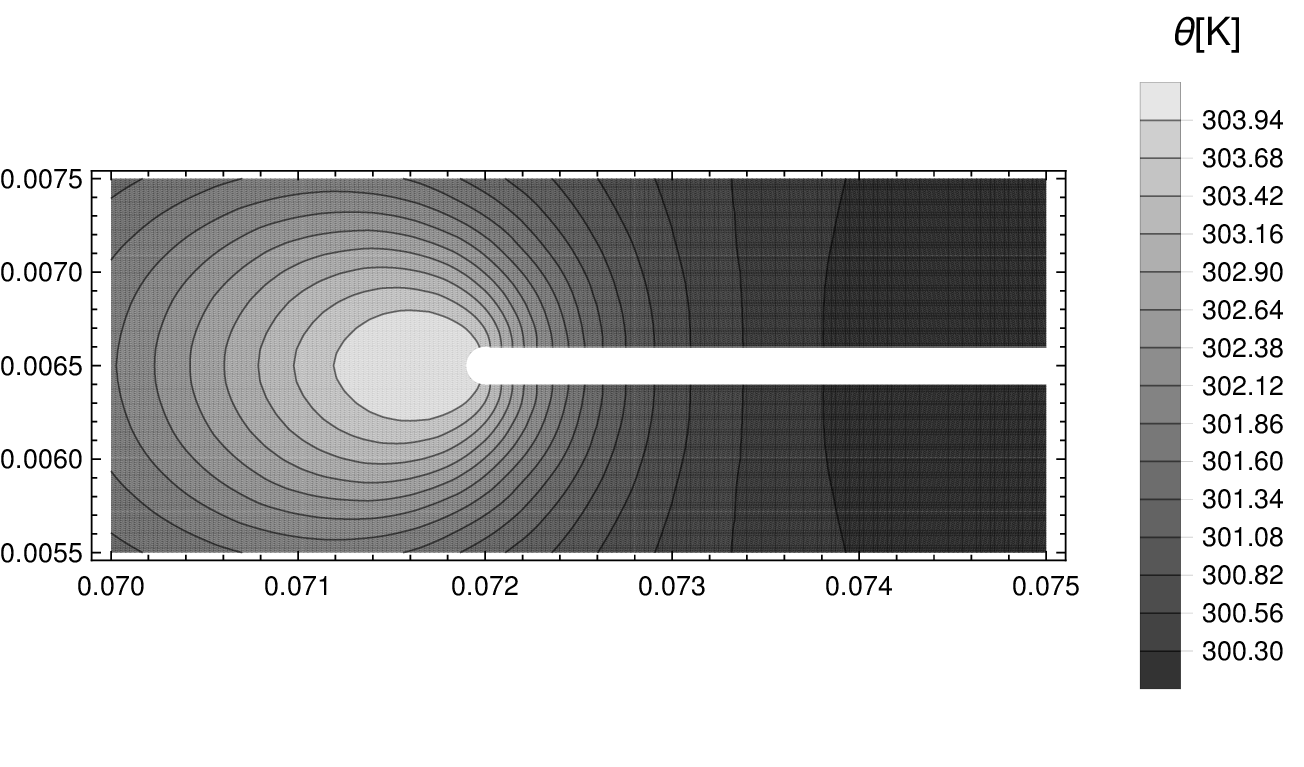}}
  \\
  \subfloat[$t=2.5$]{\includegraphics[width=0.35\textwidth]{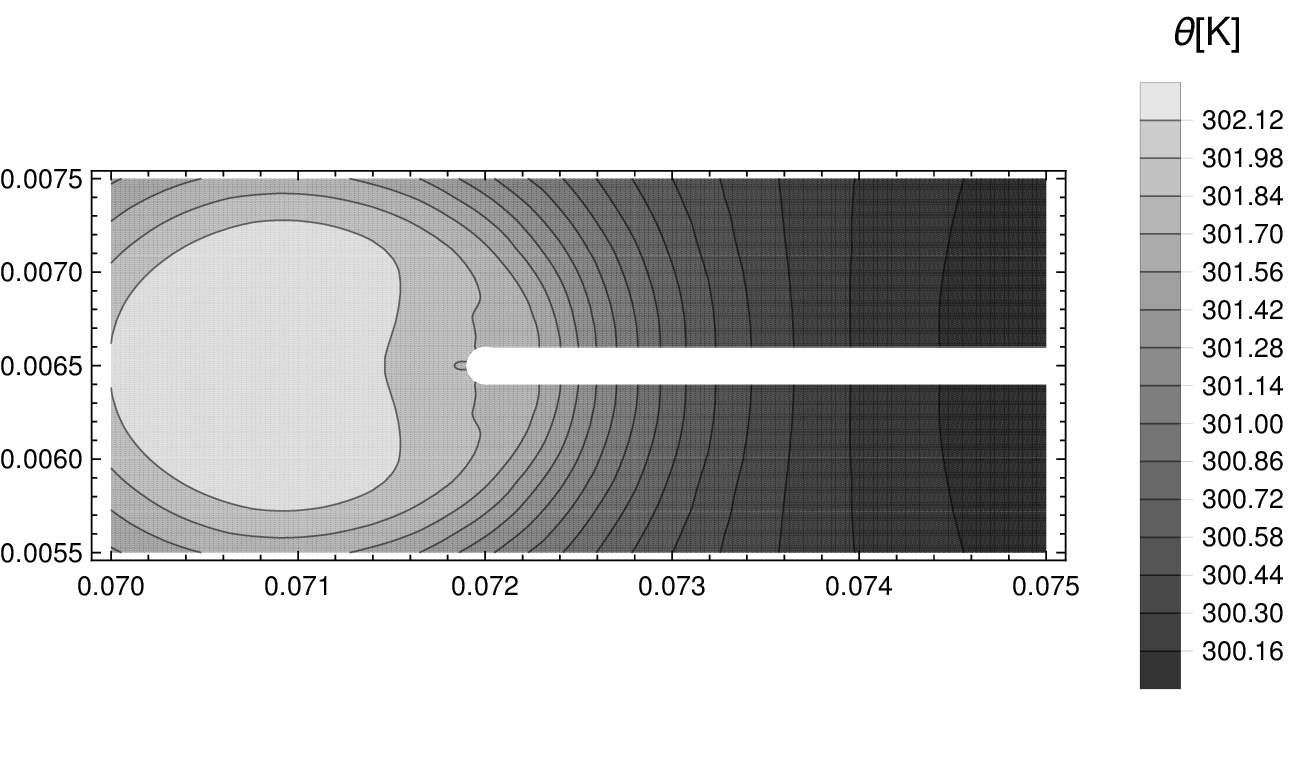}}
  \qquad
  \subfloat[$t=10$]{\includegraphics[width=0.35\textwidth]{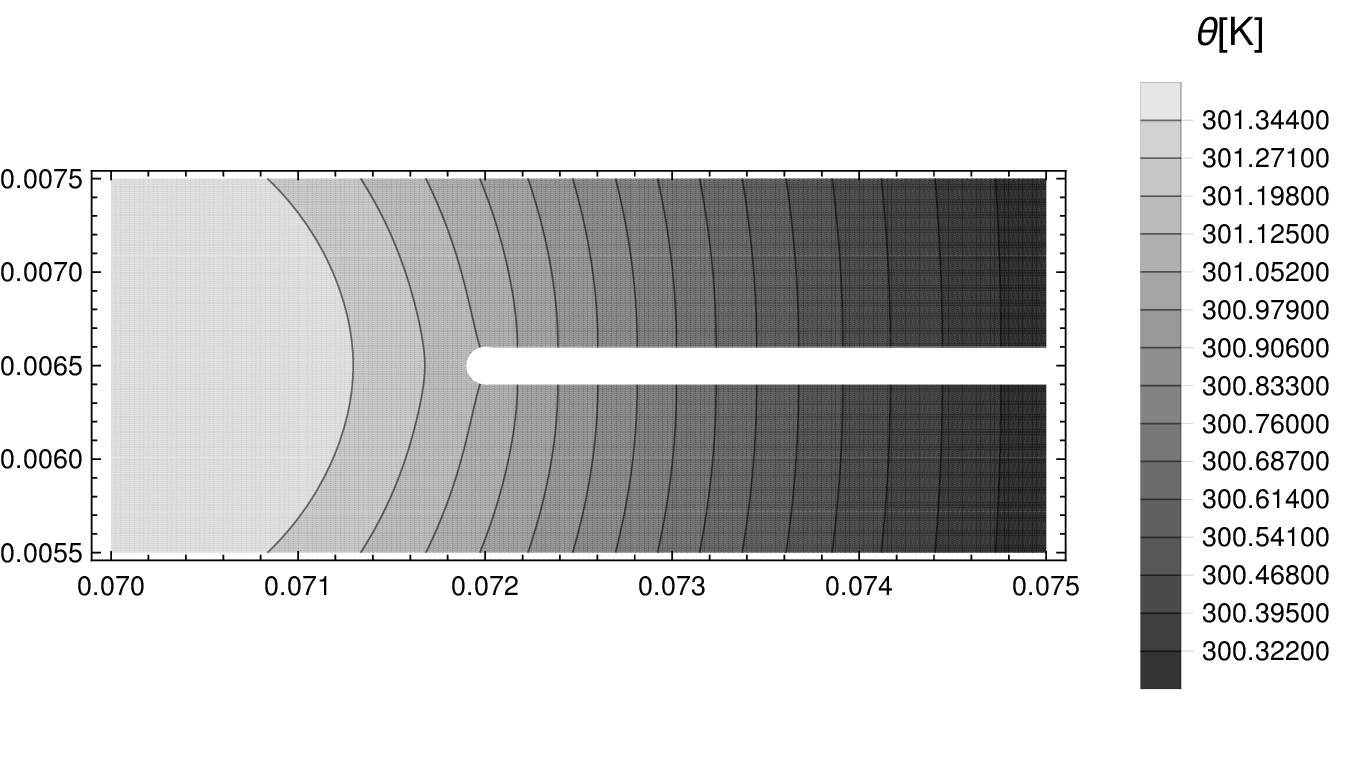}}
  \caption{Temperature field close to the cutout tip, \emph{temperature dependent} elastic moduli, \emph{x-large viscosity} value. Nomenclature for viscosity values is described in Table~\ref{tab:parameter-values-viscosity}, remaining material parameters are given in Table~\ref{tab:parameter-values} and Table~\ref{tab:parameter-values-elastic}.}
  \label{fig:temperature-field-whole-specimen-linear-parameters-x-large-viscosity}
\end{figure}

Finally, we report results regarding the influence of the chosen viscosity values on the evolution of the corresponding temperature field in the sample, see Figure~\ref{fig:temperature-field-whole-specimen-constant-parameters-small-viscosity}--Figure~\ref{fig:temperature-field-whole-specimen-linear-parameters-x-large-viscosity}. The highest temperature changes clearly take place at the vicinity of the measurement site C (cutout tip), see also Figure~\ref{fig:temperature-sites-constant-various-viscosities} and \ref{fig:temperature-sites-linear-various-viscosities}. Interestingly, the choice of ``s-large'' viscosity allows one to get temperature changes that are of the same order of magnitude as that reported in~\cite{martinez.jrs.toussaint.e.ea:heat}. Note, however, that we are not claiming the exact match with the experimental data, our numerical experiments only resemble the genuine experimental setting. For example we for simplicity use temperature boundary condition~\eqref{eq:44} which leaves much to be desired from the experimental setting perspective.

The fact that the highest temperature changes are observed at the vicinity of the tip cutout is not surprising since the highest stress variations and the biggest deformation are known to take place in this region. (Recall the classical stress concentration phenomenon.) The structure of the temperature field can be however quite complex. Interestingly, a small region in the vicinity of the measurement site C (cutout tip) can be at certain time interval \emph{colder} than its neighborhood, see for example Figure~\ref{fig:temperature-field-whole-specimen-linear-parameters-small-viscosity-c}. This is a consequence of the temperature dependent elastic moduli, no such phenomenon takes place in the material with constant elastic moduli -- compare Figure~\ref{fig:temperature-field-whole-specimen-constant-parameters-small-viscosity-c} and Figure~\ref{fig:temperature-field-whole-specimen-linear-parameters-small-viscosity-c}. The locally colder region in the in the vicinity of the measurement site C (cutout tip), then progressively disappears even if one considers the models with temperature dependent material moduli, but the viscosity must be high enough -- compare for example Figure~\ref{fig:temperature-field-whole-specimen-linear-parameters-s-large-viscosity-c} (\emph{s-large} viscosity) and Figure~\ref{fig:temperature-field-whole-specimen-linear-parameters-small-viscosity-c} (\emph{tiny} viscosity).

%
%
%
%
%
%
%

\begin{figure}[h]
  \subfloat[$t=1.0$]{\includegraphics[width=0.35\textwidth]{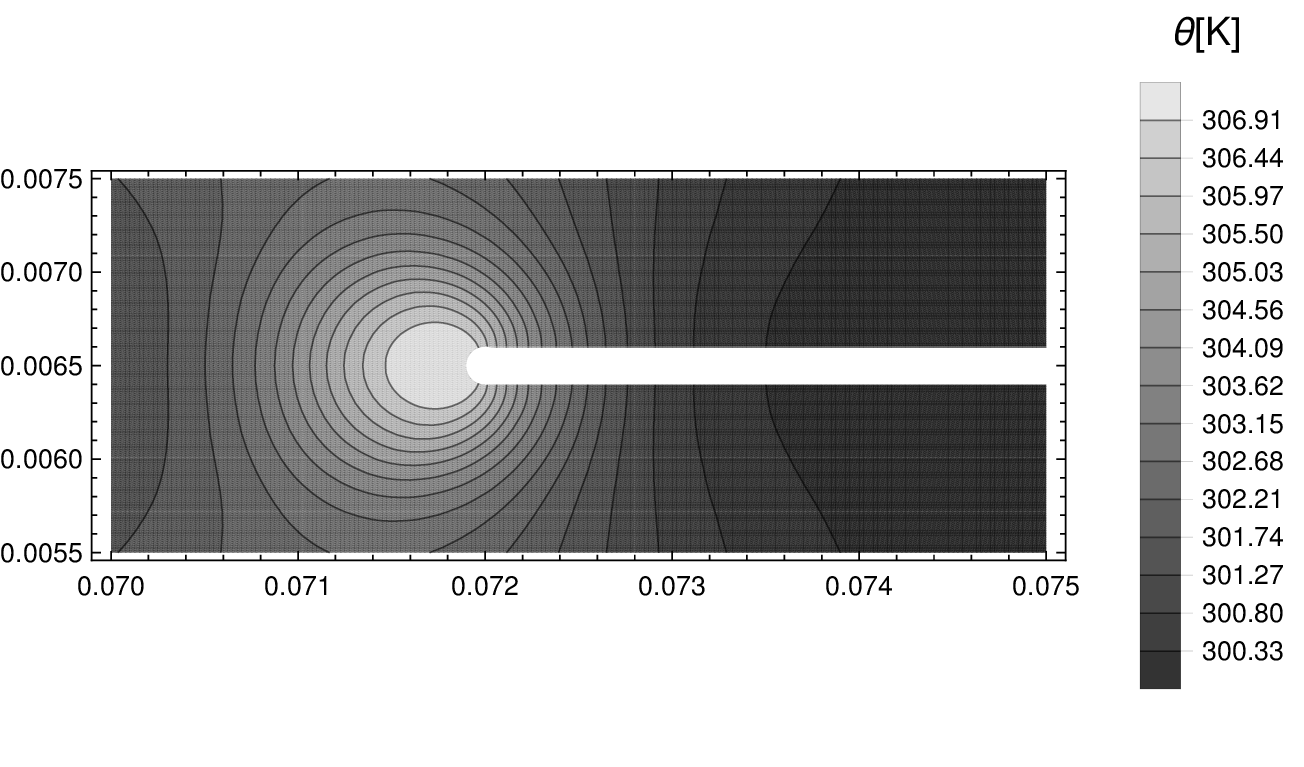}}
  \qquad
  \subfloat[$t=2.0$]{\includegraphics[width=0.35\textwidth]{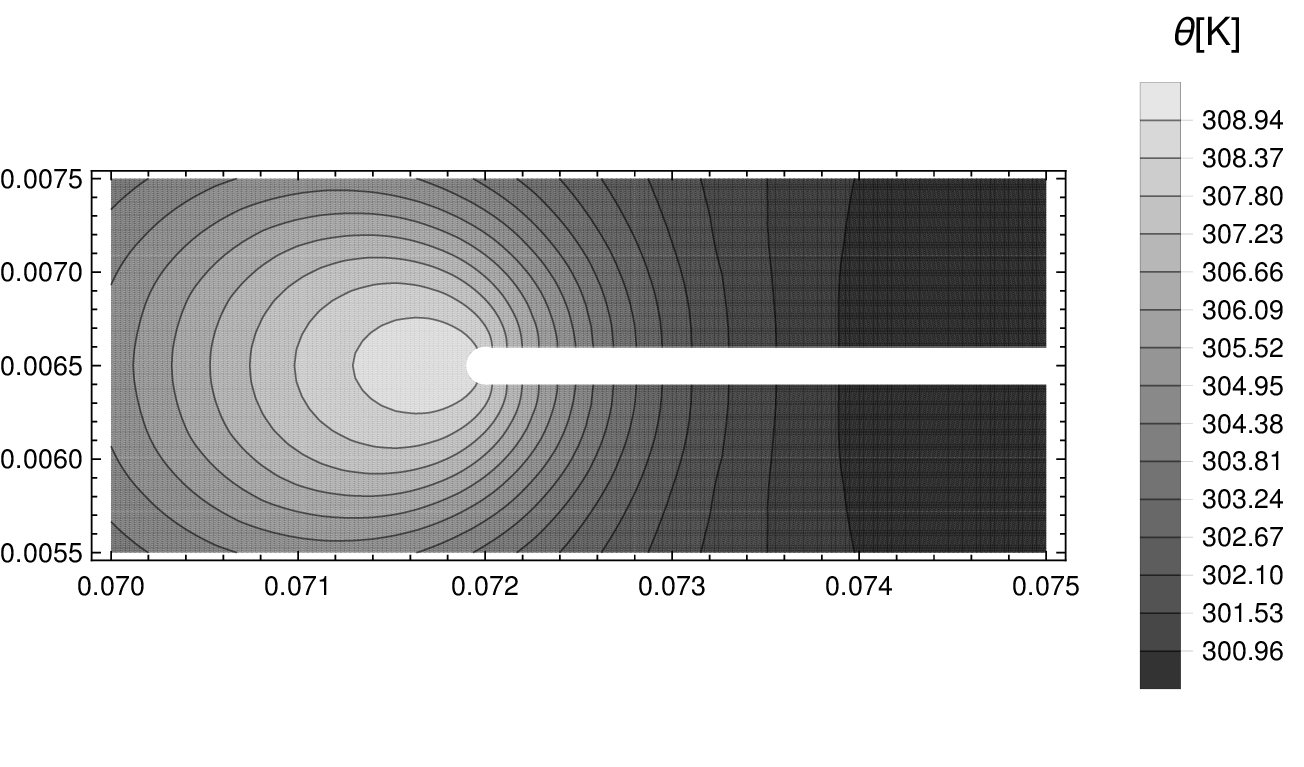}}
  \\
  \subfloat[$t=2.5$]{\includegraphics[width=0.35\textwidth]{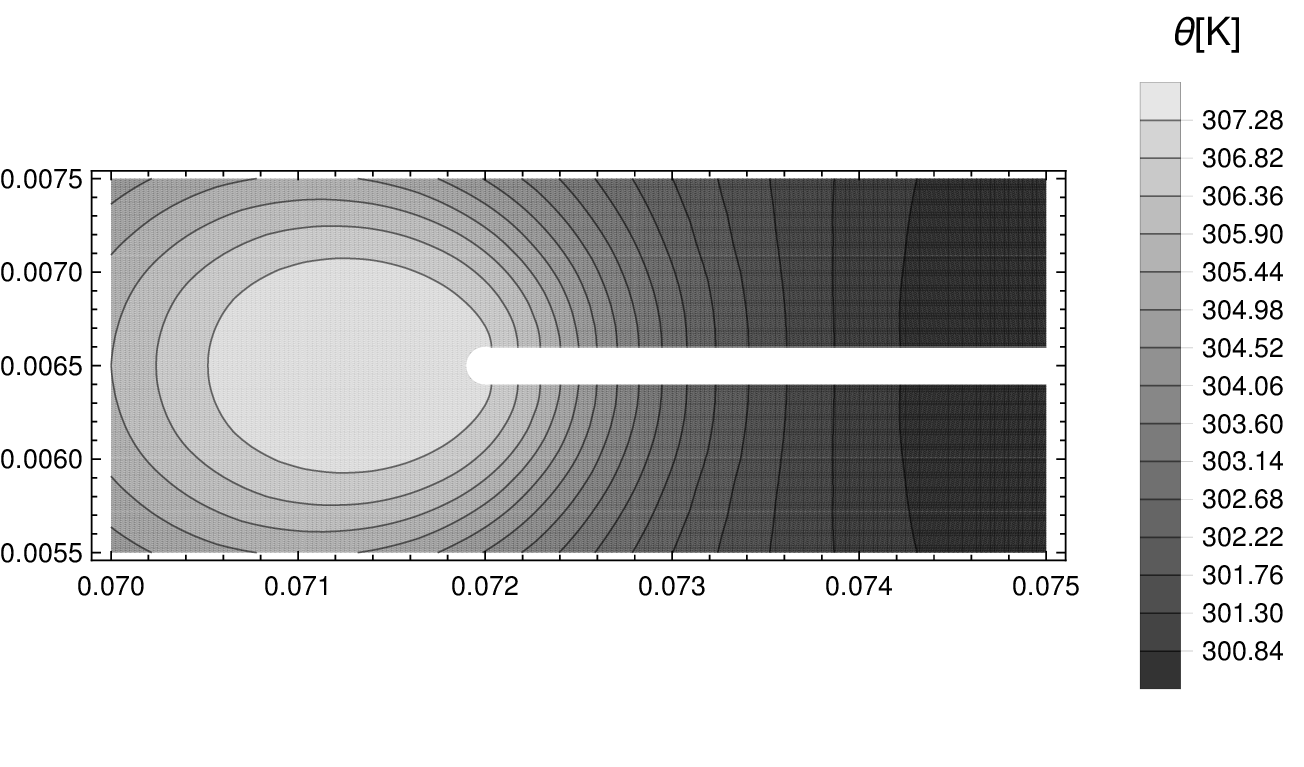}}
  \qquad
  \subfloat[$t=10$]{\includegraphics[width=0.35\textwidth]{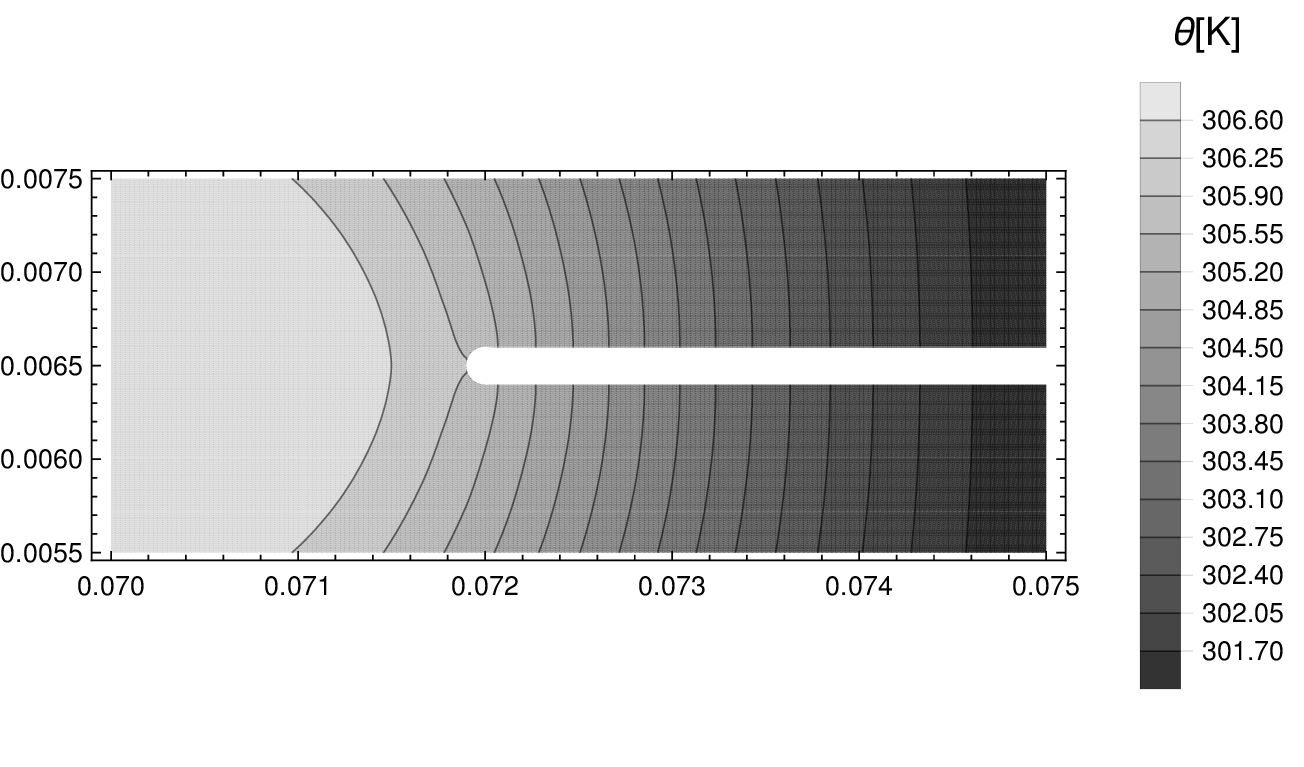}}
  \caption{Temperature field close to the cutout tip, \emph{temperature dependent} elastic moduli, \emph{xx-large viscosity} value. Nomenclature for viscosity values is described in Table~\ref{tab:parameter-values-viscosity}, remaining material parameters are given in Table~\ref{tab:parameter-values} and Table~\ref{tab:parameter-values-elastic}.}
  \label{fig:temperature-field-whole-specimen-linear-parameters-xx-large-viscosity}
\end{figure}

\FloatBarrier

\subsection{Oscillatory loading--unloading}
\label{sec:oscill-load-unlo}
\begin{figure}[h]
  \subfloat[Measurement site A.]{\includegraphics[width=0.32\textwidth]{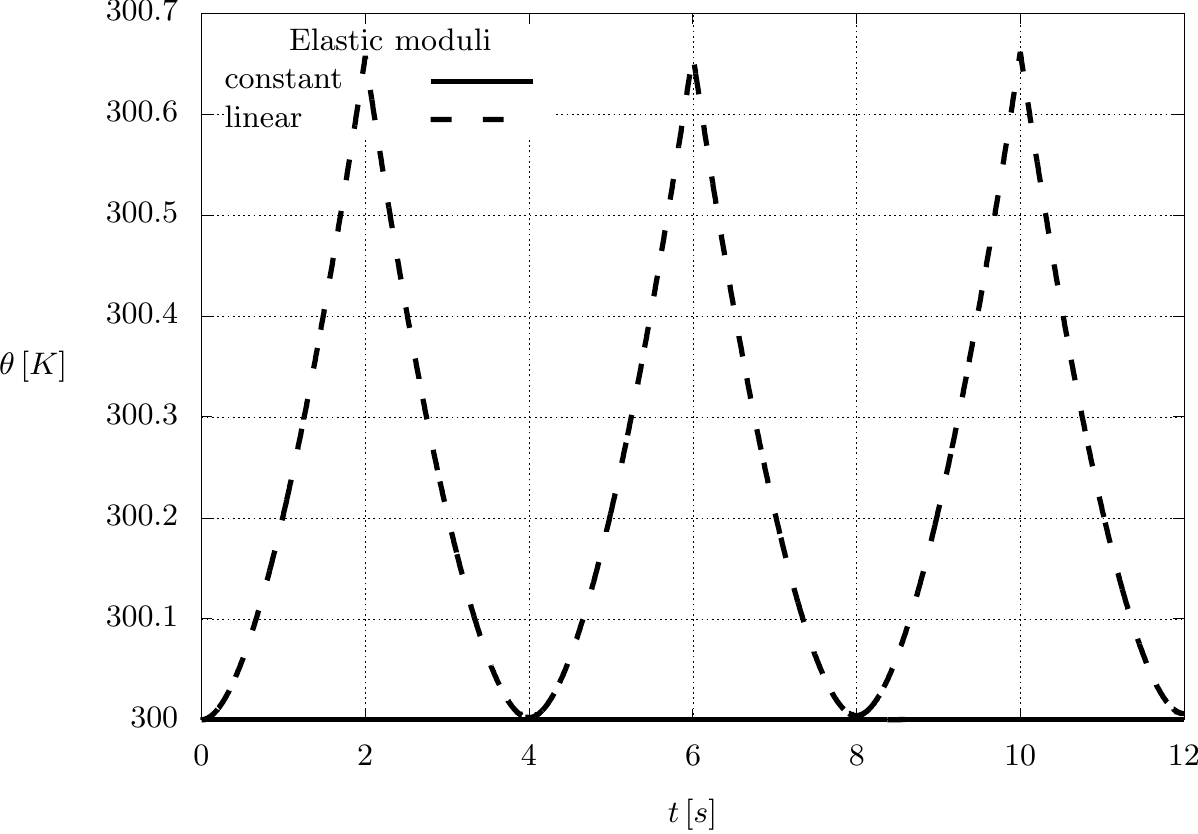}}
  \quad
  \subfloat[Measurement site B.]{\includegraphics[width=0.32\textwidth]{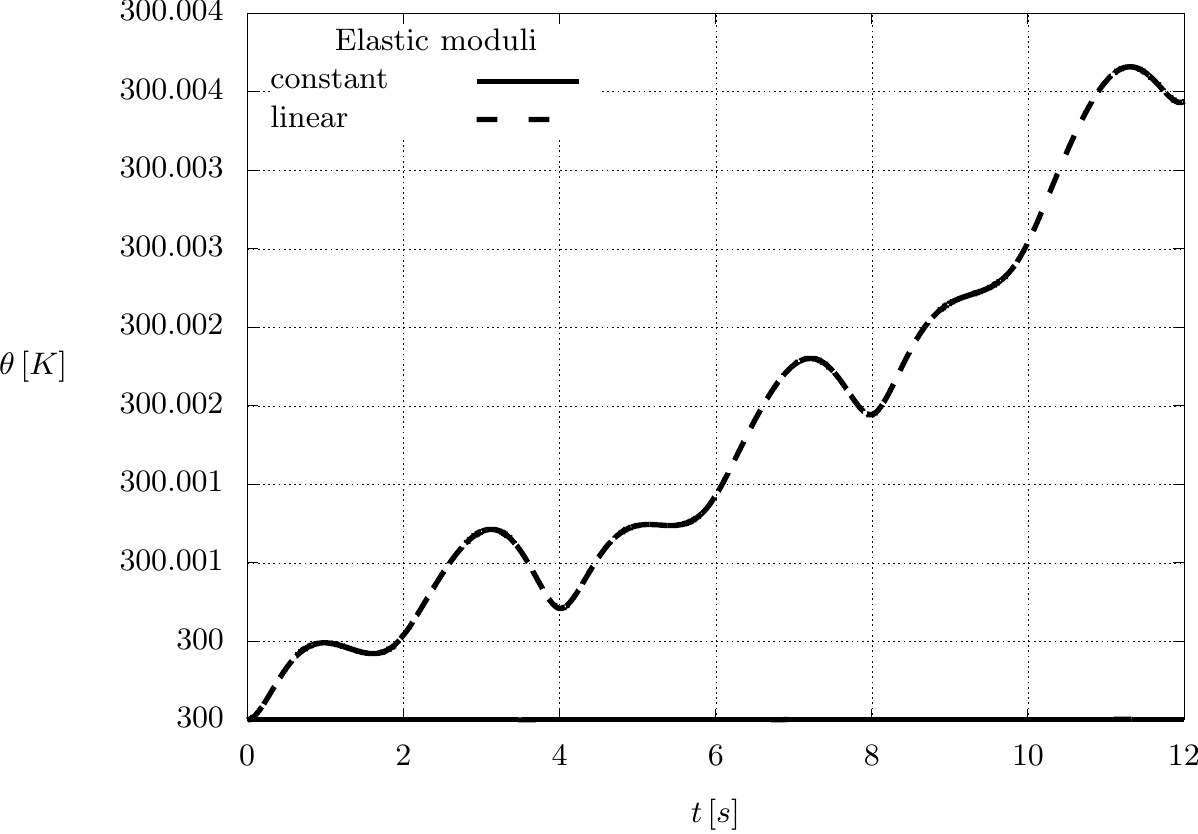}}
  \quad
  \subfloat[Measurement site C.]{\includegraphics[width=0.32\textwidth]{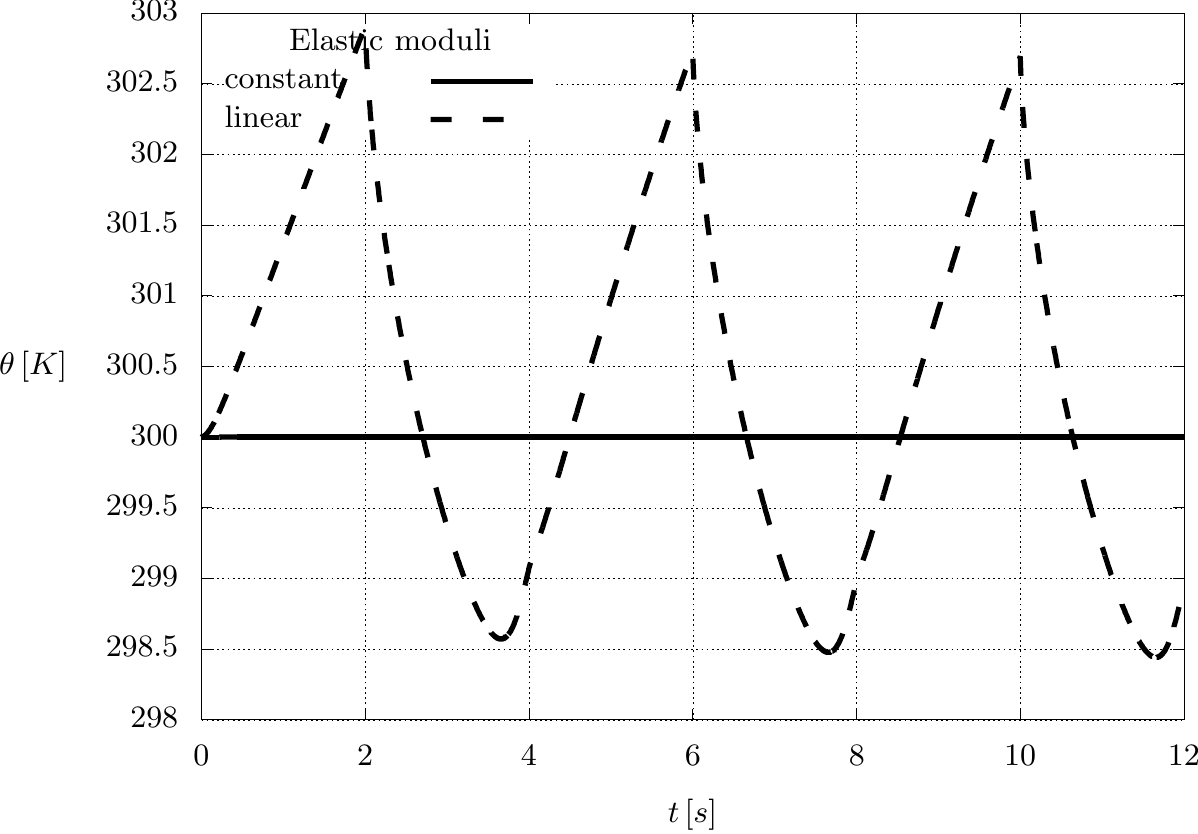}}
  \caption{Oscillatory loading--unloading; temperature values at given measurement sites. Comparison of models with \emph{constant} elastic moduli and \emph{temperature dependent} elastic moduli; \emph{tiny viscosity}. Nomenclature for viscosity values is described in Table~\ref{tab:parameter-values-viscosity}, remaining material parameters are given in Table~\ref{tab:parameter-values} and Table~\ref{tab:parameter-values-elastic}.}
  \label{fig:temperature-sites-comparison-linear-constant-oscillations}
\end{figure}

The results regarding the oscillatory loading--unloading protocol are in line with the findings regarding the single loading--unloading protocol. First we observe that the difference between the models with \emph{constant elastic moduli} and models with \emph{temperature dependent} (linear) elastic moduli remains substantial, see~Figure~\ref{fig:temperature-sites-comparison-linear-constant-oscillations}.

\begin{figure}[h]
  \subfloat[Measurement site A.]{\includegraphics[width=0.32\textwidth]{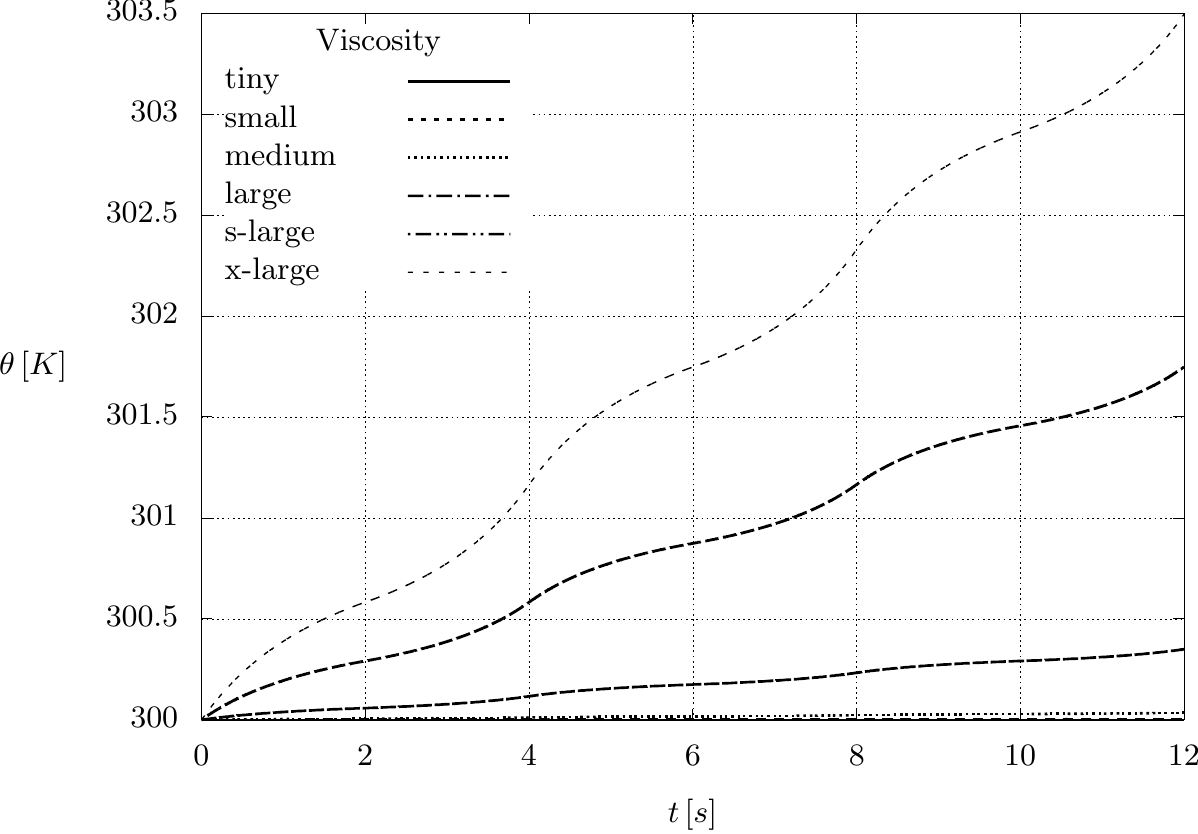}}
  \quad
  \subfloat[Measurement site B.]{\includegraphics[width=0.32\textwidth]{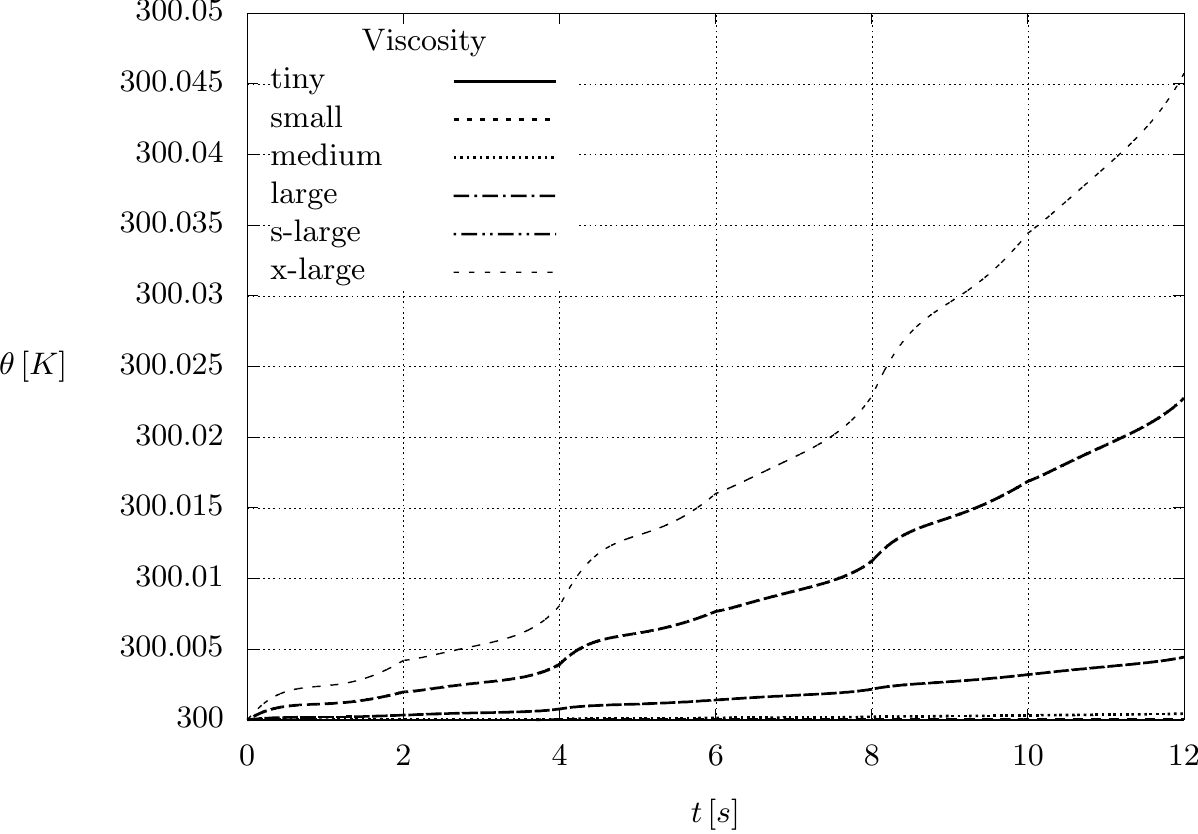}}
  \quad
  \subfloat[Measurement site C.]{\includegraphics[width=0.32\textwidth]{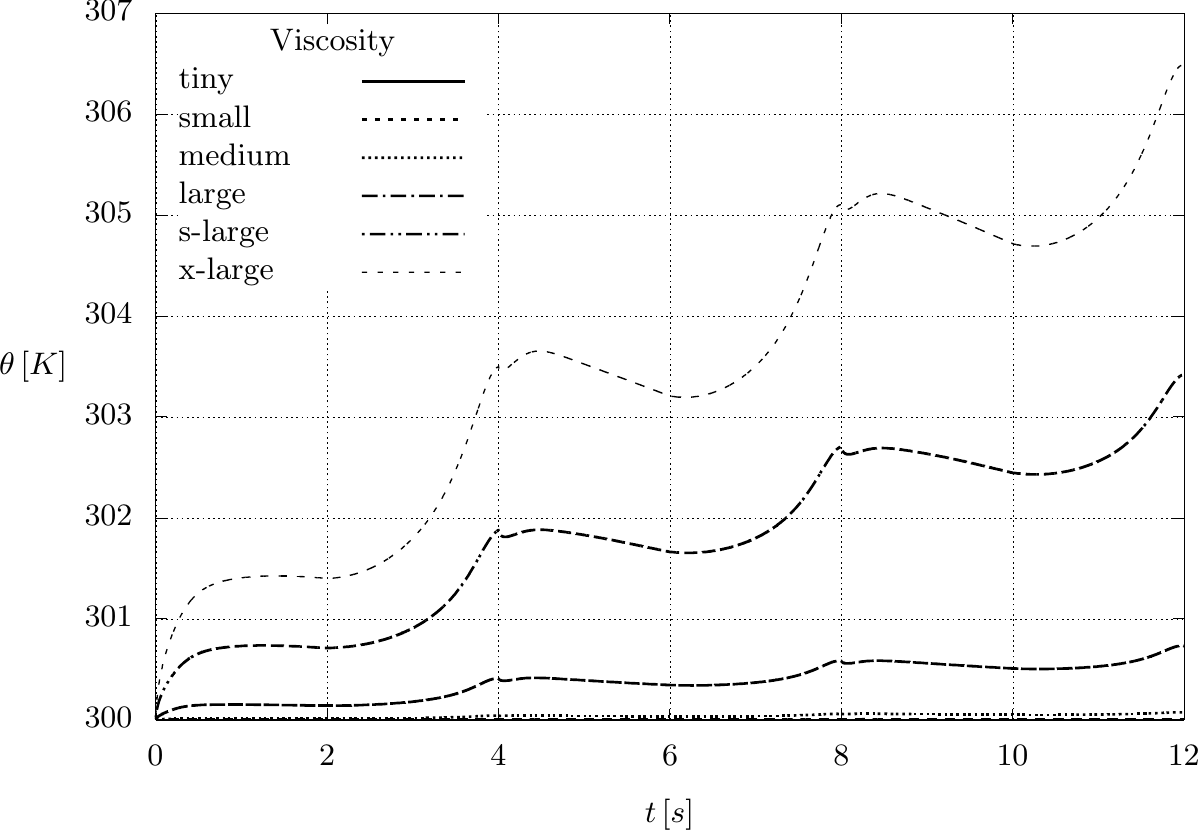}}
  \caption{Oscillatory loading--unloading; temperature values at given measurement sites. Model with \emph{constant} elastic moduli, comparison of models with different viscosity values. Nomenclature for viscosity values is described in Table~\ref{tab:parameter-values-viscosity}, remaining material parameters are given in Table~\ref{tab:parameter-values} and Table~\ref{tab:parameter-values-elastic}.}
  \label{fig:temperature-sites-constant-oscillations}
\end{figure}

\begin{figure}[h]
  \subfloat[Measurement site A.\label{fig:temperature-sites-linear-oscillations-a}]{\includegraphics[width=0.32\textwidth]{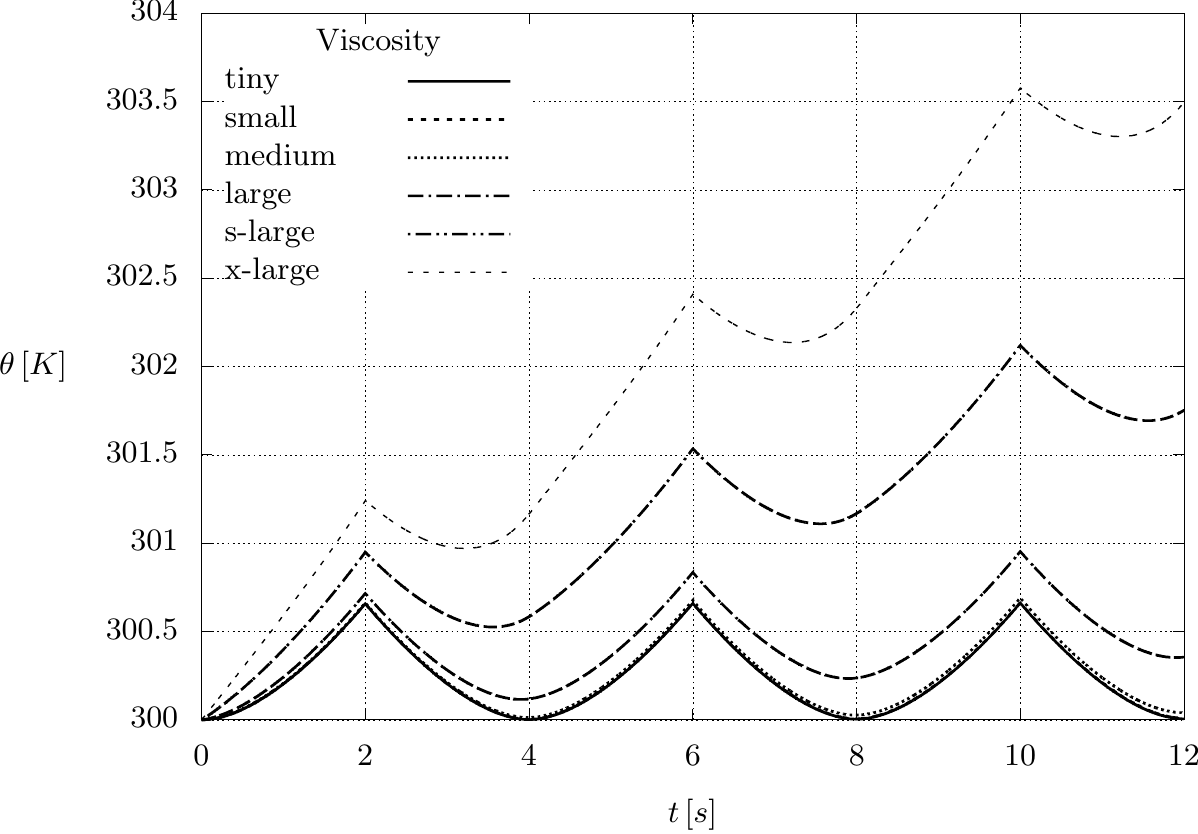}}
  \quad
  \subfloat[Measurement site B.]{\includegraphics[width=0.32\textwidth]{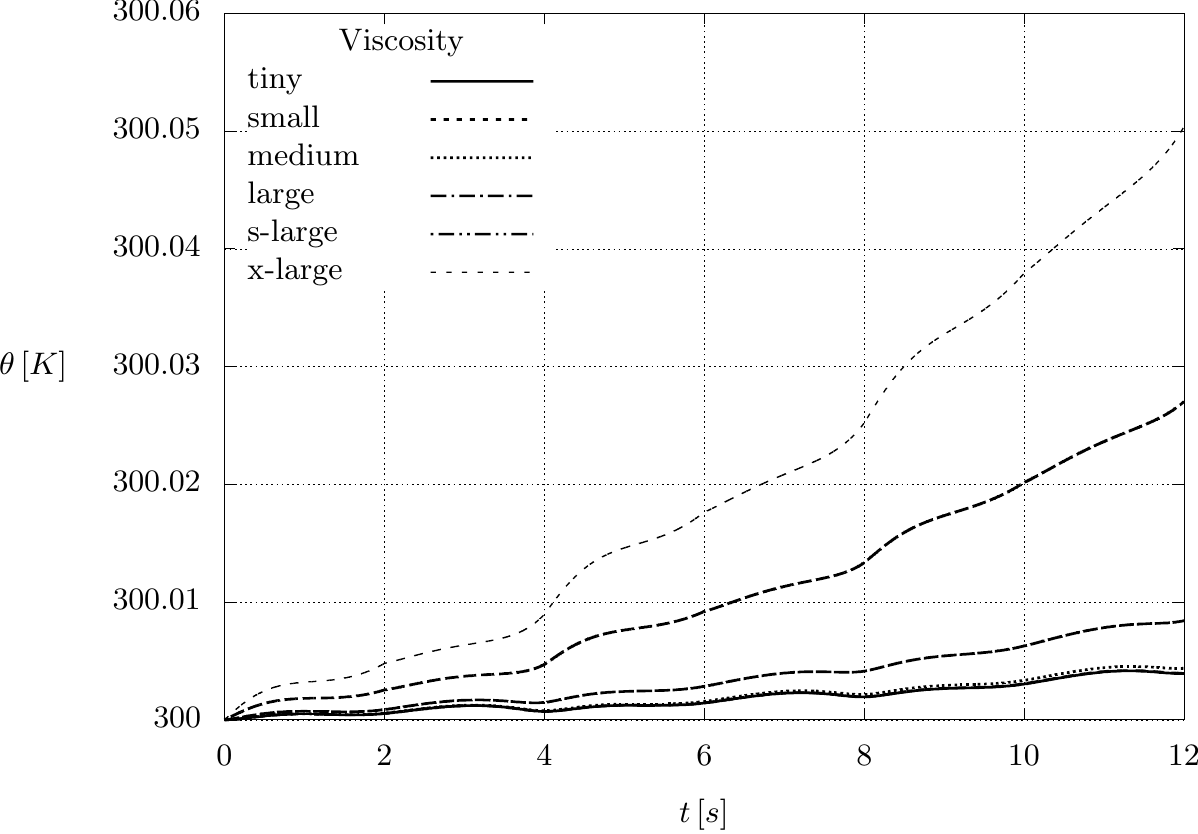}}
  \quad
  \subfloat[Measurement site C.\label{fig:temperature-sites-linear-oscillations-c}]{\includegraphics[width=0.32\textwidth]{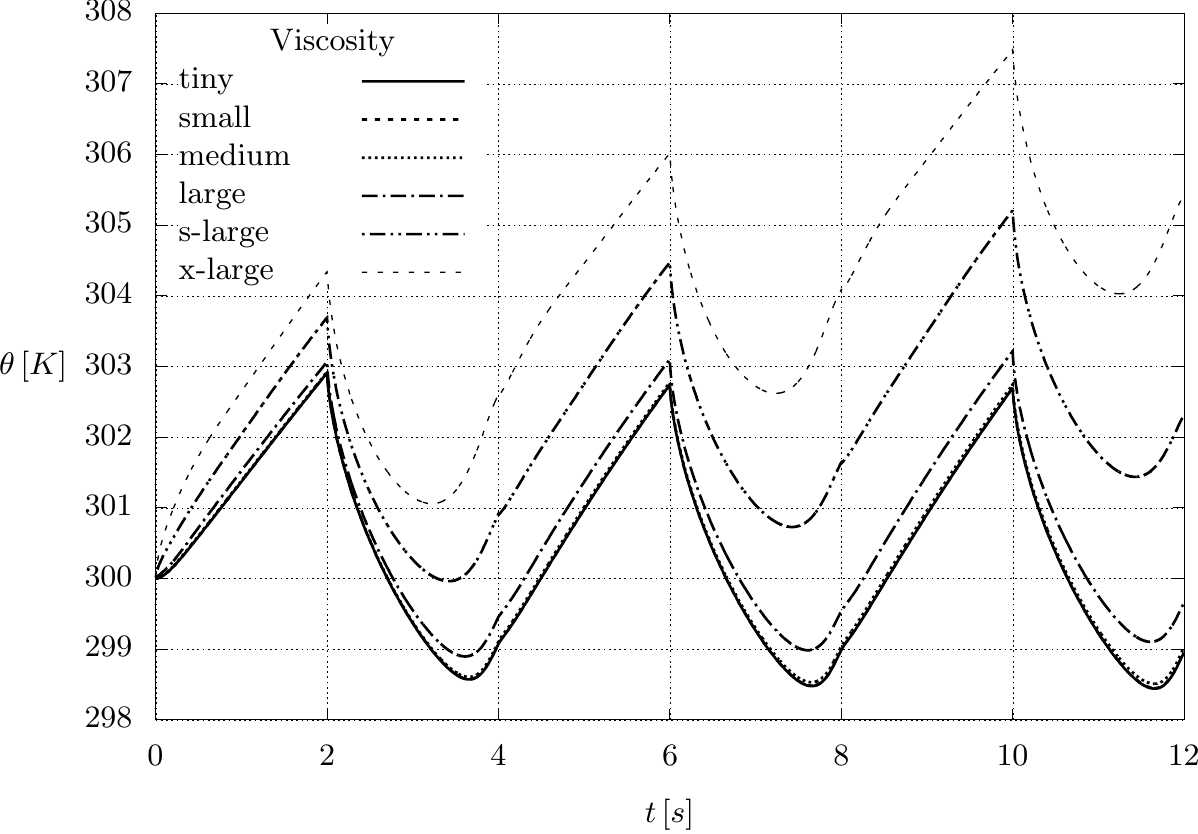}}
  \caption{Oscillatory loading--unloading; temperature values at given measurement sites. Model with \emph{temperature dependent} elastic moduli, comparison of models with different viscosity values. Nomenclature for viscosity values is described in Table~\ref{tab:parameter-values-viscosity}, remaining material parameters are given in Table~\ref{tab:parameter-values} and Table~\ref{tab:parameter-values-elastic}.}
  \label{fig:temperature-sites-linear-oscillations}
\end{figure}

The oscillatory nature of the loading--unloading however leads to the more pronounced temperature increase even for small viscosity values, see Figure~\ref{fig:temperature-sites-constant-oscillations} and Figure~\ref{fig:temperature-sites-linear-oscillations}. This is an expected result. Since we are working with a thermally isolated system, the thermal energy gained by the dissipation can not flow out of the system, hence the temperature of the sample must in average grow in time. In other words each loading--unloading cycle leads to the dissipation and the generated heat is gradually accumulated in the sample. The predicted temperature can therefore substantially increase even in the case of models with constant material moduli and low viscosity---provided that the sample undergoes high number of loading--unloading cycles.   

In Figure~\ref{fig:temperature-sites-linear-oscillations}, see especially Figure~\ref{fig:temperature-sites-linear-oscillations-a} and Figure~\ref{fig:temperature-sites-linear-oscillations-c}, we can see that if the viscosity is small, that is if we are dealing with almost purely elastic material, then the temperature is almost periodic in time. This corresponds to the fact that the processes in the elastic material are ``reversible''. (We see only conversion of thermal and mechanical energy, and no entropy is generated.) 

\section{Conclusion}
\label{sec:conclusion}
Using numerical simulations we have investigated temperature field in the vicinity of a cutout in a viscoelastic solid undergoing loading--unloading process, while the viscoelastic solid has been described by a generalised Kelvin--Voigt type model. The numerical simulations allowed us to monitor the temperature field and the (finite) deformation of the material sample.

For the material parameter values investigated in this study we have found that the models with temperature dependent elastic moduli predict \emph{substantial temperature changes in the vicinity of the cutout tip}. This observation has been found true even for models with temperature dependent elastic moduli and \emph{almost no viscosity}. On the other hand the models with constant elastic moduli predict substantial temperature changes only for high viscosity values. The temperature dependent material moduli that are necessary for the modelling of the classical Gough--Joule effect are therefore also essential in the modelling of the temperature field evolution in viscoelastic bodies, however in the viscoelastic bodies the effects due to temperature dependent elastic moduli are coupled with the effects due to viscosity. In particular, we have found that if the viscosity is high enough, the interplay between elastic/viscous effects can be quite complex and sensitive to the choice of parameter values, see for example Section~\ref{sec:comp-temp-valu} and the discussion of the \emph{temperature drop} effect.

The numerical simulations have shown that the \emph{strongest temperature variations are localised in the vicinity of the cutout tip}, and one might speculate that this feature is common to even more complex viscoelastic rate-type models or for that matter to more complex models for an inelastic response of solids. (See for example models introduced in~\cite{rajagopal.kr.srinivasa.ar:implicit*1}, \cite{devendiran.vk.mohankumar.kv.ea:thermodynamically} and \cite{devendiran.vk.mohankumar.kv.ea:validation}, \cite{bustamante.r.rajagopal.kr.ea:implicit*1,bustamante.r.rajagopal.kr.ea:implicit}  to name a few recent contributions in this field. Regarding the heat transport in viscoelastic solids see also \cite{nieto-simavilla.d.schieber.jd.ea:evidence} and~\cite{venerus.dc.simavilla.dn.ea:thermal}.) This finding is in qualitative agreement with the available experimental data by~\cite{martinez.jrs.toussaint.e.ea:heat}, whose experimental work provided us motivation for the current computational study.

Interestingly, material behaviour at the cutout tip has been---in the isothermal regime---subject to intensive research in solid mechanics because the cutout tip is the place where the crack(s) typically start to grow. (See for example~\cite{kulvait.malek.j.ea:anti-plane}, \cite{bridges.c.rajagopal.kr:implicit}, \cite{zappalorto.m.berto.f.ea:on} or \cite{shyamkumar.r.mohankumar.kv.ea:stress} and \cite{alagappan.p.kannan.k.ea:on,alagappan.p.rajagopal.kr.ea:damage} for some recent contributions in this field.) The fact that the viscoelastic rate-type solids can generate substantial amount of heat at the cutout tip, or in other words produce substantial amount of entropy at the cutout tip, might be of interest in these investigations as well; in this context see especially recent investigations by~\cite{naderi.m.amiri.m.ea:on} and \cite{hajshirmohammadi.b.khonsari.mm:on}.



\bibliographystyle{chicago}
\bibliography{vit-prusa}

\addtocontents{toc}{\protect\end{multicols}} 
\end{document}